\documentclass{conm-p-l}

\copyrightinfo{2006}{}

\usepackage{graphicx}

\newtheorem{theorem}{Theorem}[section]

\newtheorem{corollary}[theorem]{Corollary}

\theoremstyle{definition}

\theoremstyle{remark}
\newtheorem{remark}[theorem]{Remark}

\numberwithin{equation}{section}

\newcommand{\hk}{\hat{k}}
\newcommand{\Z}{\mathbb{Z}/\{0\}}
\newcommand{\ZZ}{\mathbb{Z}^2/\{0\}}
\newcommand{\ZZZ}{\mathbb{Z}^3/\{0\}}

\newcommand{\e}{\epsilon}
\newcommand{\ve}{\varepsilon}

\newcommand{\tdl}{\tilde{\delta}}

\newcommand{\C}{{\mathcal C}}
\newcommand{\D}{{\mathcal D}}

\renewcommand{\k}{\kappa}
\newcommand{\ga}{\gamma}
\newcommand{\Ga}{\Gamma}

\newcommand{\Dl}{\Delta}
\renewcommand{\th}{\theta}
\newcommand{\ra}{\rightarrow}
\newcommand{\lra}{\longrightarrow}
\newcommand{\al}{\alpha}
\newcommand{\be}{\beta}
\newcommand{\sg}{\sigma}
\newcommand{\Sg}{\Sigma}

\newcommand{\pa}{\partial}

\newcommand{\la}{\lambda}
\newcommand{\bq}{\bar{q}}

\newcommand{\nid}{\noindent}

\newcommand{\om}{\omega}
\newcommand{\Om}{\Omega}
\newcommand{\na}{\nabla}

\newcommand{\tom}{\tilde{\omega}}

\renewcommand{\O}{{\mathcal O}}

\newcommand{\vphi}{\varphi}

\newcommand{\non}{\nonumber}

\newcommand{\tk}{\tilde{k}}

\begin{document}

\title[Navier-Stokes and Euler Equations]
{On the Dynamics of Navier-Stokes and Euler Equations}

\author{Yueheng Lan}
\address{Department of Chemistry, University of North Carolina, 
Chapel Hill, NC 27599-3290 USA}
\curraddr{}
\email{ylan@email.unc.edu}
\thanks{}

\author{Y. Charles Li}
\address{Department of Mathematics, University of Missouri, 
Columbia, MO 65211, USA}
\curraddr{}
\email{cli@math.missouri.edu}
\thanks{}

\subjclass{Primary 35, 37; Secondary 34, 46}
\date{}

\dedicatory{}


\begin{abstract}
This is a rather comprehensive study on the dynamics of Navier-Stokes and 
Euler equations via a combination of analysis and numerics. We focus upon two 
main aspects: (a). zero viscosity limit of the spectra of linear Navier-Stokes operator,
(b). heteroclinics conjecture for Euler equation, its numerical verification, 
Melnikov integral, and simulation and control of chaos. Besides Navier-Stokes and 
Euler equations, we also study two models of them.   
\end{abstract}

\maketitle

\tableofcontents








\section{Introduction}

The dynamics of Navier-Stokes and Euler equations is a challenging problem. In 
particular, such dynamics can be chaotic or turbulent. The main challenge 
comes from the large dimensionality of the phase space where the Navier-Stokes 
and Euler equations pose extremely intricate flows. Here the dynamics we refer to 
is the so-called Eulerian, in contrast to the so-called Lagrangian, dynamics of 
fluids. The Eulerian dynamics is a dynamics in an infinite dimensional phase space
(e.g. a Banach space) posed by the Cauchy problem of either the Navier-Stokes or the 
Euler equations as partial differential equations. The Lagrangian dynamics of 
fluid particles is a dynamics of a system of two or three ordinary differential 
equations with vector fields given by fluid velocities. The Lagrangian dynamics of 
fluid particles in three dimensions can be chaotic even when the Eulerian dynamics is 
steady (e.g. the ABC flow \cite{Dom86}). Nevertheless, as shown in Appendix B, 
the Lagrangian dynamics of 2D inviscid fluid particles is always integrable. 

Even though the global well-posedness of 3D Navier-Stokes and Euler equations 
is still an interesting open mathematical problem, 3D Navier-Stokes and Euler 
equations have local well-posedness which is often enough for a dynamical system 
study in the phase space. To begin such a dynamical system study, one needs to 
understand the spectra of the linear Navier-Stokes and/or Euler operators \cite{Li05}.
The spectra of the linear Navier-Stokes operators consist of eigenvalues, 
whereas the spectra of the linear Euler operators contain continuous spectra. 
Existence of invariant manifolds can be proved for Navier-Stokes equations \cite{Li05},
but is still open for Euler equations. The size of the invariant manifolds for 
Navier-Stokes equations tends to zero in the zero viscosity limit \cite{Li05}.
We find that 
the spectra of the linear Navier-Stokes and Euler operators can be classified into four 
categories in the zero viscosity limit: 
\begin{enumerate}
\item {\em Persistence:} These are the eigenvalues that persist and approach 
to the eigenvalues of the corresponding linear Euler operator when the viscosity 
approaches zero.
\item {\em Condensation:} These are the eigenvalues that approach and form 
a continuous spectrum for the corresponding linear Euler operator when the viscosity 
approaches zero.
\item {\em Singularity:} These are the eigenvalues that approach to a 
set that is not in the spectrum of the corresponding linear Euler operator when 
the viscosity approaches zero.
\item {\em Addition:} This is a subset of the spectrum of the linear Euler operator,
which has no overlap with the zero viscosity limit set of the spectrum of the linear 
NS operator.
\end{enumerate} 
We also find that as the viscosity approaches zero, the spectrum of the linear Navier-Stokes 
operator undergoes a fascinating deformation. Focusing upon the persistent unstable 
eigenvalue, we propose a heteroclinics conjecture, i.e. there should be a heteroclinic 
orbit (in fact heteroclinic cycles) associating to the instability for Euler equation.
We will present both analytical and numerical study upon this heteroclinics conjecture. 
Then we conduct a Melnikov integral calculation along the numerically obtained 
approximate heteroclinic orbit. We also compare the Melnikov prediction with the 
numerical simulation and control of chaos for the Navier-Stokes equations. Numerically
we mainly use the Liapunov exponent as a measure of chaos. In some case, we also 
plot the Poincar\'e return map. We realize that the size of Galerkin truncations for 
the full Navier-Stokes equations is limited by the computer ability. Thus we propose 
two simpler models of the Navier-Stokes equations. For the so-called line model, we 
obtain numerically exact heteroclinic orbits for any size of Galerkin truncations. 
We also realize that due to viscosity, the chaos in Galerkin truncations of 
Navier-Stokes equations is often 
transient chaos, i.e. the chaos has a finite life time. Infinite life time chaos can 
be observed in Galerkin truncations of Euler equations.

Chaos and turbulence have no good averages \cite{Li06i}. The matter is more fundamental 
than just poor understanding of averages. The very mechanism of chaos leads to the 
impossibility of a good average \cite{Li04}. On the other hand, chaos and turbulence are 
ubiquitous. In high dimensional systems, there exists tubular chaos \cite{Li04} 
\cite{Li03a} \cite{Li03b} \cite{Li04c} which further confirms that there is no good 
average. The hope is that chaos and turbulence can be controlled. Two aspects of control 
are practically important in applications: Taming and enhancing. When an airplane meets 
turbulence, it will be nice, safer and economic if we can tame the turbulence into a 
laminar flow or a less turbulent flow \cite{ABT01} \cite{Kim03} \cite{LB98}. In a 
combustor, enhancing turbulence can get the fuel and oxidant mixed and burned more 
efficiently \cite{LB98} \cite{Kim03}. Theoretically, one can also make use of the 
ergodicity of chaos to gear an orbit to a specific target \cite{OGY90}. Many other 
possibilities of applications of control can be designed too. An advantage of the 
control is that it can be done in a trial-correction manner without a detailed knowledge 
of turbulence.

Clearly, control of chaos and turbulence has great industrial value. From a mathematical 
point of view, the question is how much mathematis is in this control theory. So far, the 
mathematical merit of the theory of control of chaos is not nearly as great as proving 
the existence of chaos \cite{Li04}. Obviously, a lot of good numerics is in this control 
theory. In this article, we will address this control theory from a mathematical 
perspective, and try to formulate some good mathematical problems. One can add a control 
to any equation. But the only meaningful controls are the ones that are practical. 
Consider the 3D Navier-Stokes equations for example
\[
u_{i,t} +u_j u_{i,j} = -p_{,i} + \e  u_{i,jj} + f_i + \C_i \ ,
\]
defined on a spatial domain $\D$ with appropriate boundary conditions, where $\e =1 / 
\text{Re}$ is the inverse of the
Reynolds number, and $f_i=f_i(t,x)$ is the external force. Assume that without the 
control $\C_i$, the solutions are turbulent. The goal is to find a practical control 
to either tame or enhance turbulence. For instance, a practical control $\C_i = 
\C_i(t,x)$ should be spatially localized (perhaps near the boundary).

Recently, there has been quite amount of works on numerical simulations of chaos in 
Navier-Stokes equations \cite{FS95a} \cite{FS95b} \cite{KK01} \cite{MCH05} \cite{TOAG05} 
\cite{SYE06}. Here we try to combine numerics with analysis in terms of Melnikov integrals. 
Unlike the sine-Gordon system studied in Appendix A, analytical calculation of the 
Melnikov integrals is not feasible at this moment for Navier-Stokes equations. So we 
will resort to numerical calculations. It is an interesting open mathematical problem 
that whether or not 2D Euler equation is integrable as a Hamiltonian system in the 
Liouville sense. Since 2D Euler equation possesses infinitely many constants of motion,
it is tempting to conjecture that 2D Euler equation is integrable. Another support to 
such a conjecture is that both 2D and 3D Euler equations have Lax pairs \cite{Li01}
\cite{LY03} \cite{Li05L}. In fact, it is even rational to conjecture that 3D Euler equations
are integrable. As mentioned above, we propose the so-called heteroclinics conjecture
for Euler equations, i.e. there exist heteroclinic cycles for Euler equations. We numerically 
simulate the heteroclinic orbits and use the numerical results to conduct numerical calculations 
on Melnikov integrals. In these numerical simulations, it is crucial to
make use of known results on the spectra of linear 
Navier-Stokes and Euler operators \cite{Li00} \cite{LLM04} \cite{Li05}. We use the 
numerical Melnikov integral as a tool for both predicting and controling chaos. As a 
measure of chaos, we calculate the Liapunov exponents. We find that the calculated 
Liapunov exponents depend on the computational time interval and the precision of the 
computation. Nevertheless, as a measure of chaos, Liapunov exponents prove to be very robust.
Since the chaos is often transient, we make comparison on the base of fixed time interval
and fixed precision of computation.

Our numerics resorts to Galerkin truncations. But Galerkin truncations are somewhat 
singular perturbations of Euler equations. Higher single Fourier modes have more unstable 
eigenvalues. Therefore, it is difficult to derive dynamical pictures for Euler equations
from their Galerkin truncations. On the other hand, higher single Fourier modes have more
dissipation under Navier-Stokes flows. So Galerkin truncations perform better for 
Navier-Stokes equations than Euler equations. Today's computer ability still limits the 
size of the Galerkin truncations. With better future computer ability, Galerkin truncations
will paint better and better pictures of Navier-Stokes and Euler equations. It seems also 
important to design special models that can picture special aspects of the dynamics of 
Navier-Stokes and Euler equations.

\section{Zero Viscosity Limit of the Spectrum of 2D Linear Navier-Stokes Operator}

We will study the following form of 2D Navier-Stokes (NS) equation with a control,
\begin{equation}
\pa_t \Om + \{ \Psi, \Om \} = \e [\Dl \Om + f(t,x) +b\tdl (x)] \ ,
\label{2DNS}
\end{equation}
where $\Om$ is the vorticity which is a real scalar-valued function
of three variables $t$ and $x=(x_1, x_2)$, the bracket $\{\ ,\ \}$ 
is defined as
\[
\{ f, g\} = (\pa_{x_1} f) (\pa_{x_2}g) - (\pa_{x_2} f) (\pa_{x_1} g) \ ,
\]
where $\Psi$ is the stream function given by,
\[
u_1=- \pa_{x_2}\Psi \ ,\ \ \ u_2=\pa_{x_1} \Psi \ ,
\]
the relation between vorticity $\Om$ and stream 
function $\Psi$ is,
\[
\Om =\pa_{x_1} u_2 - \pa_{x_2} u_1 =\Dl \Psi \ ,
\]
and $\e = 1/\text{Re}$ is the inverse of the Reynolds number, $\Dl$ is the 
2D Laplacian, $f(t,x)$ is the external force, $b\tdl (x)$ is the spatially 
localized control, and $b$ is the control parameter. We pose the periodic 
boundary condition
\[
\Om (t, x_1 +2\pi , x_2) = \Om (t, x_1 , x_2) = \Om (t, x_1, x_2 +2\pi /\al ),
\]
where $\al$ is a positive constant, i.e. the 2D NS is defined on the 2-torus $\mathbb{T}^2$. 
We require that $\Psi$, $f$ and $\tdl$ have mean zero
\[
\int_{\mathbb{T}^2} \Psi dx = \int_{\mathbb{T}^2} f dx = 
\int_{\mathbb{T}^2} \tdl dx = 0\ .
\]
Of course $\Om$ always has zero mean. In this case, $\Psi = \Dl^{-1} \Om $.

In both 2D and 3D, the linear NS operator obtained by linearizing NS 
at a fixed point has only point spectrum consisting of 
eigenvalues lying in a parabolic region \cite{Li05}. On the other hand, the 
corresponding linear Euler can have continuous spectrum besides 
point spectrum \cite{Li05}. The spectra of the linear NS and Euler operators can 
be classified into four classes in the zero viscosity limit:
\begin{enumerate}
\item {\em Persistence:} These are the eigenvalues that persist and approach 
to the eigenvalues of the corresponding linear Euler operator when the viscosity 
approaches zero.
\item {\em Condensation:} These are the eigenvalues that approach and form 
a continuous spectrum for the corresponding linear Euler operator when the viscosity 
approaches zero.
\item {\em Singularity:} These are the eigenvalues that approach to a 
set that is not in the spectrum of the corresponding linear Euler operator when 
the viscosity approaches zero.
\item {\em Addition:} This is a subset of the spectrum of the linear Euler operator,
which has no overlap with the zero viscosity limit set of the spectrum of the linear 
NS operator.
\end{enumerate} 

\subsection{A Shear Fixed Point}

For the external force $f= \Ga \cos x_1$ ($b=0$), $\Om = \Ga \cos x_1$ is a shear fixed 
point, where $\Ga$ is an arbitrary real nonzero constant. Choose $\al \in (0.5, 0.84)$. There is a 
$\e_* > 0$ such that when $\e > \e_*$, the fixed point has no eigenvalue 
with positive real part, and when $\e \in [0, \e_*)$, the fixed point has 
a unique positive eigenvalue \cite{Li05}. 
Notice that this unique eigenvalue persists even for linear Euler 
($\e =0$). In fact, for linear Euler ($\e =0$), there is a pair of 
eigenvalues, and the other one is the negative of the above eigenvalue. 
Precise statements on such results are given in the theorem below. Later we will 
discover numerically that
some of the rest eigenvalues of the linear Navier-Stokes operator somehow form the 
continuous spectrum of linear Euler ($\e =0$) as $\e \ra 0$, while others
do not converge to the spectrum of linear Euler ($\e =0$) at all \cite{Li05} 
\cite{Li05a}. Using the Fourier series 
\[
\Om = \sum_{k \in \ZZ} \om_k e^{i(k_1 x_1 + \al k_2 x_2)}\ , 
\]
where $\om_{-k} = \overline{\om_k}$ (in fact, we always work in the 
subspace where all the $\om_k$'s are real-valued), one gets
the spectral equation of the linearized 2D Navier-Stokes operator at the 
fixed point $\Om = 2 \cos x_1$, 
\begin{equation}
A_{n-1} \om_{n-1} -\e |\hk +np|^2 \om_n - A_{n+1} \om_{n+1} = \la \om_n \ ,
\label{le}
\end{equation}
where $\hk \in \ZZ$, $p=(1,0)$, $\om_n = \om_{\hk +np}$, $A_n = A(p, \hk +np)$, and
\[
A(q,r) = \frac{\al}{2}\left [ \frac{1}{r_1^2+(\al r_2)^2} - 
\frac{1}{q_1^2+(\al q_2)^2}\right ]\left | \begin{array}{lr} 
q_1 & r_1 \\ q_2 & r_2 \\ \end{array} \right | \ .
\]
(In fact, the $A_n$'s should be counted twice due to switching $q$ and $r$, but the 
difference is only a simple scaling of $\e$ and $\la$.)
Thus the 2D linear NS decouples according to lines labeled by $\hk$. The following 
detailed theorem on the spectrum of the 2D linear NS at the fixed point 
$\Om = 2 \cos x_1$ was proved in \cite{Li05}.
\begin{theorem}[The Spectral Theorem \cite{Li05}]
The spectra of the 2D linear NS operator (\ref{le}) have the following 
properties.
\begin{enumerate}
\item $(\al \hk_2)^2+(\hk_1+n)^2 > 1$, $\forall n \in \Z$. 
When $\e > 0$, there is no eigenvalue of non-negative real part. 
When $\e = 0$, the entire spectrum is the continuous spectrum
\[
\left [ -i\al |\hk_2|, \ i\al |\hk_2| \right ]\ .
\]
\item $\hk_2 = 0$, $\hk_1 = 1$. The spectrum consists of the eigenvalues 
\[
\la = - \e n^2 \ , \quad n \in \Z \ .
\]
The eigenfunctions are the Fourier modes
\[
\tom_{np} e^{inx_1} + \ \mbox{c.c.}\ \ , \quad \forall \tom_{np} \in 
\C\ , \quad n \in \Z \ .
\]
As $\e \ra 0^+$, the eigenvalues are dense on the negative half of the real 
axis $(-\infty, 0]$. Setting $\e =0$, the only eigenvalue is $\la = 0$ of 
infinite multiplicity with the same eigenfunctions as above.
\item $\hk_2 = -1$, $\hk_1 = 0$. (a). $\e >0$. For any $\al \in (0.5, 0.95)$,
there is a unique $\e_*(\al)$,
\begin{equation}
\frac{\sqrt{32-3\al^6-17\al^4-16\al^2}}{2(\al^2+1)(\al^2+4)} < 
\e_*(\al) < \frac{1}{(\al^2+1)} \sqrt{\frac{1-\al^2}{2}}\ ,
\label{nuda}
\end{equation}
where the term under the square root on the left is positive for 
$\al \in (0.5, 0.95)$, and the left term is always less than the right term.
When $\e > \e_*(\al)$, there is no eigenvalue of non-negative real part. 
When $\e = \e_*(\al)$, $\la =0$ is an eigenvalue, and all the rest 
eigenvalues have negative real parts. When $\e < \e_*(\al)$, there is 
a unique positive eigenvalue $\la (\e )>0$, and all the rest 
eigenvalues have negative real parts. $\e^{-1} \la (\e )$ is a strictly 
monotonically decreasing function of $\e$. When $\al \in (0.5, 0.8469)$,
we have the estimate
\begin{eqnarray*}
& & \sqrt{\frac{\al^2(1-\al^2)}{2(\al^2+1)}-\frac{\al^4 (\al^2+3)}{4
(\al^2+1)(\al^2+4)}} - \e (\al^2+1) < \la (\e ) \\
& & < \sqrt{\frac{\al^2(1-\al^2)}{2(\al^2+1)}}- \e \al^2 \ ,
\end{eqnarray*}
where the term under the square root on the left is positive for 
$\al \in (0.5, 0.8469)$.
\[
\sqrt{\frac{\al^2(1-\al^2)}{2(\al^2+1)}-\frac{\al^4 (\al^2+3)}{4
(\al^2+1)(\al^2+4)}} \leq \lim_{\e \ra 0^+} \la (\e )  \leq 
\sqrt{\frac{\al^2(1-\al^2)}{2(\al^2+1)}} \ .
\]
In particular, as $\e \ra 0^+$, $\la (\e ) =\O (1)$.

(b). $\e =0$. When $\al \in (0.5, 0.8469)$, we have only two eigenvalues
$\la_0$ and $-\la_0$, where $\la_0$ is positive,
\[
\sqrt{\frac{\al^2(1-\al^2)}{2(\al^2+1)}-\frac{\al^4 (\al^2+3)}{4
(\al^2+1)(\al^2+4)}} < \la_0 <
\sqrt{\frac{\al^2(1-\al^2)}{2(\al^2+1)}} \ .
\]
The rest of the spectrum is a continuous spectrum $[-i\al , \ i\al ]$.

(c). For any fixed $\al \in (0.5, 0.8469)$,
\begin{equation}
\lim_{\e \ra 0^+} \la (\e ) = \la_0 \ .
\label{pet1}
\end{equation}
\item Finally, when $\e = 0$, the union of all the above pieces of 
continuous spectra is the imaginary axis $i\mathbb{R}$.
\end{enumerate}
\label{PET}
\end{theorem}
\begin{remark}
In the current periodic boundary condition case, viscosity does not destablize 
the flow in contrast to the non-slip boundary condition case \cite{Lin45}. 
The Orr-Sommerfeld equation and Rayleigh equation have the same periodic boundary 
condition in the former case, and different number of boundary conditions in the 
latter case.
\end{remark}
The following invariant manifold theorem of the 2D NS at the fixed point 
$\Om = 2 \cos x_1$ was also proved in \cite{Li05}.
\begin{theorem}[Invariant Manifold Theorem \cite{Li05}]
For any $\al \in (0.5, 0.95)$, and $\e \in (0, \e_*(\al ))$ where 
$\e_*(\al ) > 0$ satisfies (\ref{nuda}), in a neighborhood of $\Om = 2 \cos x_1$ 
in the Sobolev space $H^\ell (\mathbb{T}^2)$ ($\ell \geq 3$),
there are an $1$-dimensional $C^\infty$ unstable manifold and an 
$1$-codimensional $C^\infty$ stable manifold.
\end{theorem}

One of the goals of the work \cite{Li05} is to study the zero viscosity limit of 
the invariant manifolds of the 2D NS. For this study, it is crucial to understand 
the deformation of the linear spectra as $\e \ra 0^+$. Below we will study this 
numerically.

When $\hk_1=0$ and $\hk_2=1$, $\al =0.7$, the unique $\e_*$ in (\ref{nuda}) 
belongs to the interval $0.332 < \e_* < 0.339$,
such that when $\e < \e_*$, a positive eigenvalue appears. We test this criterion 
numerically and find that it is very sharp even when the truncation of 
the linear system (\ref{le}) is as low as $|n| \leq 100$. 
As $\e \ra 0^+$, we tested the truncation of 
the linear system (\ref{le}) up to $|n| \leq 1024$ for $\al =0.7$, the patterns are 
all the same. Below we present the case $|n| \leq 200$ for which the pattern is more clear. 
\begin{figure}
\includegraphics[width=4.0in,height=4.0in]{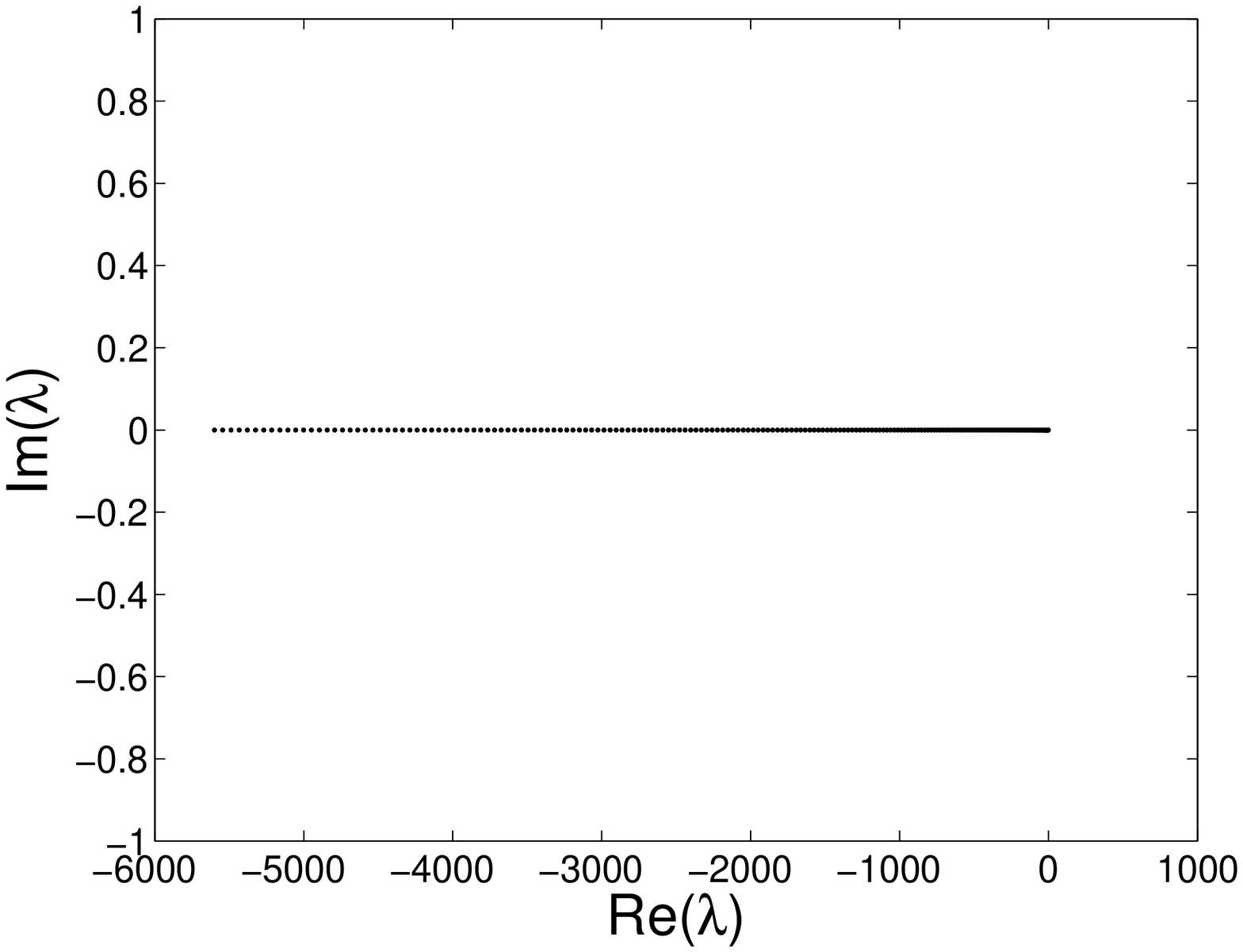}
\caption{The eigenvalues of the linear system (\ref{le}) when $\hk_1=0$ and $\hk_2=1$, 
$\al =0.7$, and $\e =0.14$.}
\label{ge1}
\end{figure}
\begin{figure}
\includegraphics[width=4.0in,height=4.0in]{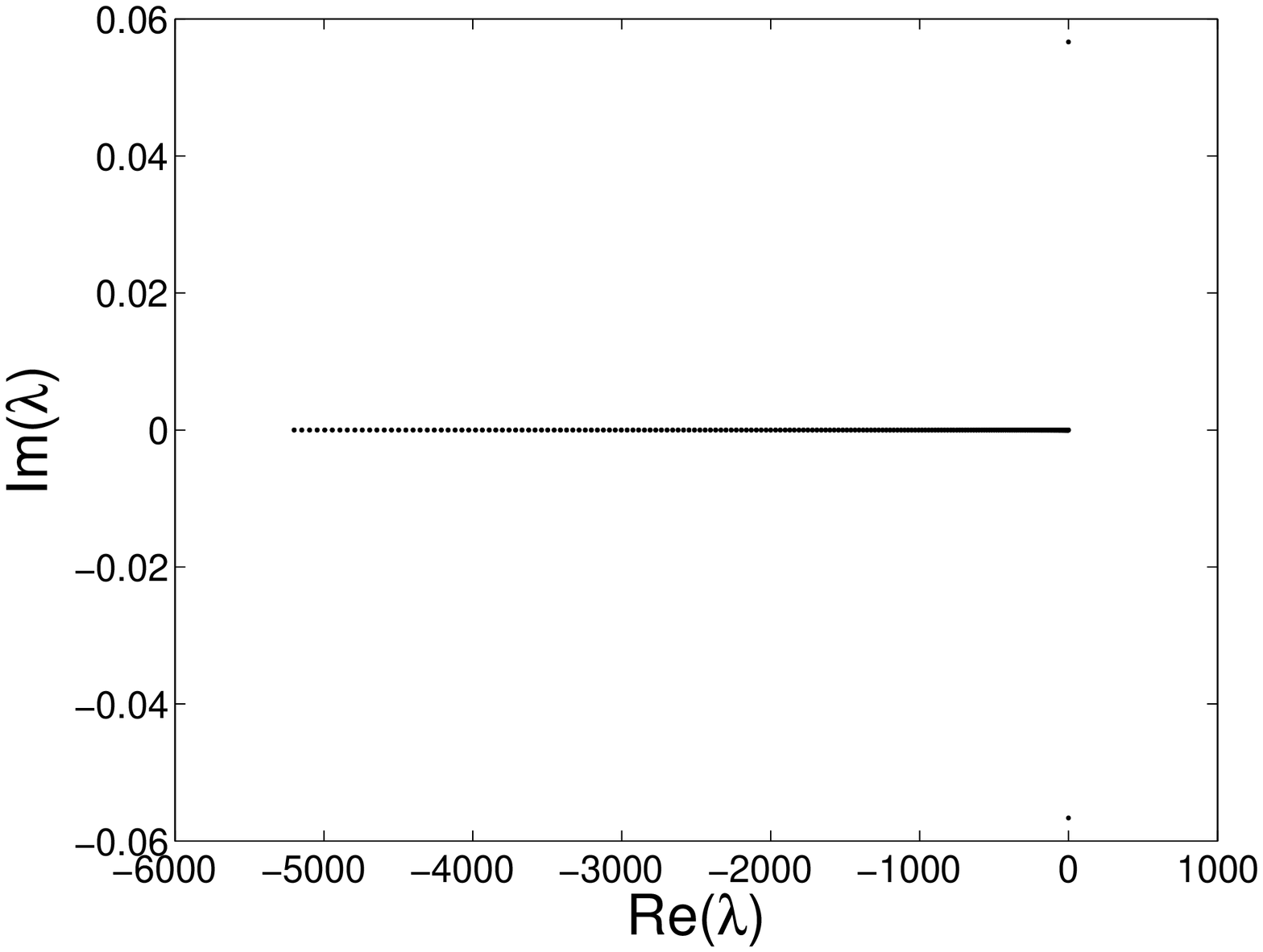}
\caption{The eigenvalues of the linear system (\ref{le}) when $\hk_1=0$ and $\hk_2=1$, 
$\al =0.7$, and $\e =0.13$.}
\label{ge2}
\end{figure}
\begin{figure}
\includegraphics[width=4.0in,height=4.0in]{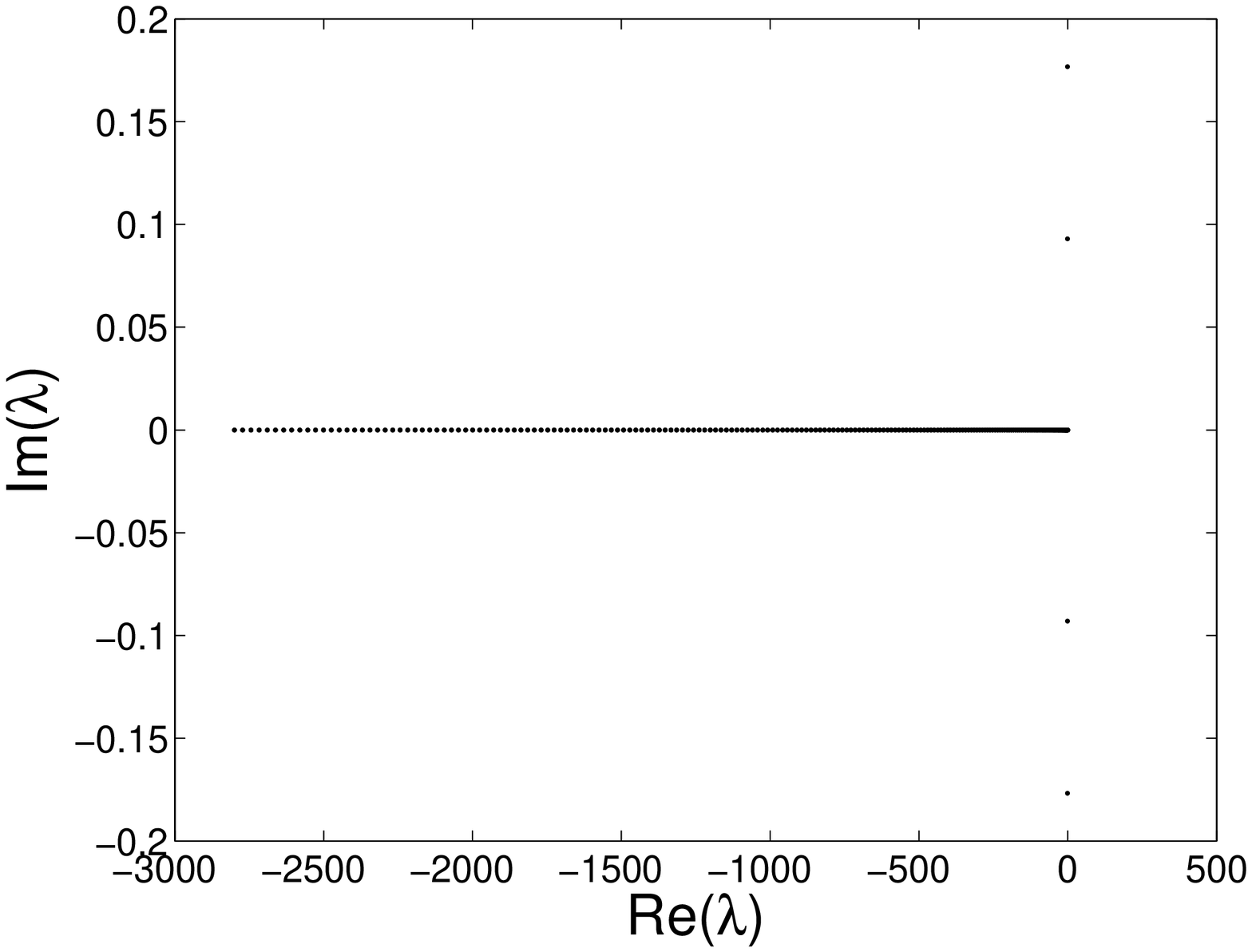}
\caption{The eigenvalues of the linear system (\ref{le}) when $\hk_1=0$ and $\hk_2=1$, 
$\al =0.7$, and $\e =0.07$.}
\label{ge3}
\end{figure}
\begin{figure}
\includegraphics[width=4.0in,height=4.0in]{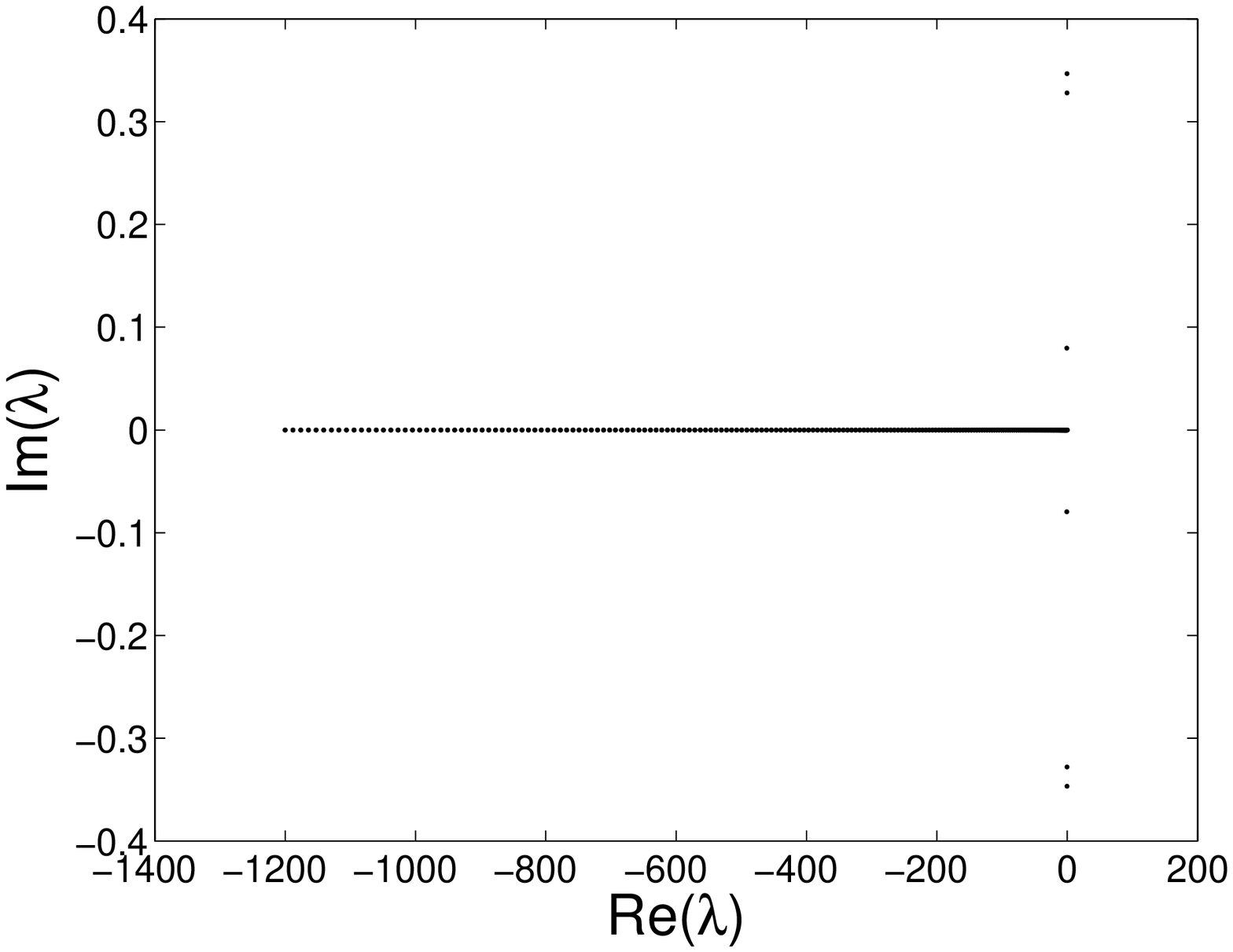}
\caption{The eigenvalues of the linear system (\ref{le}) when $\hk_1=0$ and $\hk_2=1$, 
$\al =0.7$, and $\e =0.03$.}
\label{ge4}
\end{figure}
\begin{figure}
\includegraphics[width=4.0in,height=4.0in]{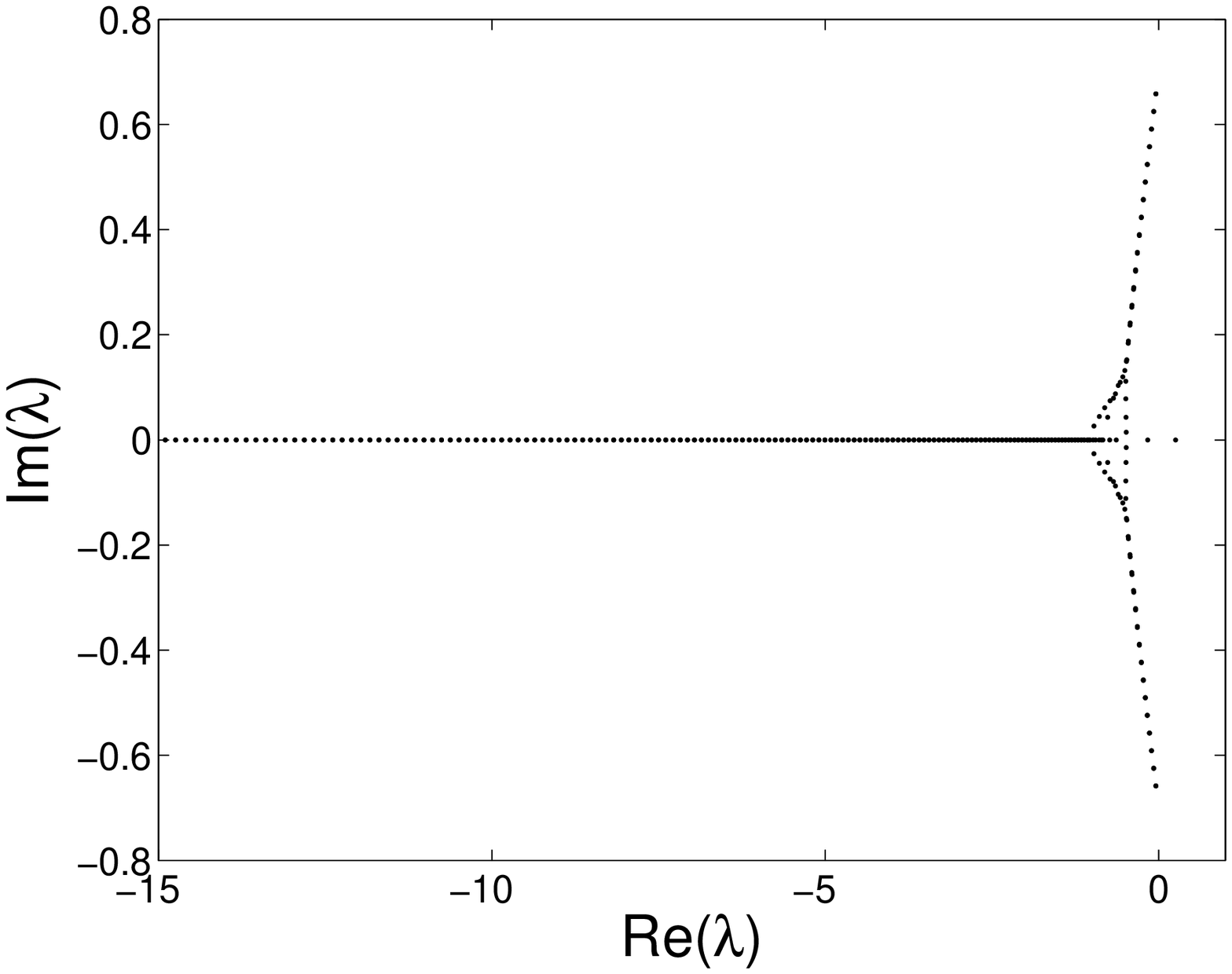}
\caption{The eigenvalues of the linear system (\ref{le}) when $\hk_1=0$ and $\hk_2=1$, 
$\al =0.7$, and $\e =0.0004$.}
\label{ge5}
\end{figure}
\begin{figure}
\includegraphics[width=4.0in,height=4.0in]{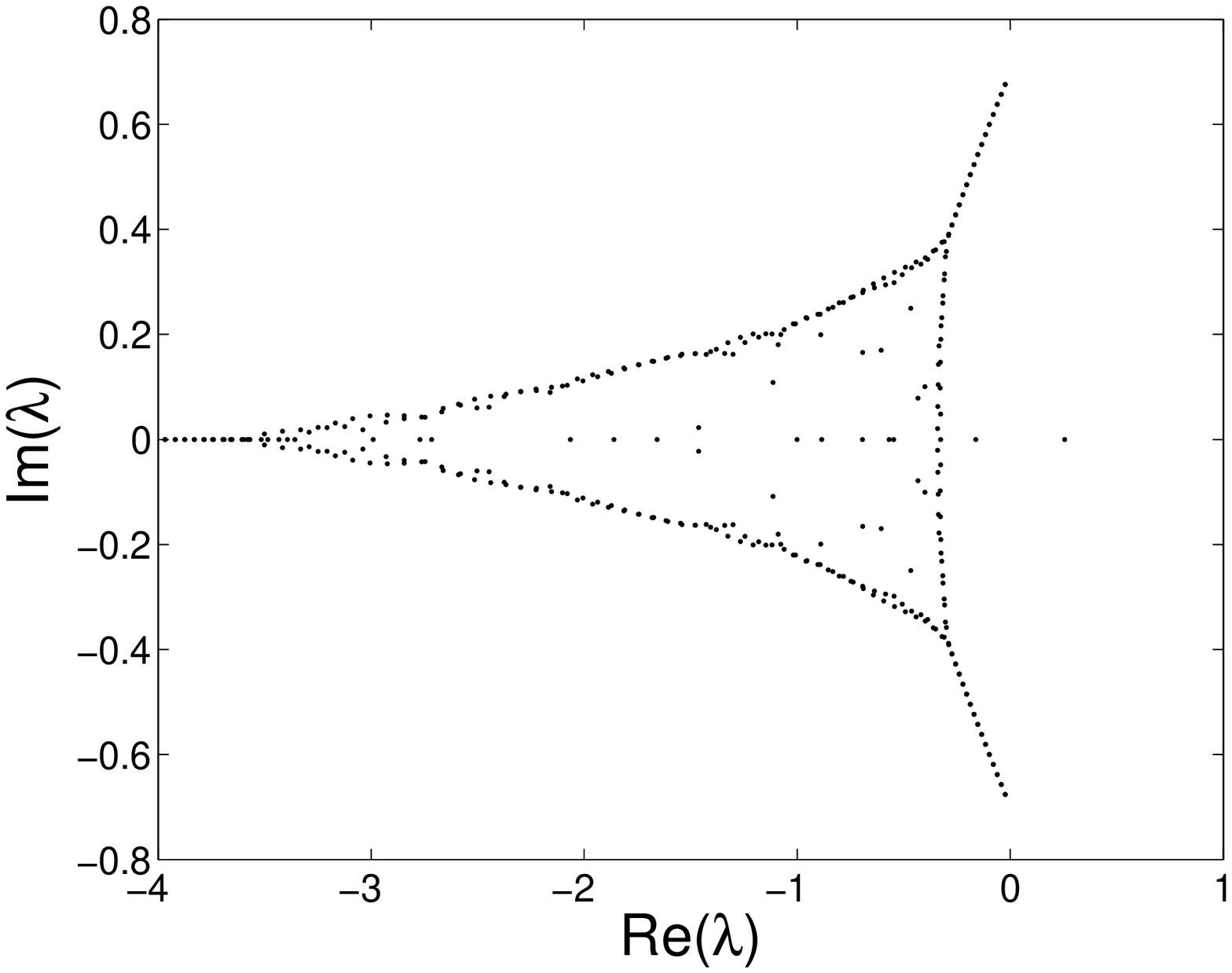}
\caption{The eigenvalues of the linear system (\ref{le}) when $\hk_1=0$ and $\hk_2=1$, 
$\al =0.7$, and $\e =0.00013$.}
\label{ge6}
\end{figure}
\begin{figure}
\includegraphics[width=4.0in,height=4.0in]{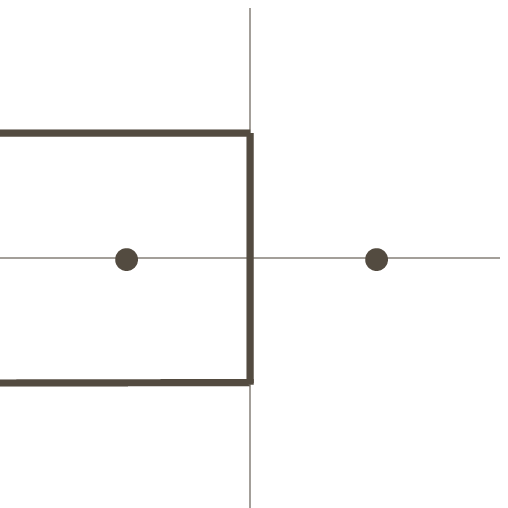}
\caption{The ($\e \ra 0^+$) limiting picture of the eigenvalues of the linear system (\ref{le}) 
when $\hk_1=0$ and $\hk_2=1$, and $\al =0.7$.}
\label{ge7}
\end{figure}
\begin{figure}
\includegraphics[width=4.0in,height=4.0in]{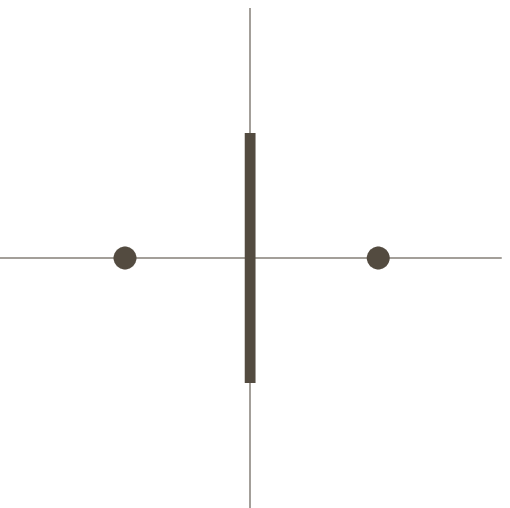}
\caption{The spectrum of the linear system (\ref{le}) when $\e = 0$, $\hk_1=0$ and $\hk_2=1$, 
and $\al =0.7$.}
\label{ge8}
\end{figure}
\begin{figure}
\includegraphics[width=4.0in,height=4.0in]{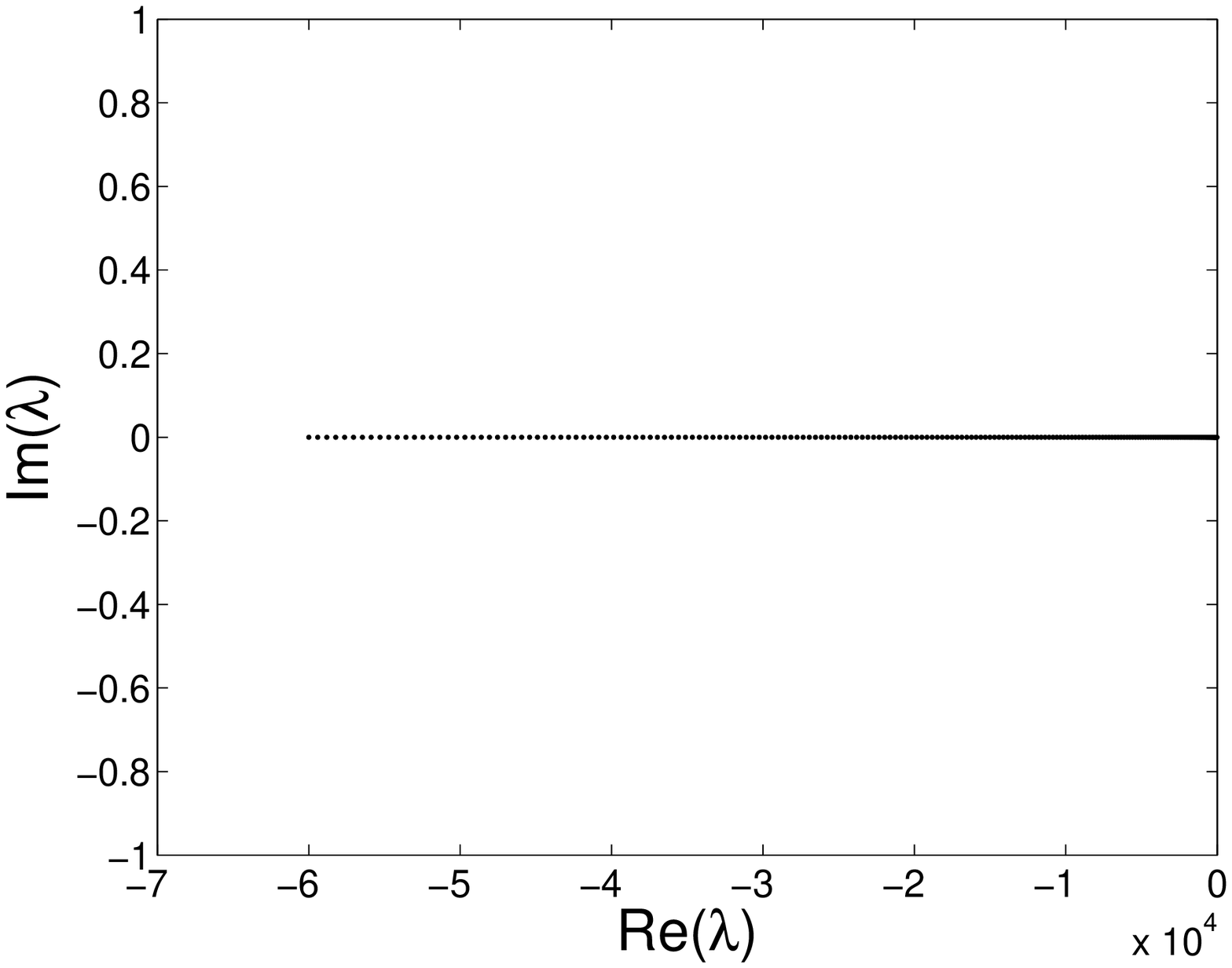}
\caption{The eigenvalues of the linear system (\ref{le}) when $\hk_1=0$ and $\hk_2=2$, 
$\al =0.7$, and $\e =1.5$.}
\label{ge9}
\end{figure}
\begin{figure}
\includegraphics[width=4.0in,height=4.0in]{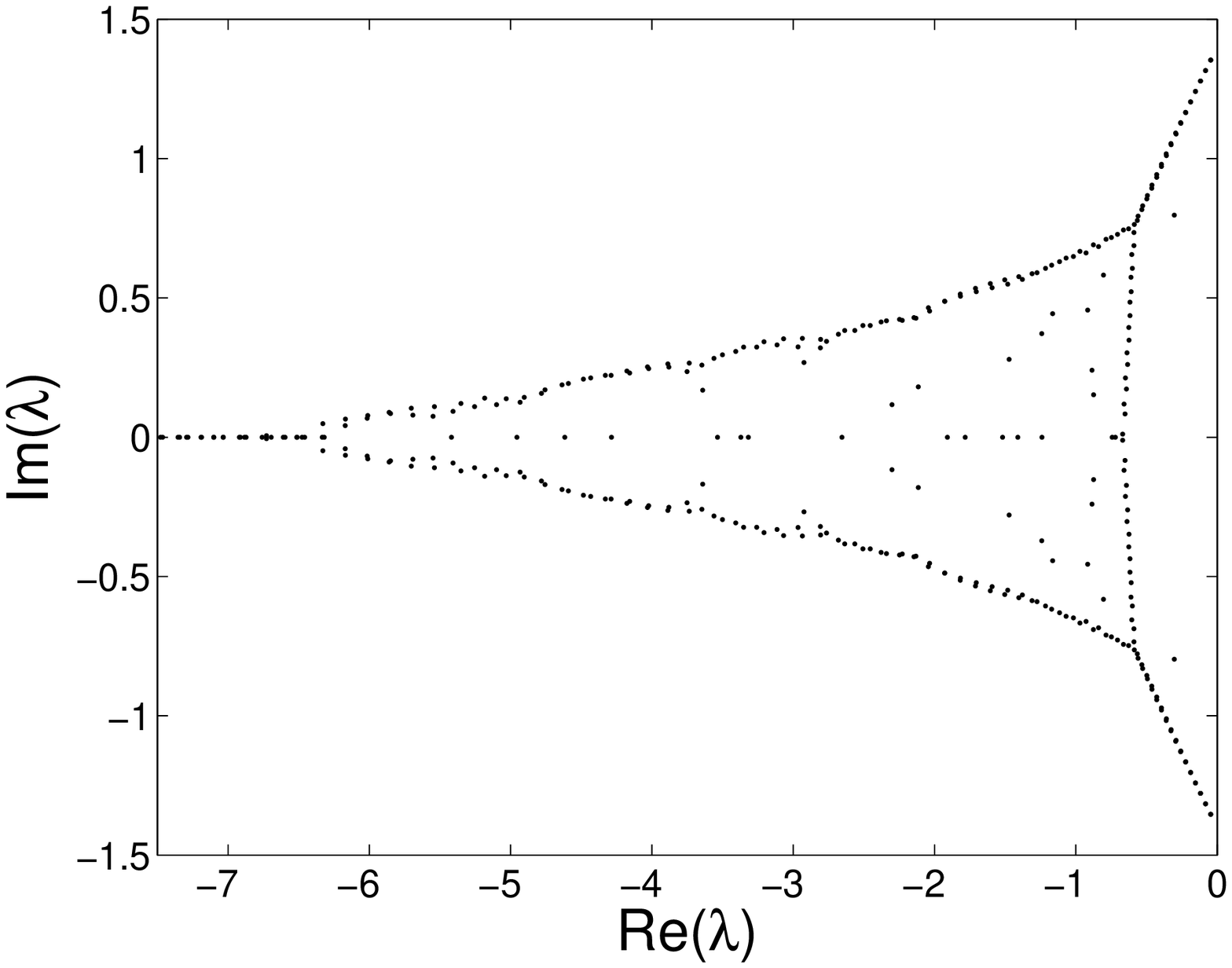}
\caption{The eigenvalues of the linear system (\ref{le}) when $\hk_1=0$ and $\hk_2=2$, 
$\al =0.7$, and $\e =0.00025$.}
\label{ge10}
\end{figure}
\begin{figure}
\includegraphics[width=4.0in,height=4.0in]{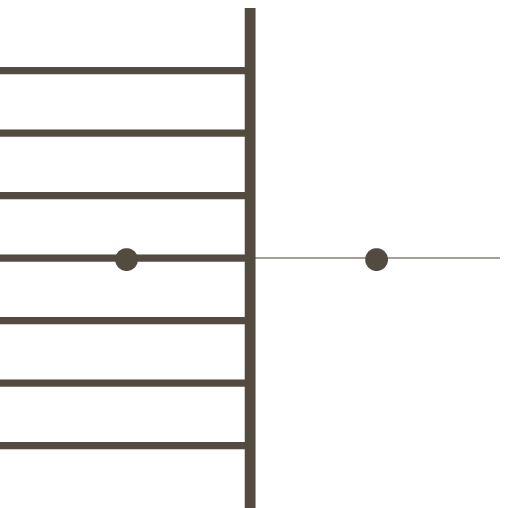}
\caption{The ($\e \ra 0^+$) limiting picture of the entire spectrum of the linear 
NS operator (\ref{le}) when $\al =0.7$.}
\label{ge11}
\end{figure}
\begin{figure}
\includegraphics[width=4.0in,height=4.0in]{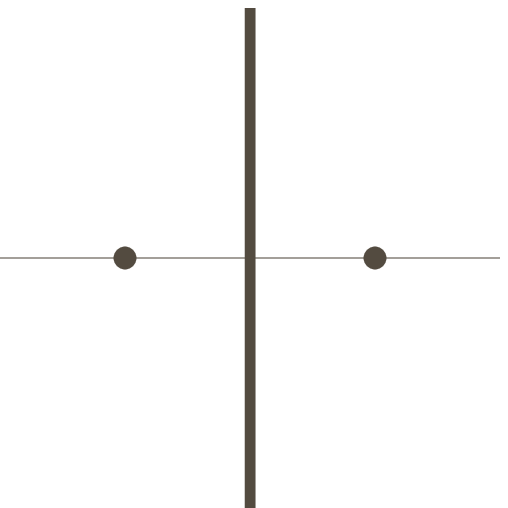}
\caption{The spectrum of the linear Euler operator (\ref{le}) where $\e = 0$ and $\al =0.7$.}
\label{ge12}
\end{figure}

\nid
Figure \ref{ge1} shows the case $\e =0.14$ where there is one positive eigenvalue 
and all the rest eigenvalues are negative. Figure \ref{ge2} shows the case $\e =0.13$ where
a pair of eigenvalues jumps off the real axis and becomes a complex conjugate pair.
Figure \ref{ge3} shows the case $\e =0.07$ where
another pair of eigenvalues jumps off the real axis and becomes a complex conjugate pair.
Figure \ref{ge4} shows the case $\e =0.03$ where
another pair of eigenvalues jumps off the real axis and becomes a complex conjugate pair,
while the former two pairs getting closer to each other. 
Figure \ref{ge5} shows the case $\e =0.0004$ where
many pairs of eigenvalues have jumped off the real axis and a bubble is formed. 
Figure \ref{ge6} shows the case $\e =0.00013$ where the bubble has expanded.
As $\e \ra 0^+$, the limiting picture is shown in Figure \ref{ge7}. Setting $\e =0$, the 
spectrum of the line $\hk_1=0$ and $\hk_2=1$ of the linear Euler operator is shown in 
Figure \ref{ge8}, where the segment on the imaginary axis is the continuous spectrum. 
Comparing Figures \ref{ge7} and \ref{ge8}, we see that the two eigenvalues represent 
``persistence'', the vertical segment represents ``condensation'', and the two horizontal 
segments represent ``singularity''. Next we study one more line: $\hk_1=0$ and $\hk_2 =2$ 
($\al =0.7$). In this case, there is no unstable eigenvalue. Figure \ref{ge9} shows the 
case $\e =1.5$ where all the eigenvalues are negative. As $\e$ is decreased, the eigenvalues 
go through the same process of jumping off the real axis and developing a bubble. 
Figure \ref{ge10} shows the case $\e =0.00025$ where the bubble has expanded. As $\e \ra 0^+$, 
the limiting picture is similar to Figure \ref{ge7} except that there is no persistent 
eigenvalue. The cases $\hk_1=0$ and $\hk_2 > 2$ ($\al =0.7$) are all the 
same with the case $\hk_1=0$ and $\hk_2 =2$ ($\al =0.7$). 
Figure \ref{ge11} shows the limiting picture of the entire spectrum of the 
linear NS operator as $\e \ra 0^+$. Figure \ref{ge12} shows the entire spectrum of the 
linear Euler operator ($\e=0$). 

The fascinating deformation of the spectra as $\e \ra 0^+$ and the limiting spectral picture 
clearly depict the nature of singular limit of the spectra as $\e \ra 0^+$. In the 
``singularity'' part of the limit, there is a discrete set of values for the imaginary 
parts of the eigenvalues, which represent decaying oscillations with a discrete set of 
frequencies. Overall, the ``singularity'' part represents the temporally irreversible 
nature of the $\e \ra 0^+$ limit, in contrast to the reversible nature of the linear 
Euler equation ($\e =0$).

\subsection{A Cat's Eye Fixed Point}

In this subsection, we will study another important fixed point -- a cat's eye fixed point.
The periodic domain now is the square, i.e. $\al =1$ (instead of $0.7$). 
The cat's eye fixed point in physical variable is given by 
\begin{equation}
\Om = 2\cos mx_1 + 2 \ga \cos mx_2 \ , 
\label{cateye}
\end{equation}
where $m$ is a positive integer, and $\ga \in (0, 1]$.
In terms of Fourier modes: Let $p=(m,0)$ and $q=(0,m)$, then the cat's eye is given by
\[
\om^*_p = 1, \ \om^*_{-p} = 1, \ \om^*_q = \ga , \ \om^*_{-q} = \ga \ ,
\]
and all other $\om^*_k$'s are zero. The spectral equation for the linear 
2D NS operator at the Cat's eye is then given by
\begin{eqnarray}
\la \om_k &=& A(p,k-p)\om_{k-p} - A(p,k+p)\om_{k+p} -\e |k|^2 \om_k \non \\
& & + \ga A(q,k-q)\om_{k-q} - \ga A(q,k+q)\om_{k+q} \ , \label{catl}
\end{eqnarray}
where 
\[
A(k,r) = \left [ \frac{1}{r_1^2+r_2^2} - 
\frac{1}{k_1^2+k_2^2}\right ]\left | \begin{array}{lr} 
k_1 & r_1 \\ k_2 & r_2 \\ \end{array} \right | \ .
\]
\begin{figure}
\includegraphics[width=4.0in,height=4.0in]{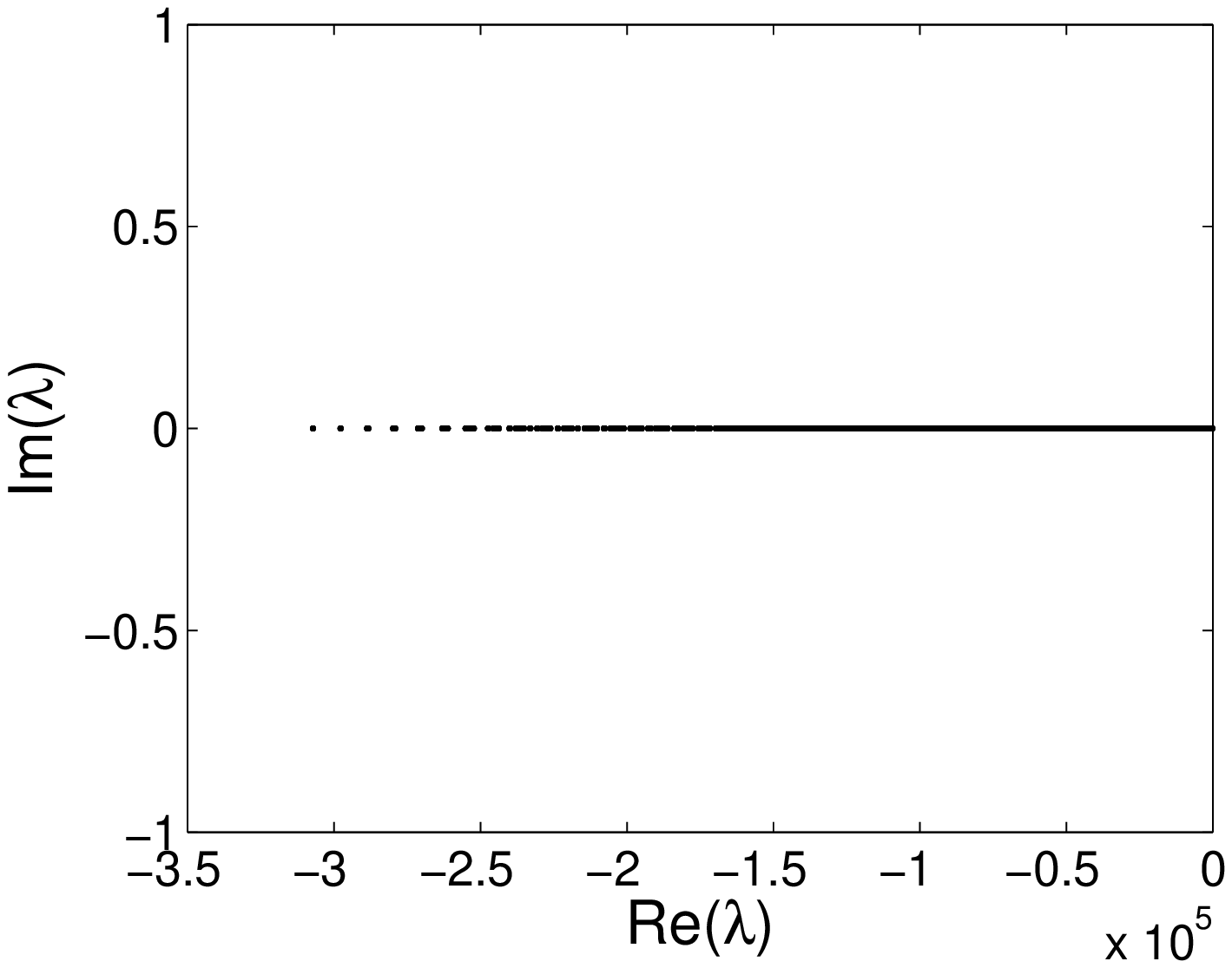}
\caption{The spectrum of the linear NS operator (\ref{catl}) where $\e = 150$, $m=1$ and $\ga =0.5$.}
\label{kat1m1}
\end{figure}
\begin{figure}
\includegraphics[width=4.0in,height=4.0in]{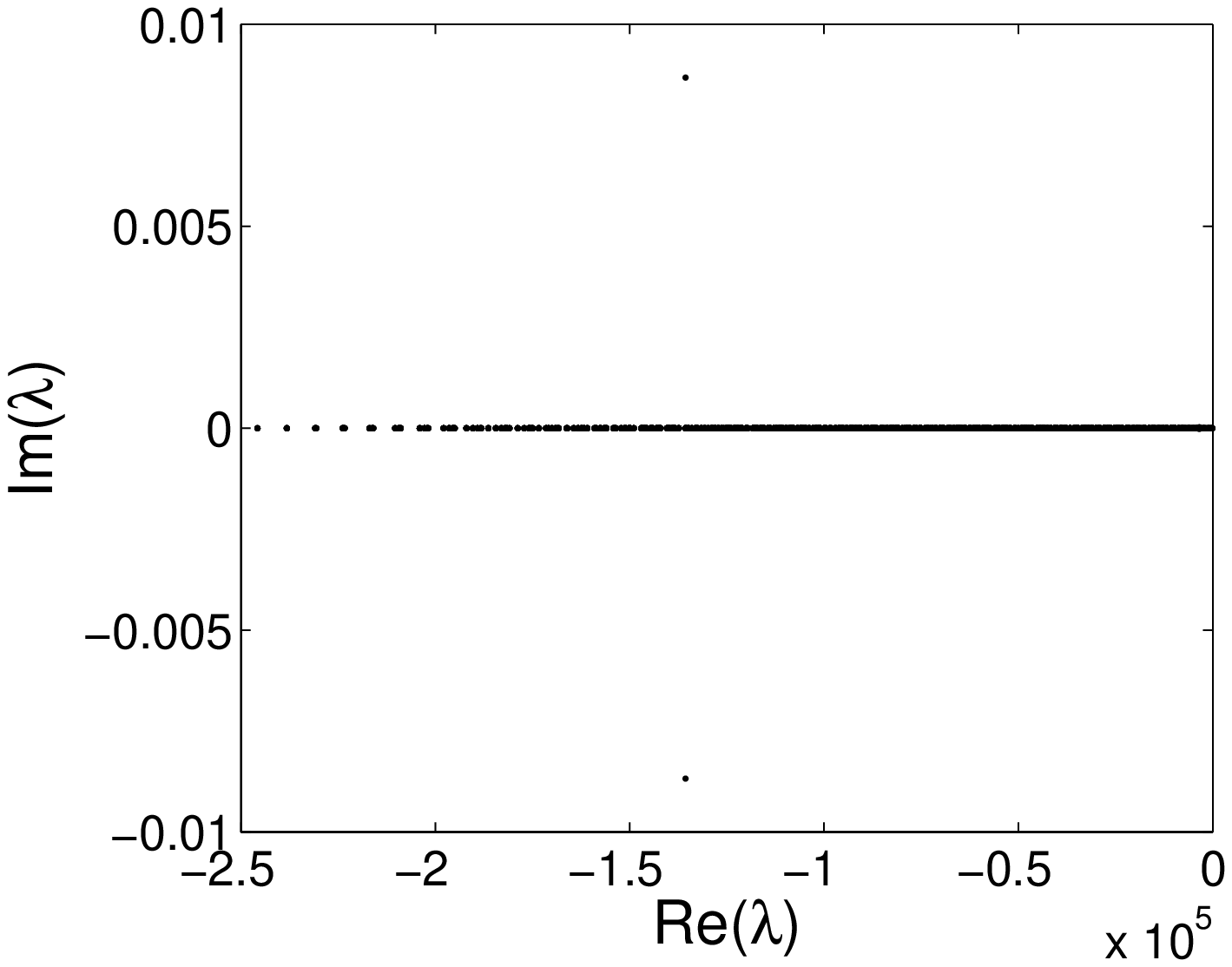}
\caption{The spectrum of the linear NS operator (\ref{catl}) where $\e = 120$, $m=1$ and $\ga =0.5$.}
\label{kat2m1}
\end{figure}
\begin{figure}
\includegraphics[width=4.0in,height=4.0in]{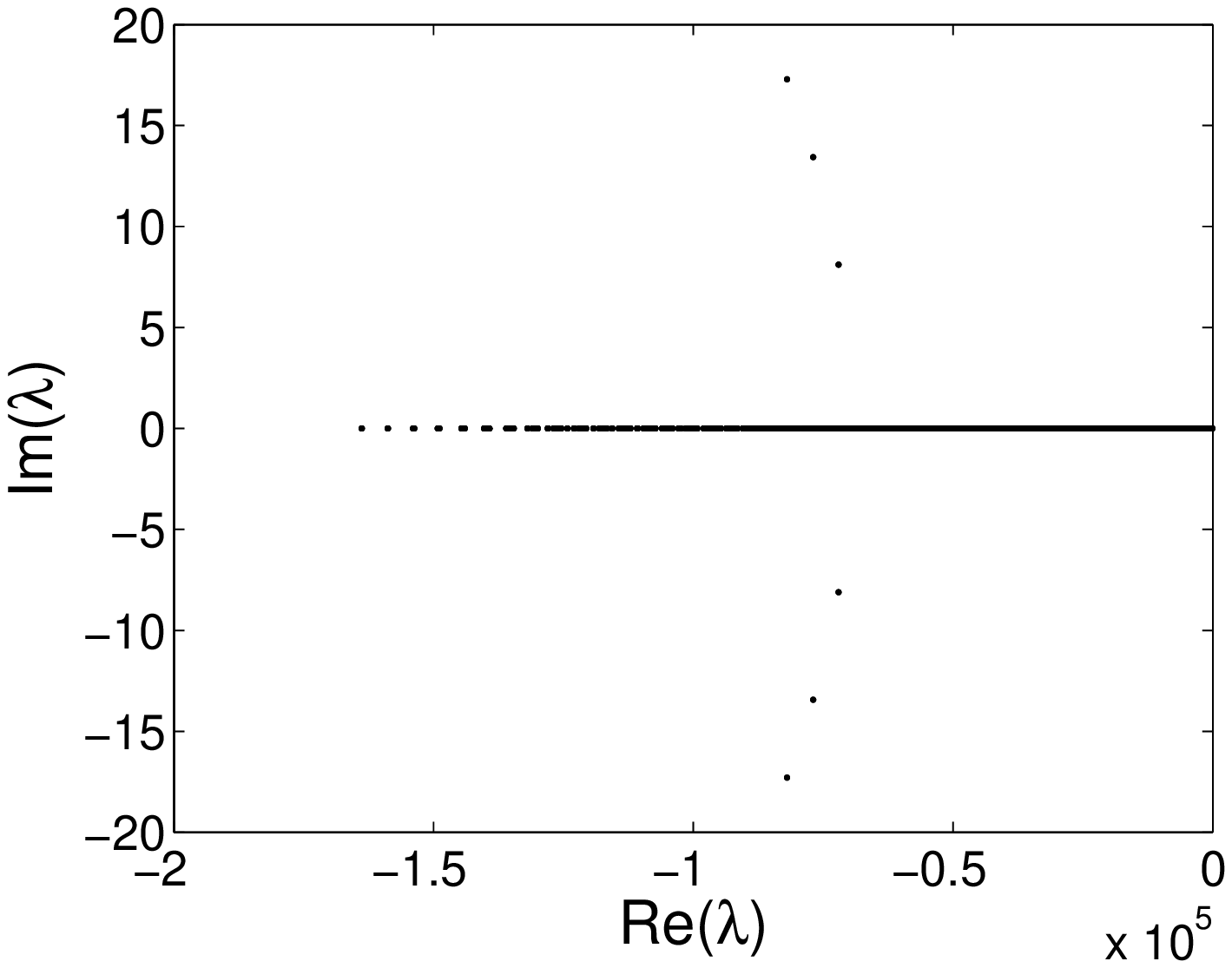}
\caption{The spectrum of the linear NS operator (\ref{catl}) where $\e = 80$, $m=1$ and $\ga =0.5$.}
\label{kat3m1}
\end{figure}
\begin{figure}
\includegraphics[width=4.0in,height=4.0in]{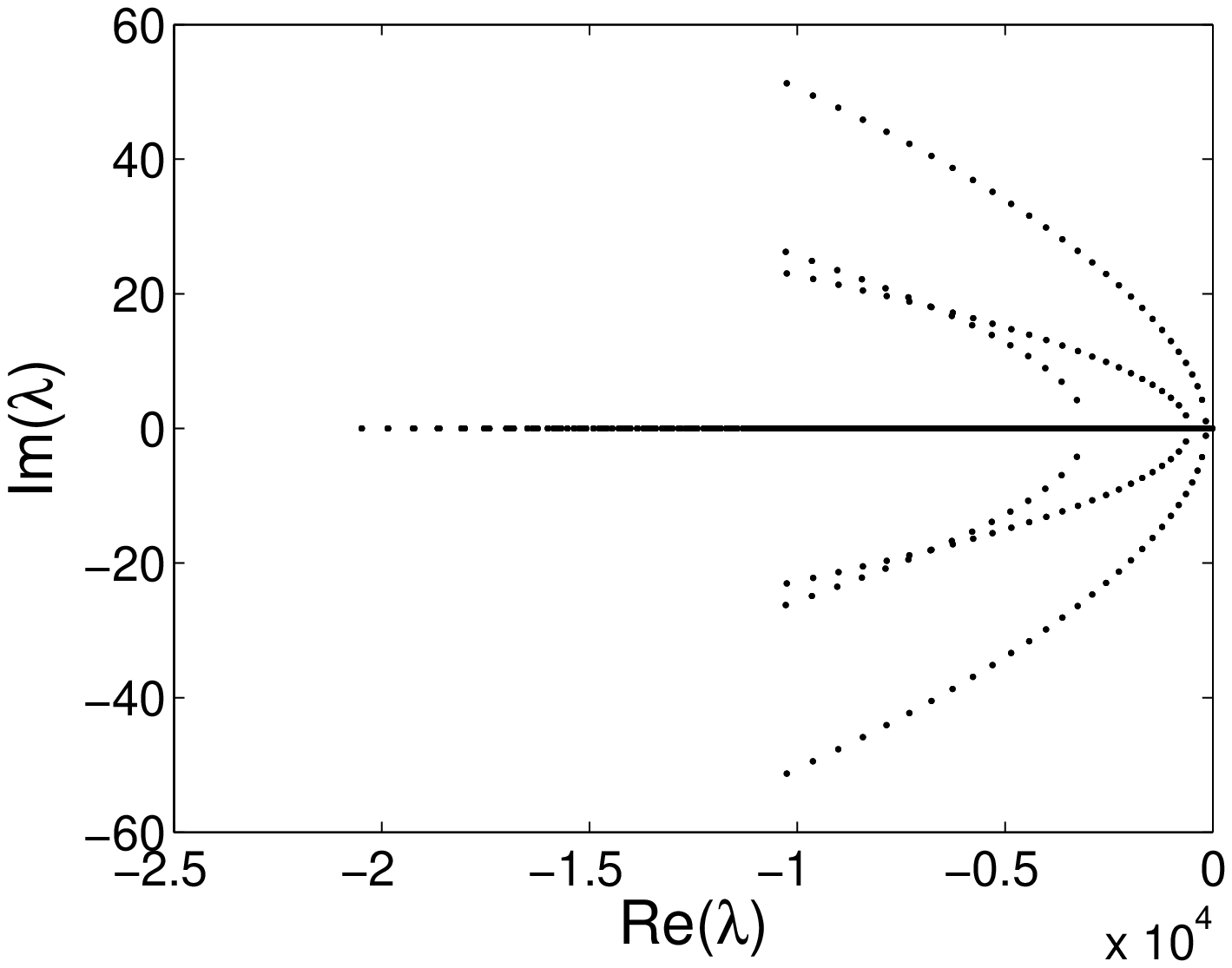}
\caption{The spectrum of the linear NS operator (\ref{catl}) where $\e = 10$, $m=1$ and $\ga =0.5$.}
\label{kat4m1}
\end{figure}
\begin{figure}
\includegraphics[width=4.0in,height=4.0in]{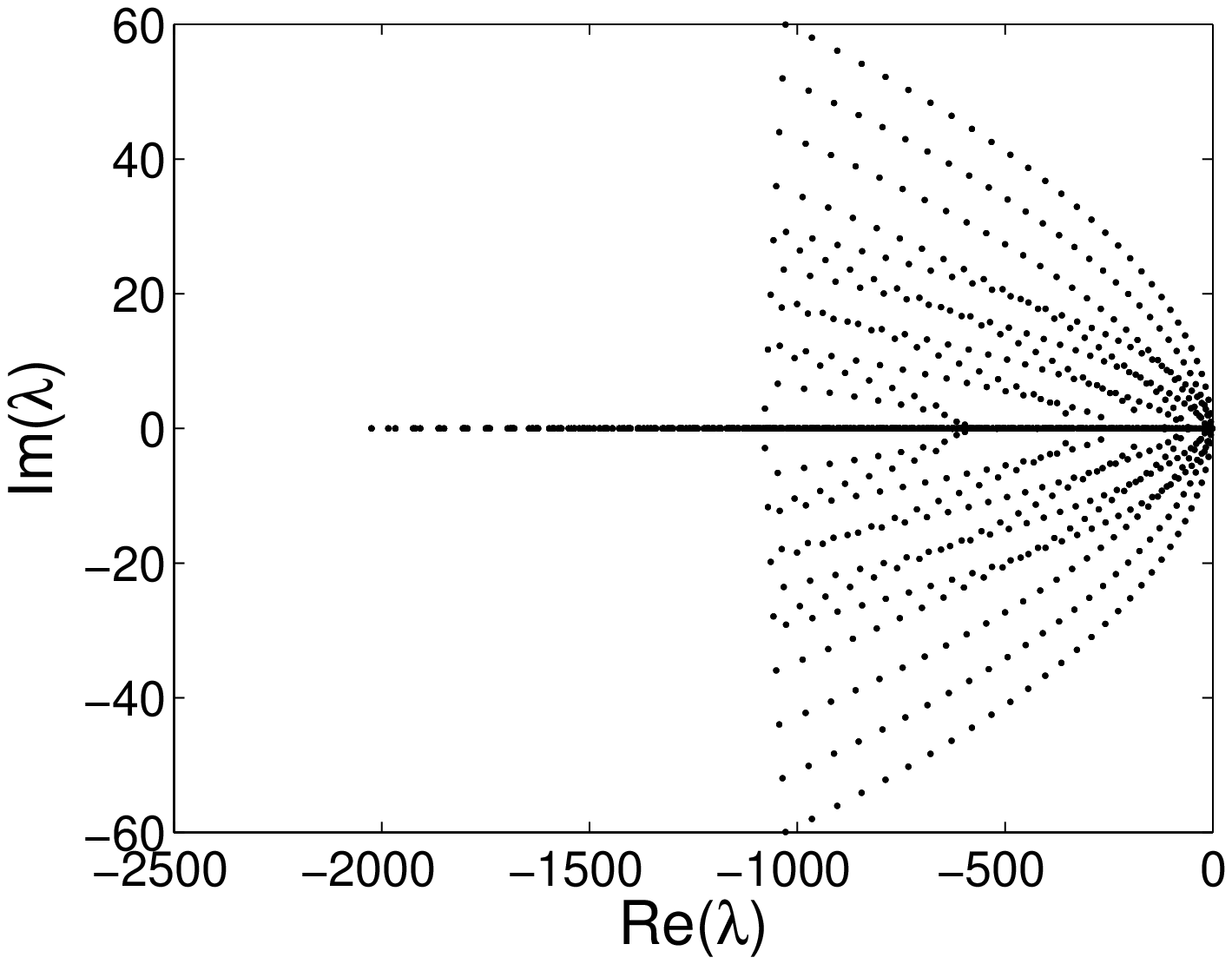}
\caption{The spectrum of the linear NS operator (\ref{catl}) where $\e = 1$, $m=1$ and $\ga =0.5$.}
\label{kat5m1}
\end{figure}
\begin{figure}
\includegraphics[width=4.0in,height=4.0in]{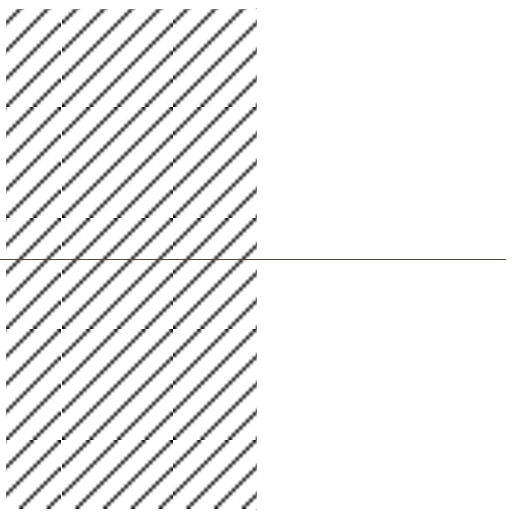}
\caption{The limiting picture of the spectrum of the linear NS operator (\ref{catl}) as $\e \ra 0$, 
where $m=1$ and $\ga =0.5$.}
\label{kat6m1}
\end{figure}
\begin{figure}
\includegraphics[width=4.0in,height=4.0in]{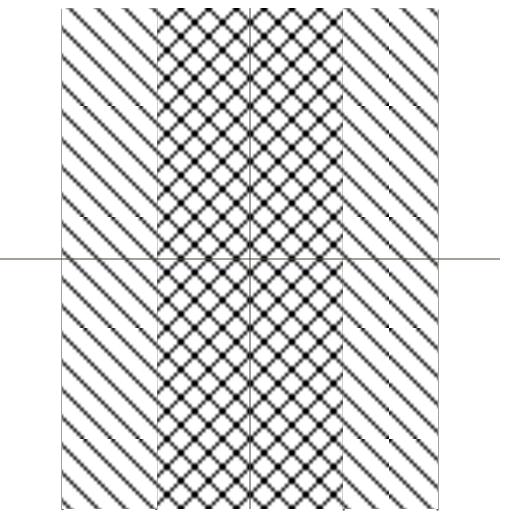}
\caption{The spectrum of the linear Euler operator (\ref{catl}) where $\e = 0$, 
$m=1$ and $\ga =0.5$.}
\label{kat7m1}
\end{figure}
First we study the case of $m=1$ and $\ga =0.5$. Changing the value of $\ga$ does not 
affect the deformation patterns of the eigenvalues of the linear NS (\ref{catl}) as 
$\e \ra 0^+$. We truncate the system (\ref{catl}) via the Galerkin truncation 
$|k_1| \leq 32$ and $|k_2| \leq 32$. This is the largest Galerkin truncation that we are 
able to compute in a reasonable time. For smaller Galerkin truncations, the deformation 
patterns of the eigenvalues are the same. When $\e = 150$, all the eigenvalues of the 
linear NS (\ref{catl}) are negative as shown in Figure \ref{kat1m1}. When $\e$ is decreased 
to $\e = 120$, a pair of eigenvalues jumps off the real axis as shown in Figure \ref{kat2m1}.
When $\e$ is decreased to $\e = 80$, three pairs of eigenvalues jump off the real axis as 
shown in Figure \ref{kat3m1}. When $\e$ is decreased to $\e = 10$, many pairs of eigenvalues 
have jumped off the real axis and form several parabolas as shown in Figure \ref{kat4m1}.
When $\e$ is decreased to $\e = 1$, many parabolas are formed as shown in Figure \ref{kat5m1}.
The $\e \ra 0^+$ limiting picture of the eigenvalues of the linear NS (\ref{catl}) is that the 
eigenvalues are dense on the entire left half plane as shown in Figure \ref{kat6m1}.
The continuous spectrum of the linear Euler, i.e. $\e =0$ in (\ref{catl}), in any Sobolev space 
$H^s(\mathbb{T}^2)$ where $s$ is a non-negative integer, is a vertical band of width 
$2s \sg$ symmetric with respect to the imaginary axis 
$\{ \la \ : \ |\text{Re}(\la )| \leq s \sg \}$ as shown in Figure \ref{kat7m1}, where 
$\sg >0$ is the largest Liapunov exponent of the vector field given by the cat's eye (\ref{cateye})
\cite{SL03}. Thus the width of the vertical band is proportional to the scale $s$ of the 
Sobolev space $H^s(\mathbb{T}^2)$. The union of all such bands for all integers $s \geq 0$ is 
the entire complex plane. The eigenfunctions of the linear NS (\ref{catl}) when $\e >0$ 
belong to $H^s(\mathbb{T}^2)$ for all integers $s \geq 0$. All the eigenvalues of the 
linear NS (\ref{catl}) condense into the entire left half plane -- ``condensation''. The 
right half plane (or right half of the vertical band corresponding to $H^s(\mathbb{T}^2)$) 
represents ``addition''. Thus the possible instability hinted by the right half band of the 
continuous spectrum of linear Euler in $H^s(\mathbb{T}^2)$ can not be realized by real 
viscous fluids.

\begin{figure}
\includegraphics[width=4.0in,height=4.0in]{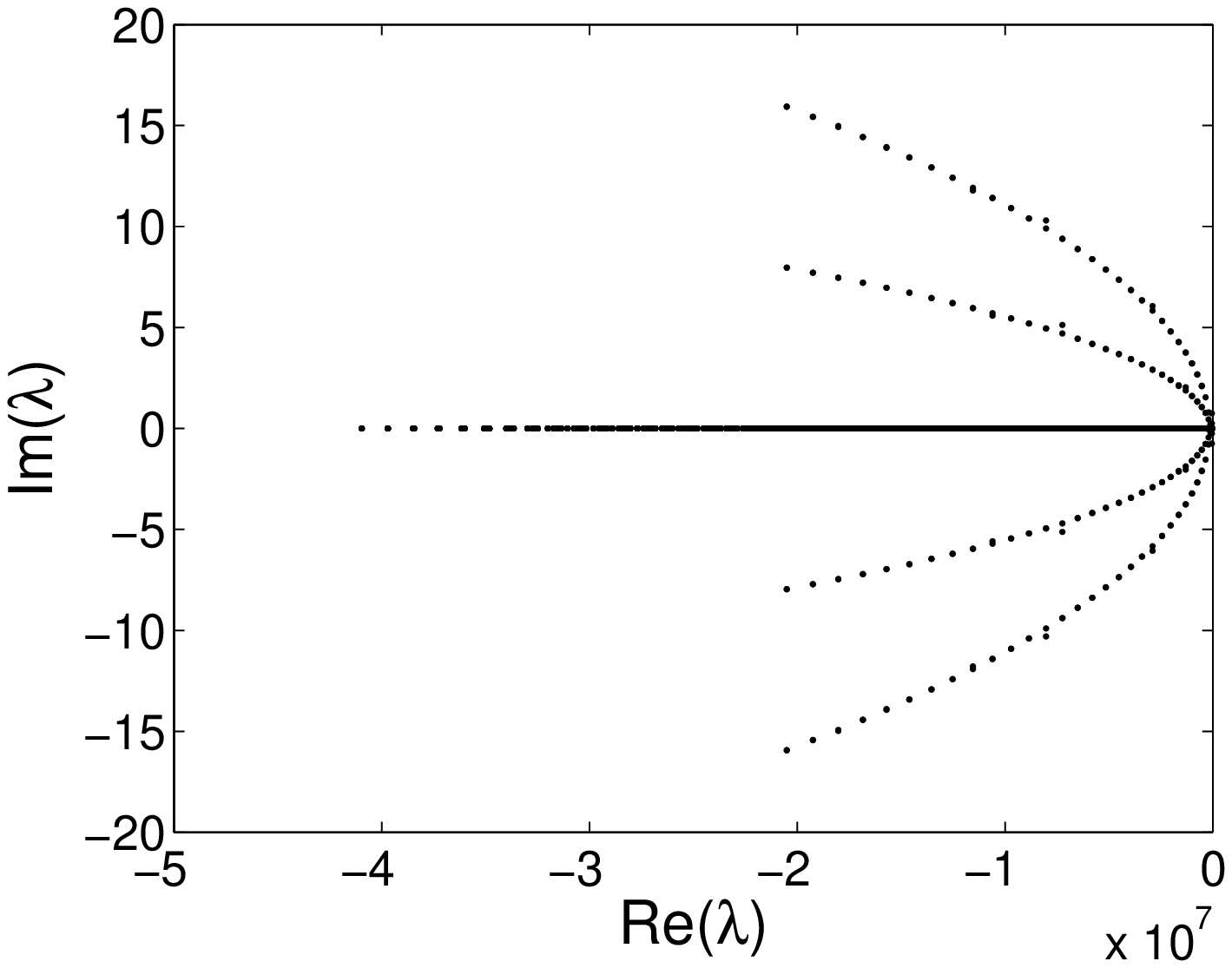}
\caption{The spectrum of the linear NS operator (\ref{catl}) where $\e = 20000$, 
$m=2$ and $\ga =0.5$.}
\label{kat1m2}
\end{figure}

\begin{figure}
\includegraphics[width=4.0in,height=4.0in]{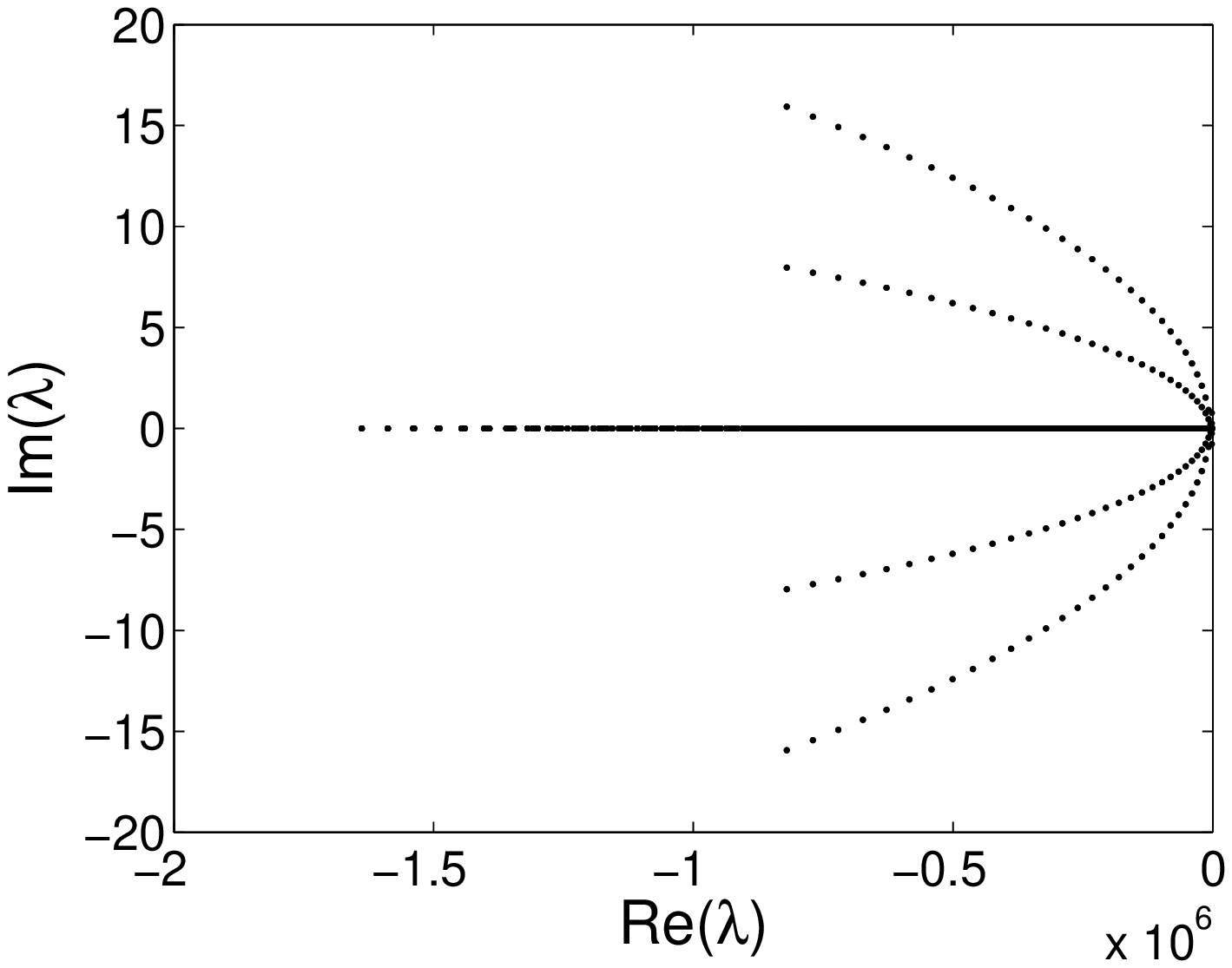}
\caption{The spectrum of the linear NS operator (\ref{catl}) where $\e = 800$, 
$m=2$ and $\ga =0.5$.}
\label{kat2m2}
\end{figure}

\begin{figure}
\includegraphics[width=4.0in,height=4.0in]{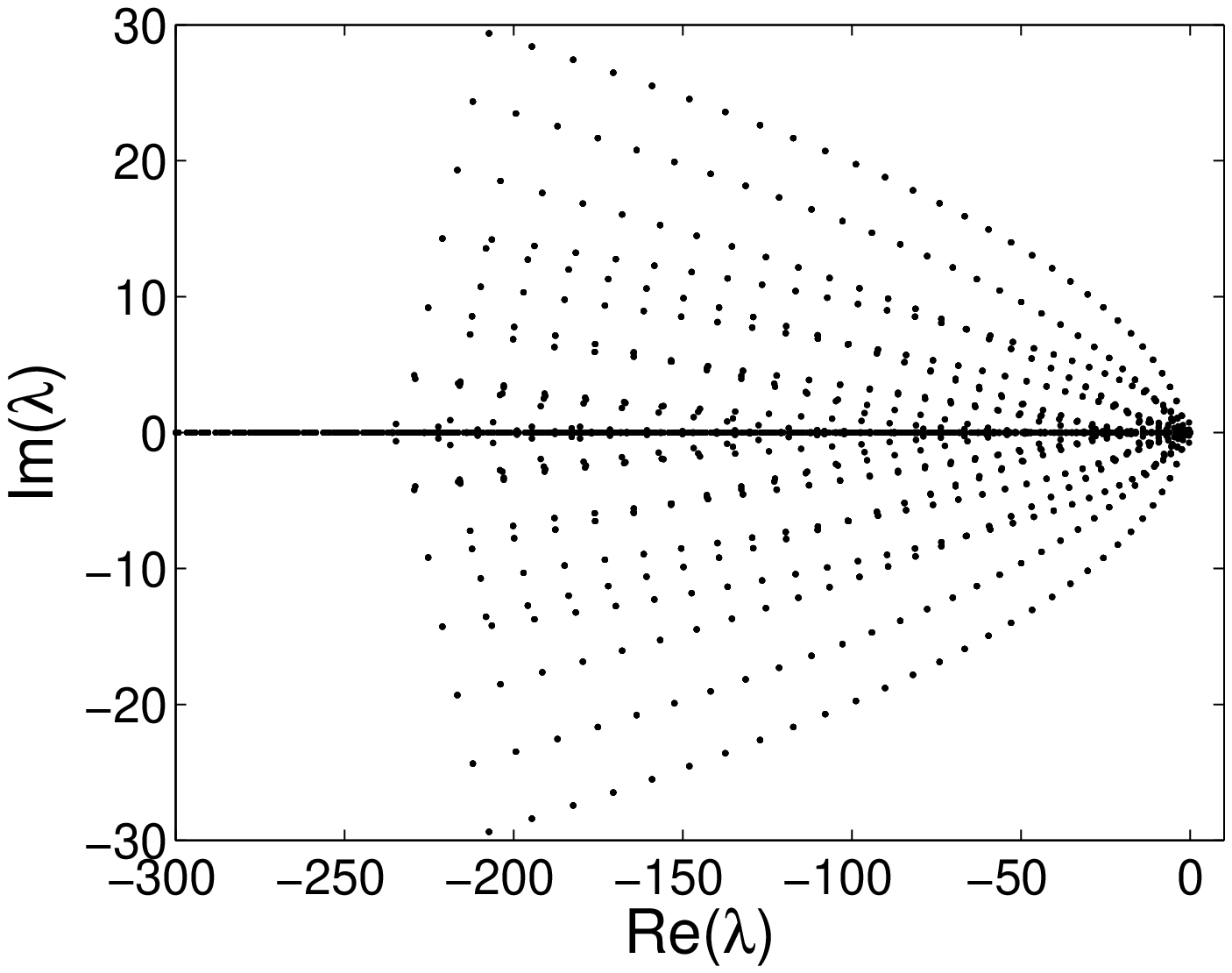}
\caption{The spectrum of the linear NS operator (\ref{catl}) where $\e = 0.2$, 
$m=2$ and $\ga =0.5$.}
\label{kat3m2}
\end{figure}

Next we study the case of $m=2$ and $\ga =0.5$. Changing the value of $\ga$ does not 
affect the deformation patterns of the eigenvalues of the linear NS (\ref{catl}) as 
$\e \ra 0^+$. We truncate the system (\ref{catl}) via the Galerkin truncation 
$|k_1| \leq 32$ and $|k_2| \leq 32$. We increase the value of $\e$ up to $2\times 10^4$, 
there are still eigenvalues with nonzero imaginary parts. These eigenvalues seem always 
complex no matter how large is $\e$. The imaginary parts of these eigenvalues are 
unchanged between $\e =2\times 10^4$ and $\e = 800$ as can be seen from Figures 
\ref{kat1m2} and \ref{kat2m2}. Decreasing $\e$, the deformation patterns are similar to 
those of $m=1$. When $\e = 0.2$, many eigenvalues have jumped off the real axis and 
form a dense parabolic region as shown in Figure \ref{kat3m2}. Decreasing $\e$ further, 
$6$ eigenvalues with positive real parts appear, two of which are real, and the rest 
four are complex. The limiting picture of the spectrum of the linear NS operator 
(\ref{catl}) as $\e \ra 0$ is the same with the $m=1$ case as shown in Figure \ref{kat6m1}
except the extra six unstable eigenvalues. The continuous spectrum of the linear Euler 
operator (\ref{catl}) where $\e = 0$ is the same with the $m=1$ case as shown in Figure 
\ref{kat7m1}. There is no analytical result on the eigenvalues of the linear Euler 
operator (\ref{catl}) where $\e = 0$. The numerics indicates that the $6$ eigenvalues 
in the right half plane and other $6$ eigenvalues in the left half plane, of the linear 
NS operator (\ref{catl}) persist as $\e \ra 0$, and result in $6$ eigenvalues 
in the right half plane and their negatives for the linear Euler operator (\ref{catl}) 
where $\e = 0$. 

\section{The Heteroclinics Conjecture for 2D Euler Equation}


Setting $\e = 0$ in the 2D Navier-Stokes equation (\ref{2DNS}), one gets the corresponding 
2D Euler equation for which one has the following constants of motion:
\[
\int_{\mathbb{T}^2} |u|^2dx\ , \quad \int_{\mathbb{T}^2} F(\Om )dx 
\]
where $F$ is an arbitrary function. Consider the simple fixed point $\Om = \Ga \cos x_1$ 
($\Ga \neq 0$ real constant). It has one unstable  and one stable real eigenvalues 
which are negative of each other. The rest of the spectrum is the entire imaginary 
axis which is a continuous spectrum \cite{Li00} \cite{Li05}. We will use the constant 
of motion 
\[
G = \int_{\mathbb{T}^2} \Om^2 dx - \int_{\mathbb{T}^2} |u|^2dx 
\]
to build a Melnikov integral for the corresponding 2D Navier-Stokes equation (\ref{2DNS}).
We will try to make use of the Melnikov integral as a measure of chaos and to conduct 
a control of chaos, around the line of fixed points $\Om = \Ga \cos x_1$ parametrized 
by $\Ga$. $G$ is a linear combination of the kinetic energy and the enstrophy. The 
gradient of $G$ in $\Om$ is given by
\[
\na_\Om G = 2(\Om +\Dl^{-1} \Om )
\]
which is zero along the line of fixed points $\Om = \Ga \cos x_1$. We define the 
Melnikov integral for the 2D NS (\ref{2DNS}) as 
\begin{eqnarray}
M &=& \frac{\al}{8\pi^2}\int_{-\infty}^{+\infty} \int_{\mathbb{T}^2} \na_\Om G 
[\Dl \Om + f(t,x) +b\tdl (x)] dxdt \label{gMel} \\
&=& M_0 + b M_c \ , \non 
\end{eqnarray}
where
\begin{eqnarray*}
M_0 &=& \frac{\al}{4\pi^2}\int_{-\infty}^{+\infty} \int_{\mathbb{T}^2} (\Om +\Dl^{-1} \Om )
[\Dl \Om + f(t,x)]dxdt \ , \\
M_c &=& \frac{\al}{4\pi^2}\int_{-\infty}^{+\infty} \int_{\mathbb{T}^2} (\Om +\Dl^{-1} \Om )
\tdl (x) dxdt \ .
\end{eqnarray*}
The question is: Where do we evaluate $M$? We propose the following conjecture.
\begin{itemize}
\item The Heteroclinics Conjecture: There is a heteroclinic orbit of the 2D Euler 
equation that connects $\Om = \Ga \cos x_1$ and $-\Om$. 
\end{itemize}
The rationality of this conjecture has been discussed in the Introduction.
If this conjecture is true, we can evaluate $M$ along the heteroclinic orbit. 
Also, under the perturbation of the $\e$ term, the heteroclinic orbits may 
break and re-connect somewhere, thereby generating the heteroclinic chaos.
In Appendix A, we show that one can use a Melnikov integral as a criterion to rigorously 
prove the existence chaos, and to conduct control of chaos. In the current case of 
2D NS, the phase space is much more complicated. We do not expect the above Melnikov 
integral to be a rigorous measure of chaos, rather we believe that it is still relevant 
to chaos and control of chaos. In fact, often the chaos in 2D NS is a transient chaos.
In general, the zeros of a Melnikov integral do not immediately imply the existence of
heteroclinic or homoclinic chaos, rather they imply the intersections of the broken 
heteroclinic orbit with certain center-stable manifold \cite{Li04}. In fact, the 
Melnikov integral represents the leading order distance between the broken 
heteroclinic orbit and the center-stable manifold \cite{Li04}. Here for 2D NS, rigorous 
justification even on the above claim is an open problem due to the fact that existence 
of invariant manifolds for 2D Euler is an open problem.

The 2D Euler equation has several symmetries:
\begin{enumerate}
\item $\Om (t, x_1, x_2) \lra \Om (t, -x_1, -x_2)$,
\item $\Om (t, x_1, x_2) \lra -\Om (-t, x_1, x_2)$,
\item $\Om (t, x_1, x_2) \lra -\Om (t, -x_1, x_2)$, or $\Om (t, x_1, x_2) \lra -\Om (t, x_1, -x_2)$,
\item $\Om (t, x_1, x_2) \lra \Om (t, x_1+\th_1, x_2+\th_2)$, $\quad \forall \th_1, \th_2$.
\end{enumerate}
The first symmetry allows us to work in an invariant subspace in which all the $\om_k$'s are 
real-valued. This corresponds to the cosine transform in (\ref{FS}). In this paper, we will 
always work in the invariant subspace where all the $\om_k$'s are real-valued.
The second symmetry 
maps the unstable manifold of $\Ga \cos x_1$ into the stable manifold of $-\Ga \cos x_1$. The
third symmetry maps the unstable manifold of $\Ga \cos x_1$ into the unstable manifold of 
$-\Ga \cos x_1$. By choosing $\th_1 =\pi$, the fourth symmetry maps the unstable manifold of 
$\Ga \cos x_1$ into the unstable manifold of $-\Ga \cos x_1$. To maintain the cosine transform, 
the $\th_1$ and $\th_2$ in the fourth symmetry can only be $\pi$ and $\pi /\al$. 

If the heteroclinics conjecture is true, there is in fact a pair of heteroclinic cycles due 
to the above symmetries. Indeed, if there is a heteroclinic orbit asymptotic to $\Ga \cos x_1$ 
and $- \Ga \cos x_1$ as $t \ra -\infty$ and $+\infty$, then the third symmetry generates 
another heteroclinic orbit asymptotic to $-\Ga \cos x_1$ and $\Ga \cos x_1$ as 
$t \ra -\infty$ and $+\infty$. Together they form a heteroclinic cycle. Finally the second 
symmetry generates another heteroclinic cycle. That is, we have a pair of heteroclinic cycles.

Using the Fourier series
\begin{equation}
\Om = \sum_{k \in \ZZ} \om_k e^{i(k_1 x_1 + \al k_2 x_2)}\ , 
\label{FS}
\end{equation}
where $\om_{-k} = \overline{\om_k}$ and $F_{-k} = \overline{F_k}$, one gets
the kinetic form of the 2D Euler equation
\[
\dot{\om}_k = \sum_{k=m+n} A(m,n) \ \om_m \om_n \ ,
\]
where 
\[
A(m,n) = \frac{\al}{2}\left [ \frac{1}{n_1^2+(\al n_2)^2} - 
\frac{1}{m_1^2+(\al m_2)^2}\right ]\left | \begin{array}{lr} 
m_1 & n_1 \\ m_2 & n_2 \\ \end{array} \right | \ .
\]
Denote by $\Sg$ the hyperplane
\[
\Sg = \left \{ \om \ | \ \om_k = 0 \ , \quad \forall \text{ even } k_2 \right \} \ .
\]
Notice that the existence of invariant manifolds around the fixed point $\Om =\Ga \cos x_1$ 
is an open problem. We have the following theorem.
\begin{theorem}
Assume that the fixed point $\Om = \Ga \cos x_1$ has a 1-dimensional 
local unstable manifold $W^u$, 
and $W^u \cap \Sg \neq \emptyset$; then the heteroclinics conjecture is true, i.e.  
there is a heteroclinic orbit to the 2D Euler equation that connects 
$\Om = \Ga \cos x_1$ and $-\Om$.
\label{hcthm}
\end{theorem}
\begin{proof}
Let $\Om (t, x_1, x_2)$ be an orbit in $W^u$ parametrized such that 
\[
\Om (0, x_1, x_2) \in \Sg \ .
\]
Then by the definition of $\Sg$,
\begin{equation}
\Om (0, x_1, x_2) = -\Om (0, x_1, x_2+\pi /\al ) \ .
\label{cpt}
\end{equation}
By the second and fourth symmetries, 
\[
-\Om (-t, x_1, x_2+\pi /\al )
\]
is in the stable manifold of $-\Om$. Thus 
\[
\Om (t, x_1, x_2) \text{  and  } -\Om (-t, x_1, x_2+\pi /\al )
\]
are connected at $t=0$, and together they form a heteroclinic orbit that connects 
$\Om = \Ga \cos x_1$ and $-\Om$.
\end{proof}

\section{Numerical Verification of the Heteroclinics Conjecture for 2D Euler Equation}

Besides the symmetries mentioned in last section, we will also make use of the 
conserved quantities: kinetic energy $E=\sum |k|^{-2} \om_k^2$ 
(where $|k|^2=k_1^2+\al^2 k_2^2$) and enstrophy $S=\sum \om_k^2$, which 
will survive as conserved quantities for any symmetric Galerkin truncation, to help us to 
track the heteroclinic orbit. We will only consider the case that all the $\om_k$'s are 
real-valued (i.e. $\cos$-transform). 

\begin{figure}
\includegraphics[width=4.0in,height=3.0in]{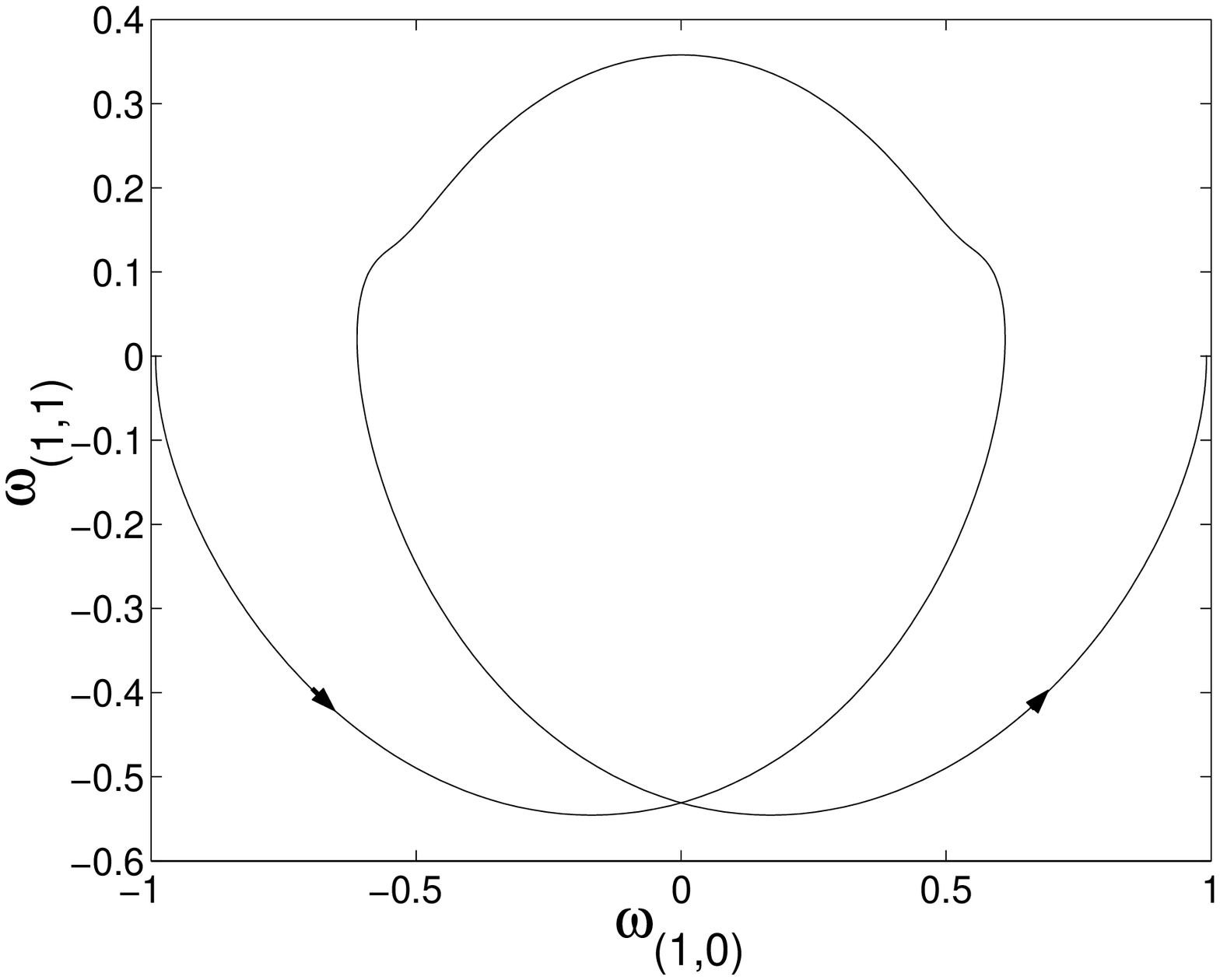}
\caption{The approximate heteroclinic orbit projected onto the ($\om_{(1,0)},\om_{(1,1)}$)-plane
in the case of the $\{ |k_1| \leq 2, |k_2| \leq 2 \}$ Galerkin truncation of the 2D Euler 
equation.}
\label{fi1}
\end{figure}

We make a Galerkin truncation by keeping modes: $\{ |k_1| \leq 2, |k_2| \leq 2 \}$, which results
in a $12$ dimensional system. We choose $\al =0.7$. After careful consideration of the above 
mentioned symmetries and conserved quantities ($E=S=1$), we discover the following initial 
condition that best tracks the heteroclinic orbit:
\begin{eqnarray}
& & \om_{(j,0)}=\om_{(j,2)} = 0 \ , \quad \forall j\ , \non \\
& & \om_{(0,1)}=0.603624\ , \quad \om_{(1,1)}=- \om_{(-1,1)}=0.357832\ , \label{IC} \\
& & \om_{(2,1)}=\om_{(-2,1)}=0.435632\ . \non 
\end{eqnarray}
Starting from this initial condition, we calculate the solution in both forward and backward time
for the same duration of $T =11.8$, and we discover the approximate heteroclinic orbit 
asymptotic to $2\cos x_1$ and $-2\cos x_1$ as $t \ra -\infty$ and $+\infty$, as shown in Figure 
\ref{fi1}. Then the third symmetry generates another heteroclinic orbit asymptotic to 
$-2\cos x_1$ and $2\cos x_1$ as $t \ra -\infty$ and $+\infty$. Together they form a heteroclinic 
cycle. Finally the second symmetry generates another heteroclinic cycle. That is, we have 
a pair of heteroclinic cycles. Notice also that the approximate heteroclinic orbit in 
Figure \ref{fi1} has an extra loop before landing near $-2\cos x_1$. This is due to the 
$k_2=2$ modes in the Galerkin truncation. For smaller Galerkin truncations, the heteroclinic orbits 
can be calculated exactly by hand and have no such extra loop \cite{Li03e} \cite{Li06e}, 
and existence of chaos generated by the heteroclinic orbit can be rigorously proved in some case \cite{Li06e}.
\begin{remark}
We have also conducted numerical experiments on Galerkin truncations by keeping more modes: 
$\{ |k_1| \leq 4, |k_2| \leq 4 \}$ and $\{ |k_1| \leq 8, |k_2| \leq 8 \}$. We found orbits 
that have similar behavior as the approximate heteroclinic orbit in Figure \ref{fi1}, but 
their approximations to heteroclinics are not as good as the one in Figure \ref{fi1}. 
\end{remark}

\section{Melnikov Integral and Numerical Simulation of Chaos in 2D Navier-Stokes Equation}

Without the control ($b=0$), using Fourier series for the 2D NS equation (\ref{2DNS}),
\[
\Om = \sum_{k \in \ZZ} \om_k e^{i(k_1 x_1 + \al k_2 x_2)}\ , \quad 
f = \sum_{k \in \ZZ} F_k e^{i(k_1 x_1 + \al k_2 x_2)} \ , 
\]
where $\om_{-k} = \overline{\om_k}$ and $F_{-k} = \overline{F_k}$ (in fact, we always work in the 
subspace where all the $\om_k$'s and $F_k$'s are real-valued), one gets
the kinetic form of 2D NS
\[
\dot{\om}_k = \sum_{k=m+n} A(m,n) \ \om_m \om_n  +\e \left ( -\left [k_1^2 + (\al k_2)^2\right ]
\om_k +F_k \right ) \ ,
\]
where 
\[
A(m,n) = \frac{\al}{2}\left [ \frac{1}{n_1^2+(\al n_2)^2} - 
\frac{1}{m_1^2+(\al m_2)^2}\right ]\left | \begin{array}{lr} 
m_1 & n_1 \\ m_2 & n_2 \\ \end{array} \right | \ .
\]

\begin{figure}
\includegraphics[width=4.0in,height=4.0in]{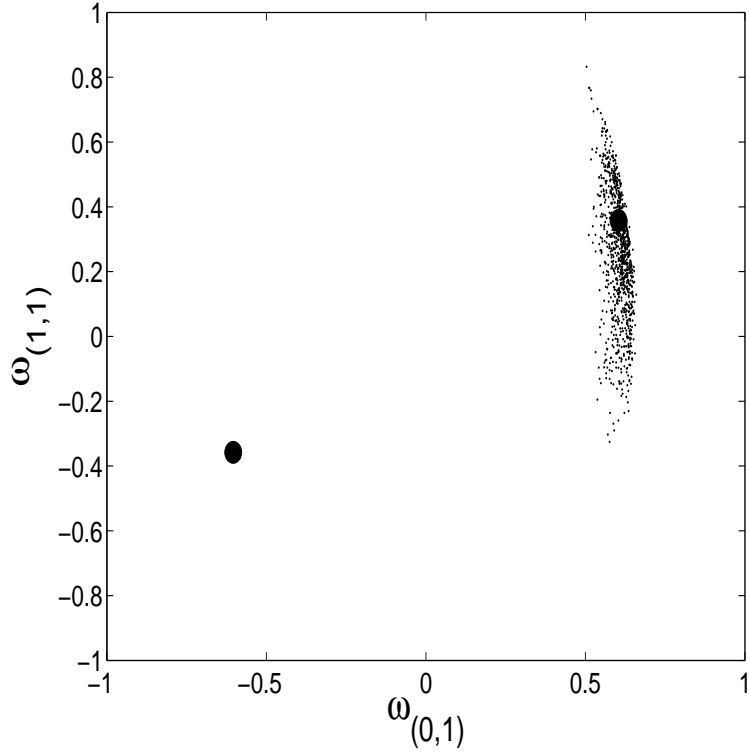}
\caption{The Poincar\'e return map plot projected onto the plane ($\om_{(0,1)},\om_{(1,1)}$)-plane,
in the case of the $\{ |k_1| \leq 2, |k_2| \leq 2 \}$ Galerkin truncation of the 2D Euler 
equation (i.e. $\e =0$), where $t \in [0, T]$, $T=4\times 10^4 \pi$, only the last $1000$ 
intersections are recorded.}
\label{fi2}
\end{figure}

\begin{figure}
\includegraphics[width=4.0in,height=4.0in]{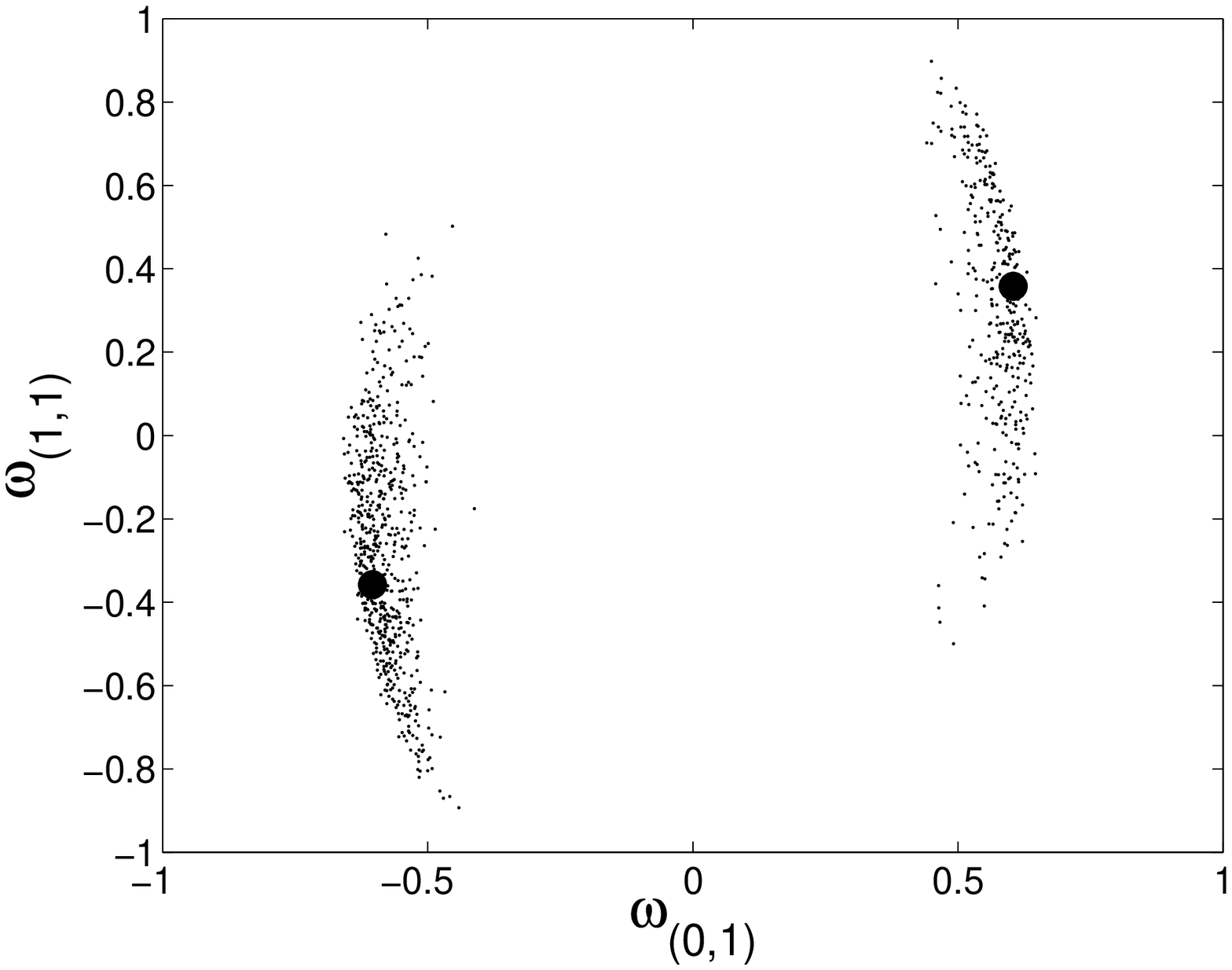}
\caption{The same as Figure \ref{fi2}, except $T=80\times 10^4 \pi$.}
\label{fi3}
\end{figure}

For the numerical simulation of chaos, we continue the study on the Galerkin truncation:
$\{ |k_1| \leq 2, |k_2| \leq 2 \}$. We will use the Melnikov integral (\ref{gMel}) to 
test the existence of chaos. We always start from the initial condition (\ref{IC}). 
We choose the external force
\begin{equation}
f = a \sin t \cos (x_1 +\al x_2)\ .
\label{EF}
\end{equation}
Then the Melnikov integral (\ref{gMel}) has the expression
\begin{equation}
M_0 = M_1 + a \sqrt{M_2^2+M_3^2} \sin (t_0 +\th )\ ,
\label{12mel}
\end{equation}
where 
\begin{eqnarray*}
& & \sin \th = \frac{M_3}{\sqrt{M_2^2+M_3^2}} \ , \quad \cos \th = \frac{M_2}{\sqrt{M_2^2+M_3^2}} \ , \\
M_1 &=& \frac{\al}{4\pi^2}\int_{-\infty}^{+\infty} \int_0^{2\pi /\al} \int_0^{2\pi}(\Om +\Dl^{-1} \Om )
\Dl \Om \ dx_1 dx_2dt \ , \\
M_2 &=& \frac{\al}{4\pi^2}\int_{-\infty}^{+\infty} \int_0^{2\pi /\al} \int_0^{2\pi}(\Om +\Dl^{-1} \Om )
\cos t \cos (x_1 +\al x_2)  \ dx_1 dx_2dt \ , \\
M_3 &=& \frac{\al}{4\pi^2}\int_{-\infty}^{+\infty} \int_0^{2\pi /\al} \int_0^{2\pi}(\Om +\Dl^{-1} \Om )
\sin t \cos (x_1 +\al x_2)  \ dx_1 dx_2dt \ ,
\end{eqnarray*}
where $\Om (t)$ is the approximate heteroclinic orbit in Figure \ref{fi1} with $\Om (0)$ given by 
(\ref{IC}). The time integral is in fact over the interval [$-11.8 ,11.8$] rather than ($-\infty , 
\infty$). Direct numerical computation gives that 
\[
M_1 =  -29.0977\ , \quad M_2 = -0.06754695 \ , \quad M_3 = 0 \ .
\]
Setting $M_0 = 0$ in (\ref{12mel}), we obtain that
\[
\sin (t_0 +\pi ) =  \frac{430.77741}{a} \ .
\]
Thus, when 
\begin{equation}
|a| > 430.77741\ ,
\label{MC}
\end{equation}
there are solutions to $M_0 = 0$. Next we will test the Melnikov criterion (\ref{MC}) and see if it is 
related to chaos. We define 
an average Liapunov exponent $\sg$ in the following manner: For a large time interval $t \in [0, T]$,
let $t_0 = T$ and 
\[
t_n = T + n 2\pi \ , \quad \text{ where } 0 \leq n \leq N \ , \text { and } 
N = 10^3 \text{ or } 2\times 10^3\ .
\]
We define
\[
\sg_n = \frac{1}{2\pi} \ln \frac{\| \Dl \om (t_n + 2\pi )\|}{\| \Dl \om (t_n )\|} \ .
\]
Then the average Liapunov exponent $\sg$ is given by
\[
\sg = \frac{1}{N} \sum_{n=0}^{N-1} \sg_n \ .
\]
We introduce the Poincar\'e return map on the section given by $\om (1,0) = 0$ and we only record 
one direction intersection (from $\om (1,0)$ positive to negative). For a large time interval 
$t \in [0, T]$, we only record the last $1000$ intersections and we use $\bullet$ to denote the 
intersections with the two heteroclinic cycles. All the numerical simulations start from the 
initial condition (\ref{IC}). The average Liapunov exponent computed here depends on the time 
interval, the ensemble of average, and computer accuracy. It only makes sense when it is compared in 
the same setting. When $\e =0$, there is no dissipation 
and no forcing. For a large time interval $t \in [0, T]$, the average Liapunov exponent $\sg$
is as follows:
\[
\begin{array}{cccc} T= 4\times 10^4 \pi & \quad T= 8\times 10^4 \pi & \quad  T= 12 \times 10^4 \pi
& \quad T= 80 \times 10^4 \pi \ , \cr 
\sg = 0.042 & \sg = 0.0344 & \sg = 0.044 & \sg = 0.0848 \ .\cr 
\end{array}
\]
Figures \ref{fi2}-\ref{fi3} are the corresponding Poincar\'e return map plots. The dynamics is chaotic.
It seems that the life time of the chaos is infinite (i.e. non-transient chaos). 

\begin{figure}
\includegraphics[width=4.0in,height=4.0in]{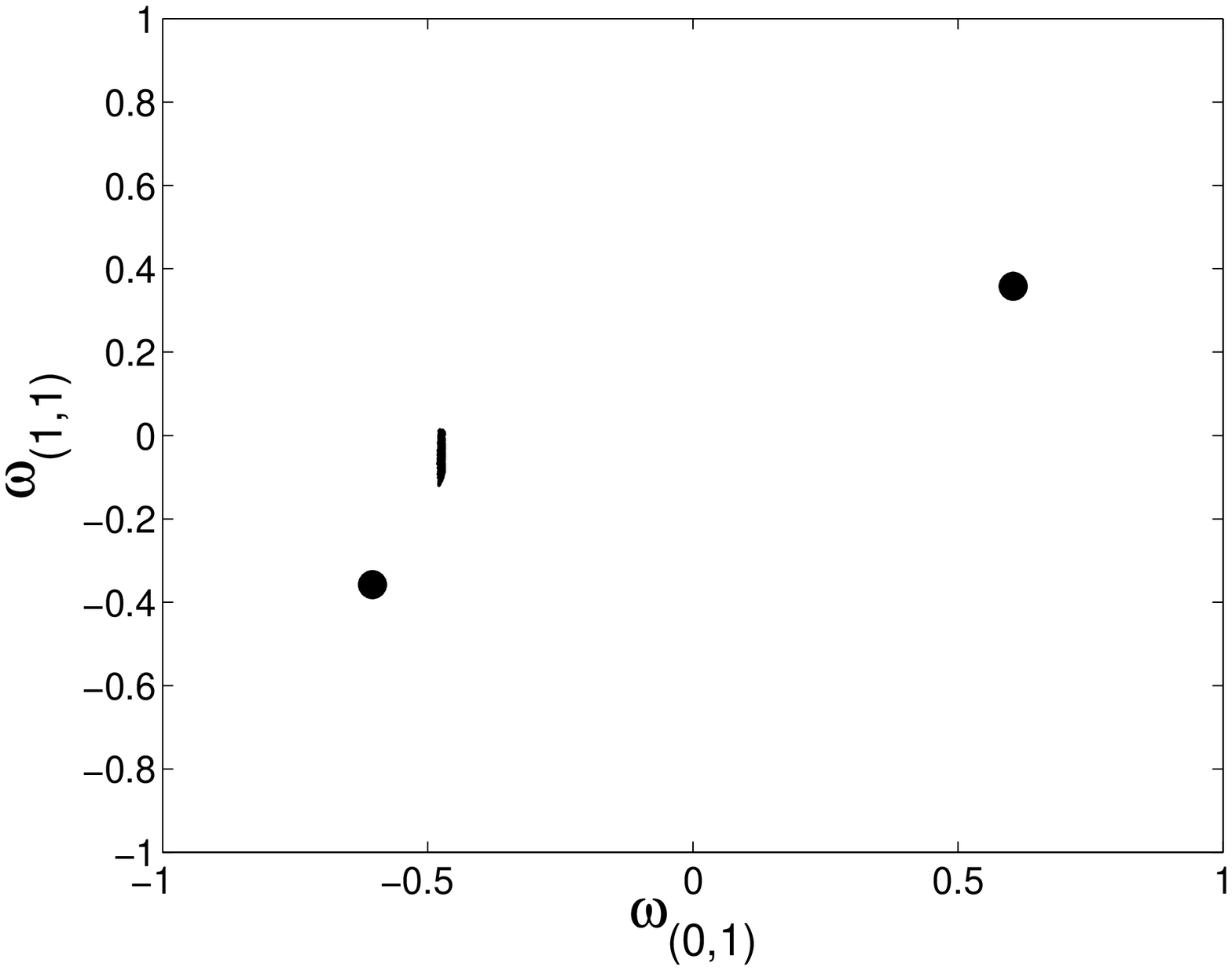}
\caption{The same as Figure \ref{fi2}, except $\e = 10^{-5}$, $a =1208$, and $T=2\times 10^4 \pi$.}
\label{fi4}
\end{figure}

\begin{figure}
\includegraphics[width=4.0in,height=4.0in]{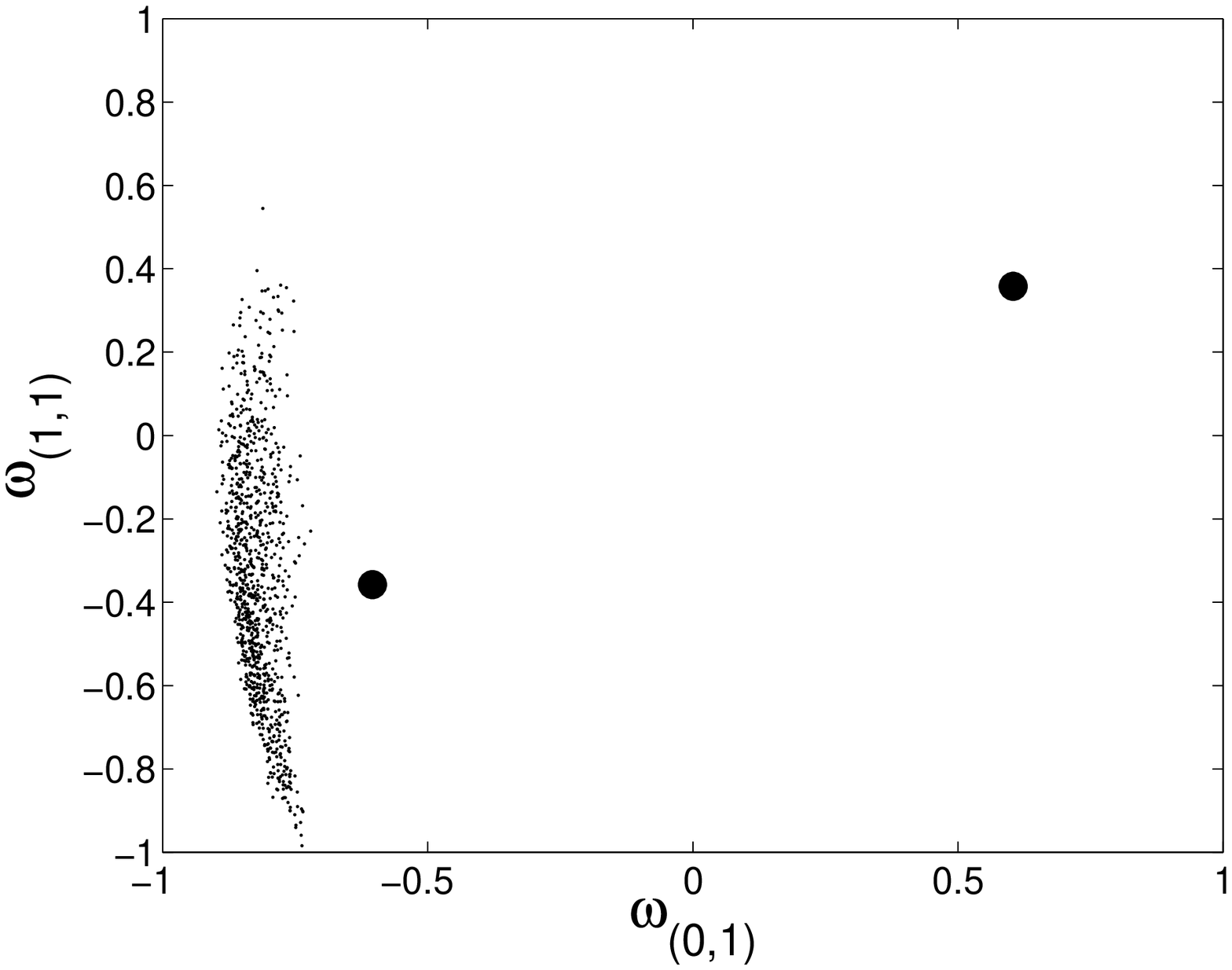}
\caption{The same as Figure \ref{fi2}, except $\e = 10^{-5}$, $a =1208.2$, and $T=2\times 10^4 \pi$.}
\label{fi5}
\end{figure}

When $\e >0$, we find that in all the cases that we tested, the dynamics is always a transient 
chaos. The Melnikov criterion is only some sort of necessary condition for the existence of 
heteroclinic chaos \cite{Li04}. When the Melnikov integral is zero, it gives an indication 
of a re-intersection of the broken heteroclinic orbit with certain large dimensional center-stable 
manifold \cite{Li04}. We believe that such a re-intersection will be reflected by the Liapunov 
exponent as inducing transient chaos. 
When $\e = 10^{-5}$, $a \in [0, 1208]$, and $T=4\times 10^4 \pi$, we find that 
\[
\sg \sim 10^{-4}\ .
\]
For instances,
\begin{equation}
\begin{array}{lllll}  a=400 &  a=430 & a=440 & a=650 & a=850 \ , \cr 
\sg = 4.9 \times 10^{-4}  & \sg = 2.6 \times 10^{-4} & \sg = 5.3 \times 10^{-4} & 
\sg = 5.9 \times 10^{-4} & \sg = 1.0 \times 10^{-4}  \ .\cr 
\end{array}
\label{noj}
\end{equation}
But we discover a sharp jump of $\sg$ near $a = 1208$ as shown below:
\begin{equation}
\begin{array}{lllll}  & T= 2\times 10^4 \pi &  T= 4 \times 10^4 \pi
& T= 8 \times 10^4 \pi & T= 8\times 10^5 \pi\ , \cr 
a=1208  & \sg = 3.6 \times 10^{-4} & \sg = 3.9 \times 10^{-4} & 
\sg = 4.1 \times 10^{-4} & \sg = 0 \ ,\cr 
a=1208.2 & \sg = 6.1 \times 10^{-2} & \sg = 6.1 \times 10^{-2} & 
\sg = 2.5 \times 10^{-2}& \sg =0 \ .\cr 
\end{array}
\label{shar}
\end{equation}
For $T= 4 \times 10^4 \pi$, the corresponding Poincar\'e return map plots are shown in 
Figures \ref{fi4}-\ref{fi5}. When $T= 80 \times 10^4 \pi$, the chaos for the $a=1208.2$ 
case also disappears. When $a > 1208.2$, $\sg$ can still be $\sim 10^{-4}$. But we 
did not observe any sharp jump of $\sg$.
We believe that the sharp jump of $\sg$ near $a=1208$ is due to the 
generation of new transient heteroclinic chaos. Our Melnikov integral calculation 
predicts when $|a| > 430.77741$, the broken heteroclinic orbit re-intersects 
with certain center-stable manifold. This does not mean that new heteroclinic 
orbit is automatically generated. Perhaps near $a=1208$, new heteroclinic cycle 
is generated and leads to transient heteroclinic chaos. Since our analysis cannot 
access such information in the phase space, our comments here are purely speculations.

\begin{remark}
We have also conducted numerical experiments on Galerkin truncations by keeping more modes: 
$\{ |k_1| \leq 4, |k_2| \leq 4 \}$ and $\{ |k_1| \leq 8, |k_2| \leq 8 \}$. We found that 
when $\e =0$, the strength of chaos increases as the modes are increased: For $T=8\times 10^3 \pi$,
\[
\begin{array}{lll} |k_1|, |k_2| \leq 2 &  |k_1|, |k_2| \leq 4 &  |k_1|, |k_2| \leq 8 \ , \cr 
\sg = 4.7 \times 10^{-2} & \sg = 1.3 \times 10^{-1} & \sg = 1.7 \times 10^{-1} \ . \cr 
\end{array}
\]
Nevertheless, this does not hint that the dynamics of 2D Euler equation is chaotic since all
Galerkin truncations are perturbations of the 2D Euler equation. In fact, higher 
single Fourier modes (as fixed points) have more eigenvalues with positive real parts. 

When $\e >0$, the strength of chaos decreases as the modes are increased. Higher modes 
have more dissipations. Also all the chaos are transient. After enough time ($\sim 
2\times 10^4 \pi$), the $\e =0$ chaos is almost smeared away by dissipation, we believe that at this stage 
the re-intersected heteroclinic orbits play a role and can enhance the transient chaos. It is this 
stage where the Melnikov calculation is effective. 
\end{remark}

\section{Melnikov Integral and Control of Chaos in 2D Navier-Stokes Equation} 

Now we turn on the control ($b \neq 0$). we continue the study on the Galerkin truncation:
$\{ |k_1| \leq 2, |k_2| \leq 2 \}$.  
We choose $\tdl (x)$ as follows
\begin{equation}
\tdl (x)= \sum_k e^{i(k_1 x_1 + \al k_2 x_2)}\ .
\label{SC}
\end{equation}
Then the Melnikov integral $M$ in (\ref{gMel}) is given by 
\begin{equation}
M=M_0 +b M_c \ ,
\label{CM}
\end{equation}
where $M_0$ is given by (\ref{12mel}) and 
\[
M_c = \frac{\al}{4\pi^2}\int_{-\infty}^{+\infty} \int_0^{2\pi /\al} \int_0^{2\pi}(\Om +\Dl^{-1} \Om )
\tdl (x) \ dx_1 dx_2dt \ , 
\]
evaluated along the approximate heteroclinic orbit in Figure \ref{fi1}. We find that
\[
M_c = -18.6884 \ .
\]
Thus
\[
M= -29.0977 -18.6884 \ b + 0.06754695 \ a \sin (t_0 +\pi )\ .
\]
When 
\[
b = - 1.557 \ ,
\]
the Melnikov integral $M$ has roots for any $a \neq 0$. 

All the numerical simulations start from the initial condition (\ref{IC}).
When $\e = 10^{-5}$, $b = - 1.557$, and $T=10^4 \pi$, we find that:
\[
\begin{array}{lllll}  a=1 &  a=10 & a=200 & a=400 & a=800 \ , \cr 
\sg = 7.1 \times 10^{-4}  & \sg = 8.8 \times 10^{-4} & \sg = 9.3 \times 10^{-4} & 
\sg = 9.0 \times 10^{-4} & \sg = 6.6 \times 10^{-4}  \ .\cr 
\end{array}
\]
\[
\begin{array}{lllll}  a=1000 &  a=1208 & a=1208.2 & a=1500 & a=3000 \ , \cr 
\sg = 8.6 \times 10^{-4}  & \sg = 8.1 \times 10^{-4} & \sg = 8.4 \times 10^{-4} & 
\sg = 8.6 \times 10^{-4} & \sg = 7.7 \times 10^{-4}  \ .\cr 
\end{array}
\]
In comparison with (\ref{noj}), the values of the Liapunov exponents under the control 
are doubled. Thus the control seems enhancing chaos but not dramatically. We did not observe 
the sharp jump of the values of $\sg$ around $a=1208$ as in the $b=0$ case (\ref{shar}).
Here the control theory is not as rigorous as that of sine-Gordon system in Appendix A. 
But we believe that the Melnikov integral can play a significant role in the control of 
chaos in NS equation. After all, the chaos in NS is generated by instabilities characterized 
by unstable eigenvalues. And these unstable eigenvalues persist for Euler equation as 
shown in a previous section. For Euler equation, these unstable eigenvalues characterize
hyperbolic structures which are very likely of heteroclinics type due to infinitely many 
constants of motion. Thus Melnikov integrals supported upon these hyperbolic structures 
should play an important role in predicting and controling chaos.

\section{Zero Viscosity Limit of the Spectrum of 3D Linear Navier-Stokes Operator}

We will study the following form of 3D Navier-Stokes equation with a control,
\begin{equation}
\pa_t \Om + (u \cdot \na) \Om - (\Om \cdot \na) u = \e [\Dl \Om + f(t,x) +b \tdl (x)] \ ,
\label{3DNS}
\end{equation}
where $u = (u_1, u_2, u_3)$ is the velocity, $\Om = (\Om_1, \Om_2, \Om_3)$
is the vorticity, $\na = (\pa_{x_1}, \pa_{x_2}, \pa_{x_3})$, 
$\Om = \na \times u$, $\na \cdot u = 0$, $\e = 1/\text{Re}$ is the inverse of 
the Reynolds number, $\Dl$ is 
the 3D Laplacian, and $f(t,x) = (f_1(t,x), f_2(t,x), f_3(t,x))$ is the external force,
$b\tdl (x)$ is the spatially localized control, and $b$ is the control parameter.
We also pose periodic boundary condition of period ($2\pi / \al , 2\pi / \be , 2\pi$), i.e.
the 3D NS is defined on the 3-torus $\mathbb{T}^3$. We require that $u$, $\Om$,
$f$ and $\tdl$ all have mean zero. In this case, $u$ can be uniquely 
determined from $\Om$ by Fourier transform:
\begin{eqnarray*}
U_1(k) &=& i |k|^{-2} [ k_2 \om_3(k) - k_3 \om_2(k)]\ , \\
U_2(k) &=& i |k|^{-2} [ k_3 \om_1(k) - k_1 \om_3(k)]\ , \\
U_3(k) &=& i |k|^{-2} [ k_1 \om_2(k) - k_2 \om_1(k)]\ ,
\end{eqnarray*}
which can be rewritten in the compact form 
\[
U_\ell (k) = i |k|^{-2} \ve_{\ell m n} k_m \om_n(k) \ ,
\]
where $\ve_{\ell m n}$ is the permutation symbol ($\ell , m, n = 1,2,3$),
\begin{eqnarray*}
& & k = (k_1,k_2,k_3)=(\al \k_1,\be \k_2, \k_3) \ , \quad \k = (\k_1,\k_2,\k_3) \ , \\
& & u_\ell (x) =\sum_{\k \in \ZZZ} U_\ell (k) e^{ik\cdot x} \ , \quad 
\Om_\ell (x) =\sum_{\k \in \ZZZ} \om_\ell (k) e^{ik\cdot x} \ .
\end{eqnarray*}
Using these Fourier transforms together with
\[
f_\ell (x) =\sum_{\k \in \ZZZ} F_\ell (k) e^{ik\cdot x} \ , \quad 
\tdl_\ell (x) =\sum_{\k \in \ZZZ} \Dl_\ell (k) e^{ik\cdot x} \ ,
\]
we can rewrite the 3D NS (\ref{3DNS}) into the kinetic form 
\begin{eqnarray}
& & \pa_t \om_\ell (k) + k_s \sum_{k=\tk + \hk}|\tk|^{-2}\tk_m \om_n(\tk )
[\ve_{\ell m n} \om_s(\hk ) - \ve_{s m n} \om_\ell (\hk )] \non \\
& & = \e [-|k|^2 \om_\ell (k) + F_\ell (k) + b \Dl_\ell (k)]\ . \label{K3D}
\end{eqnarray}
A popular example of fixed points of the 3D NS (\ref{3DNS}) is the so-called 
ABC flow \cite{Dom86}
\begin{equation}
u_1 = A \sin x_3 + C \cos x_2, \
u_2 = B \sin x_1 + A \cos x_3, \
u_3 = C \sin x_2 + B \cos x_1, 
\label{ABC}
\end{equation}
where $\al =\be =1$ and $\Om = u = f$ and $b=0$. The popularity comes from the fact that 
the Lagrangian fluid particle flow generated 
by the vector field (\ref{ABC}) can still be chaotic \cite{Dom86}. On the other hand, in 
Appendix B, we prove that the Lagrangian
flow generated by any solution to the 2D Euler equation is always integrable.

\subsection{A 3D Shear Fixed Point}

\begin{figure}
\includegraphics[width=4.0in,height=4.0in]{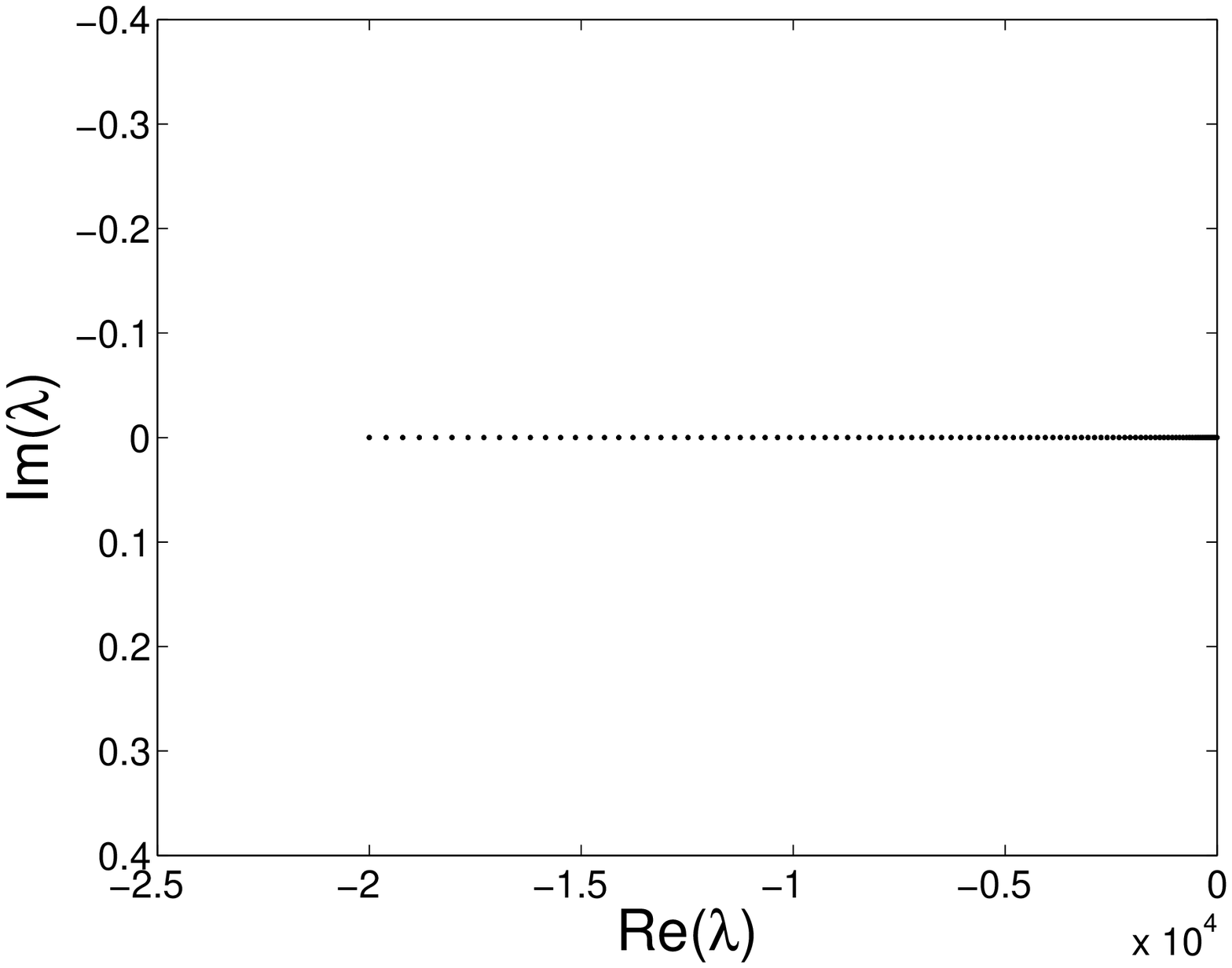}
\caption{The eigenvalues of the line $\hk = (\al ,0,0)$ are all negative when $\e =2.0$.}
\label{3ge1}
\end{figure}

\begin{figure}
\includegraphics[width=4.0in,height=4.0in]{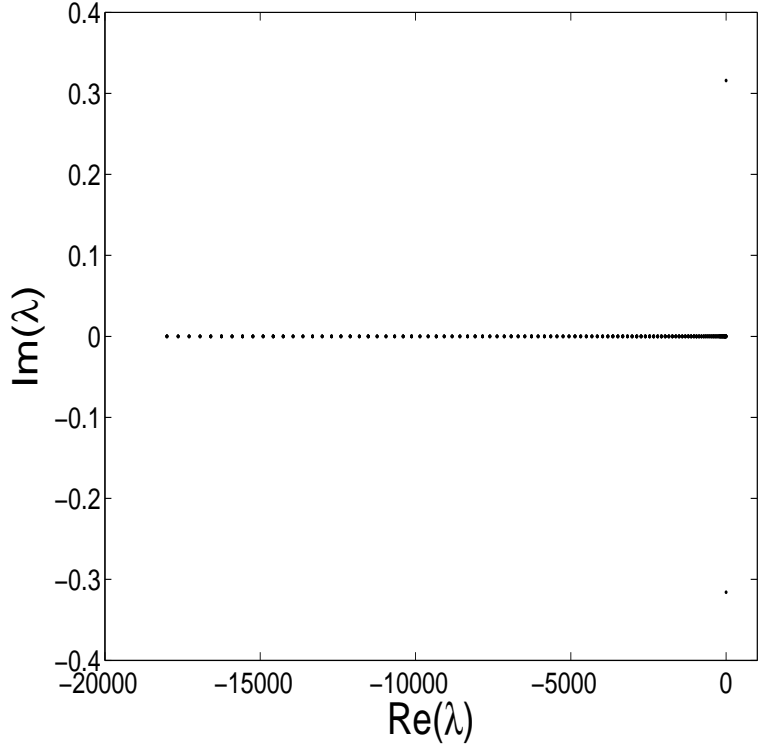}
\caption{A pair of eigenvalues of the line $\hk = (\al ,0,0)$ jumps off the real axis when $\e =1.8$.}
\label{3ge2}
\end{figure}

Below we will study the simplest fixed point -- the 3D shear flow (which is also a special 
case of the ABC flow (\ref{ABC}) where $A=2$ and $B=C=0$):
\begin{equation}
\Om_1 = 2 \sin x_3 \ , \quad 
\Om_2 = 2 \cos x_3 \ , \quad
\Om_3 = 0 \ . 
\label{SABC}
\end{equation}
Let $p=(0,0,1)$, the Fourier transform $\om^*$ of the fixed point is given by:
\[
\om^*_1(p) = - i , \ \om^*_2(p) = 1 , \ \om^*_1(-p) = i  , \ \om^*_2(-p) = 1 , 
\ \om^*_3(p) = \om^*_3(-p) = 0 ,
\]
and $\om^*_\ell (k) = 0$, $\forall k \neq p \text{ or } -p$. We choose 
$\al =0.7$ and $\be =1.3$ hoping that the fixed point $\om^*$ has only one unstable eigenvalue.
The spectral equations of the 3D linear NS operator at the fixed point $\om^*$ are given by 
\begin{eqnarray*}
& & [(k_1+ik_2)-ik_2|k-p|^{-2}]\om_1(k-p) \\
& & +[i|k-p|^{-2}k_1+i|k-p|^{-2}(k_1+ik_2)(k_3-1)]\om_2(k-p) \\
& & +[-1-ik_2(k_1+ik_2)|k-p|^{-2}]\om_3(k-p) +\e |k|^2 \om_1(k) \\
& & +[-(k_1-ik_2)-ik_2|k+p|^{-2}]\om_1(k+p)\\
& & +[i|k+p|^{-2}k_1-i|k+p|^{-2}(k_1-ik_2)(k_3+1)]\om_2(k+p) \\
& & +[-1+ik_2(k_1-ik_2)|k+p|^{-2}]\om_3(k+p) =-\la  \om_1(k) \ ,
\end{eqnarray*}
\begin{eqnarray*}
& & [k_2|k-p|^{-2}-i(k_3-1)(k_1+ik_2)|k-p|^{-2}]\om_1(k-p) \\
& & +[(k_1+ik_2)-k_1|k-p|^{-2}]\om_2(k-p) \\
& & +[-i+ik_1(k_1+ik_2)|k-p|^{-2}]\om_3(k-p) +\e |k|^2 \om_2(k) \\
& & +[-k_2|k+p|^{-2}+i(k_3+1)(k_1-ik_2)|k+p|^{-2}]\om_1(k+p) \\
& & +[-(k_1-ik_2)+k_1|k+p|^{-2}]\om_2(k+p) \\
& & +[i-ik_1(k_1-ik_2)|k+p|^{-2}]\om_3(k+p)  =-\la  \om_2(k) \ ,
\end{eqnarray*}
\begin{eqnarray*}
& & [ik_2(k_1+ik_2)|k-p|^{-2}]\om_1(k-p) 
+[-ik_1(k_1+ik_2)|k-p|^{-2}]\om_2(k-p) \\
& & +[k_1+ik_2]\om_3(k-p) +\e |k|^2 \om_3(k) 
+[-ik_2(k_1-ik_2)|k+p|^{-2}]\om_1(k+p)\\
& & +[ik_1(k_1-ik_2)|k+p|^{-2}]\om_2(k+p) 
+[-(k_1-ik_2)]\om_3(k+p) =-\la  \om_3(k) \ .
\end{eqnarray*}
Thus the 3D linear NS operator also decouples according to the lines $\hk + j p$ ($j\in 
\mathbb{Z}$). Next we study the zero viscosity limit of the spectrum of this 3D linear NS operator.
When $\hk = (\al ,0,0)$, we have tested the truncation of the line $\hk + j p$ up to $|j| \leq 400$.
The deformation pattern stays the same. Below we present the case $|j|\leq 100$ for which the 
pictures are more clear. Figure \ref{3ge1} shows the case $\e =2.0$ where all the eigenvalues 
are negative. Figure \ref{3ge2} shows the case $\e =1.8$ where a pair of eigenvalues
jumps off the real axis. When $\e \leq 0.66$, a unique positive eigenvalue appears.
Figure \ref{3ge1} shows the case $\e =0.0007$ where a bubble has developed.
\begin{figure}
\includegraphics[width=4.0in,height=4.0in]{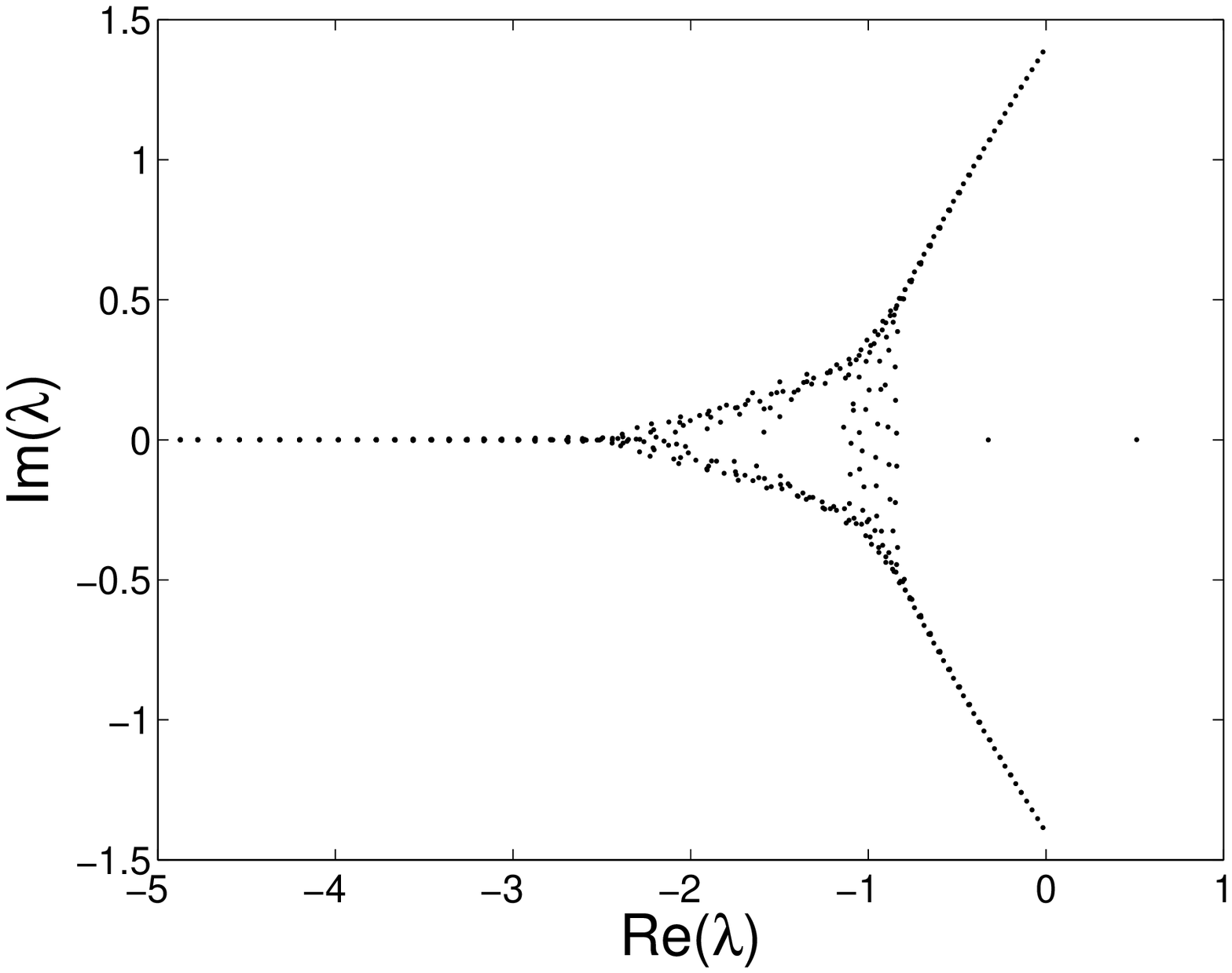}
\caption{A bubble of eigenvalues of the line $\hk = (\al ,0,0)$ has developed when $\e =0.0007$.}
\label{3ge3}
\end{figure}
As $\e \ra 0^+$, the limiting picture is the same with Figure \ref{ge7}. When 
$\e =0$, The spectrum picture is the same with Figure \ref{ge8}.
All other decoupled systems have the same bifurcation patterns but without the 
pair of persistent eigenvalues. For the entire spectrum of the 3D linear NS operator,
the limiting picture is the same with Figure \ref{ge11} as $\e \ra 0^+$; and the spectrum is the 
same with Figure \ref{ge12} when setting $\e =0$ \cite{Shv06}.
It seems that the Unstable Disk Theorem \cite{Li00} of the 2D linear Euler case is still valid: 
$|\hk +jp| < |p|$ for some $j$, implies that there is an eigenvalue of positive real part; while 
$|\hk +jp| > |p|$ for all $j$, implies that there is no eigenvalue of positive real part.

\subsection{The ABC Fixed Point}

\begin{figure}
\includegraphics[width=4.0in,height=4.0in]{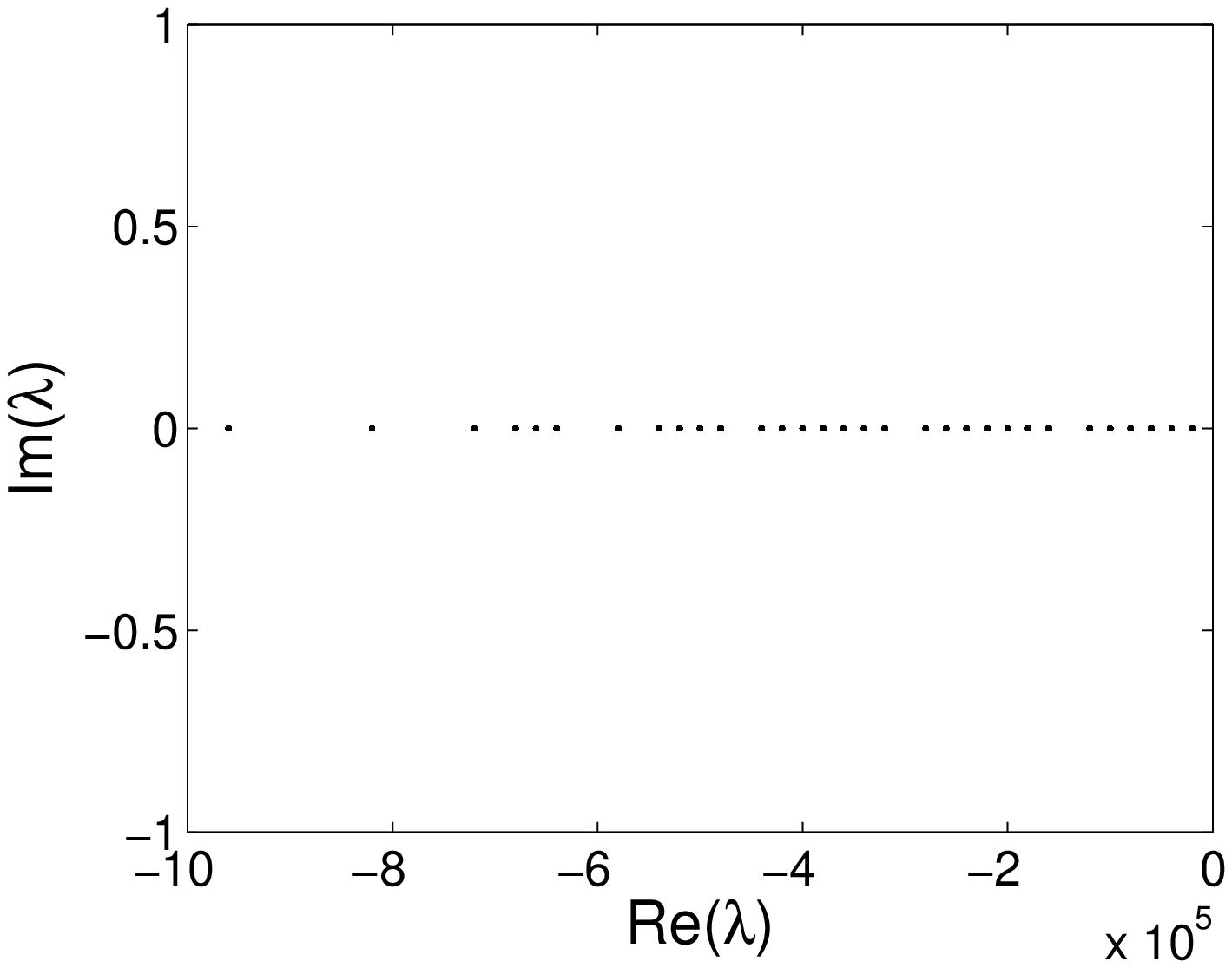}
\caption{The eigenvalues of the (Galerkin truncation of) linear NS at the ABC flow when $\e = 20000$.}
\label{eabc1}
\end{figure}

\begin{figure}
\includegraphics[width=4.0in,height=4.0in]{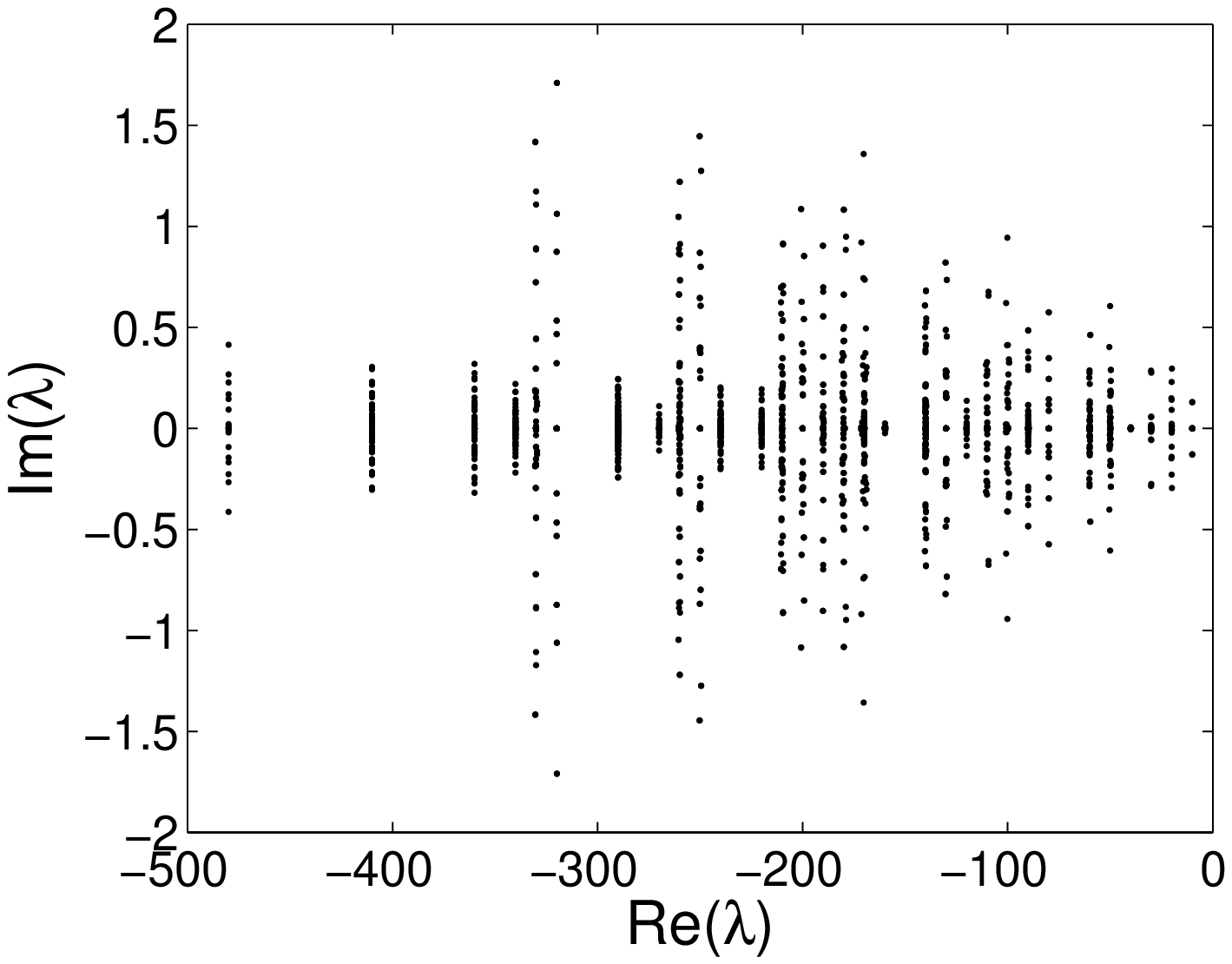}
\caption{The eigenvalues of the (Galerkin truncation of) linear NS at the ABC flow when $\e = 10$.}
\label{eabc2}
\end{figure}

\begin{figure}
\includegraphics[width=4.0in,height=4.0in]{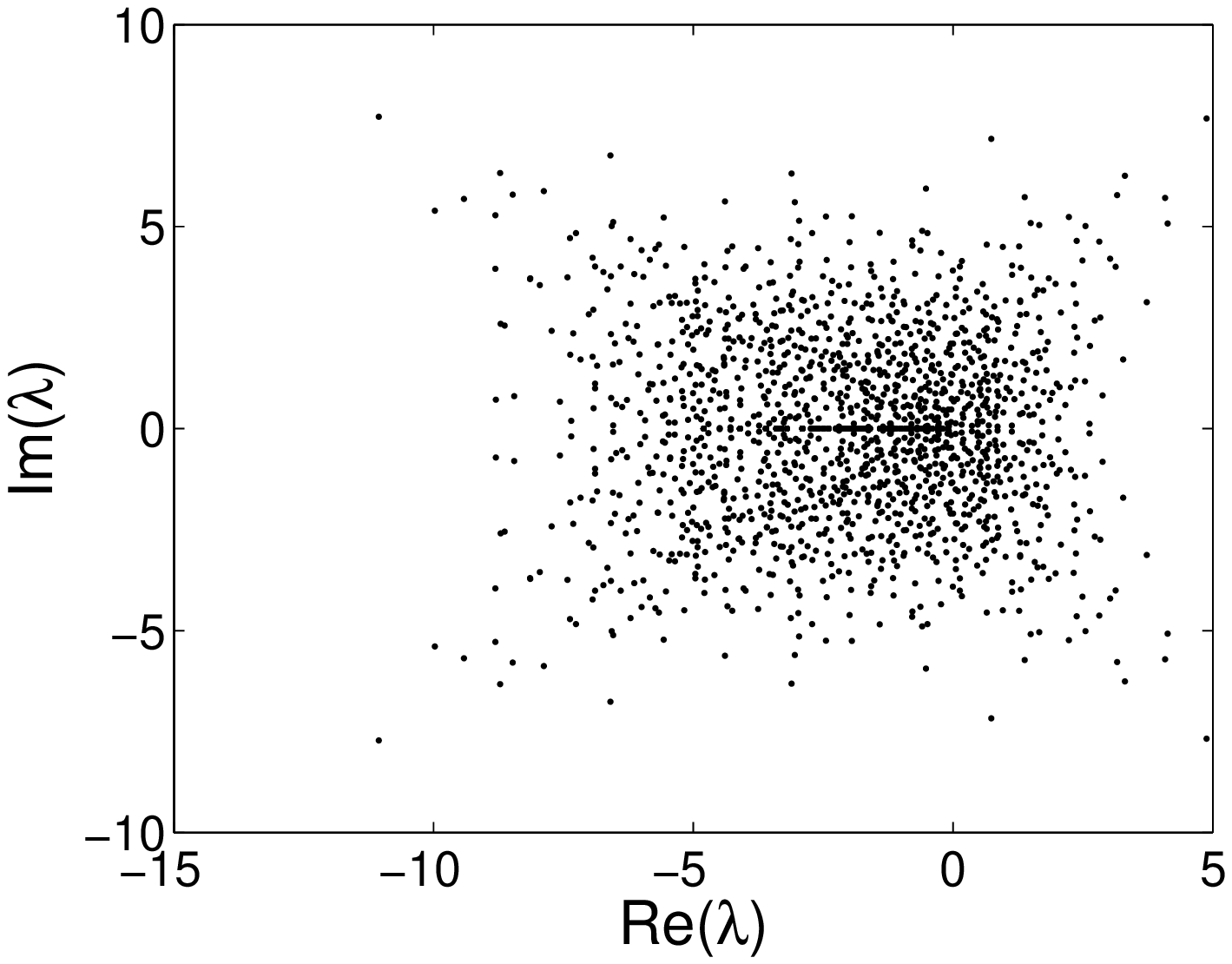}
\caption{The eigenvalues of the (Galerkin truncation of) linear NS at the ABC flow when $\e = 0.1$.}
\label{eabc3}
\end{figure}

\begin{figure}
\includegraphics[width=4.0in,height=4.0in]{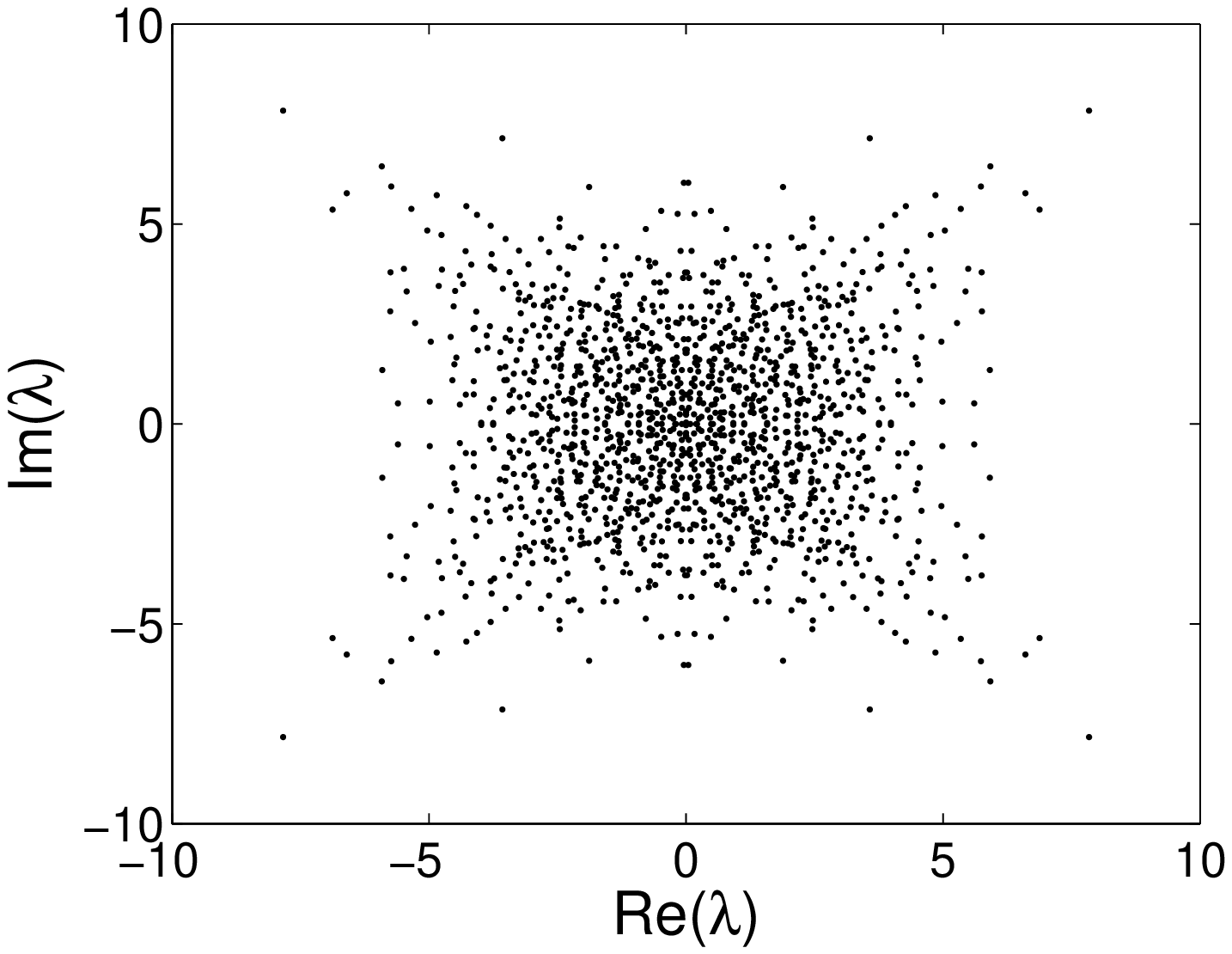}
\caption{The eigenvalues of the (Galerkin truncation of) linear NS at the ABC flow when $\e = 0$.}
\label{eabc4}
\end{figure}

In this case, the periodic domain is the cube, i.e. $\al =\be =1$. The ABC flow
is given specifically by
\[
\Om^*_1 = A \sin mx_3 + C \cos mx_2 , \ 
\Om^*_2 = B \sin mx_1 + A \cos mx_3 , \
\Om^*_3 = C \sin mx_2 + B \cos mx_1 ,  
\]
where $m$ is a positive integer, and ($A,B,C$) are real parameters.
In terms of Fourier modes: Let $p=(m,0,0)$, $q=(0,m,0)$, and $r=(0,0,m)$, 
then the ABC flow is given by
\begin{eqnarray*}
& & \om_1^*(q) = \frac{1}{2} C, \ \om_1^*(-q) = \frac{1}{2} C, \\
& & \om_1^*(r) = \frac{1}{2i} A, \ \om_1^*(-r) = -\frac{1}{2i} A, \\
& & \om_2^*(p) = \frac{1}{2i} B, \ \om_2^*(-p) = -\frac{1}{2i} B, \\
& & \om_2^*(r) = \frac{1}{2} A, \ \om_2^*(-r) = \frac{1}{2} A, \\
& & \om_3^*(p) = \frac{1}{2} B, \ \om_3^*(-p) = \frac{1}{2} B, \\
& & \om_3^*(q) = \frac{1}{2i} C, \ \om_3^*(-q) = -\frac{1}{2i} C .
\end{eqnarray*}
The spectral equation for the linear 3D NS operator at the ABC flow is 
then given by
\begin{eqnarray*}
\la \om_\ell (k) &=& -\e |k|^2 \om_\ell (k) -
k_s \sum_{k=\tk + \hk}\bigg [ |\tk|^{-2}\tk_m \om^*_n(\tk )
[\ve_{\ell m n} \om_s(\hk ) - \ve_{s m n} \om_\ell (\hk )] \\
& & +|\tk|^{-2}\tk_m \om_n(\tk )
[\ve_{\ell m n} \om^*_s(\hk ) - \ve_{s m n} \om^*_\ell (\hk )] \bigg ]. 
\end{eqnarray*}
Calculating the eigenvalues of the Galerkin truncations of this system becomes challenging.
Beyond the size $\{ |k_n| \leq 6, \ n=1,2,3 \}$, the computing time is too long. 
Below we present some pictures for the Galerkin truncation $\{ |k_n| \leq 4 , \ n=1,2,3 \}$.
We choose $m=1$, $A=1.2$, $B=0.7$ and $C=0.9$. When $\e = 20000$, all the eigenvalues 
are negative as shown in Figure \ref{eabc1}. As $\e$ is decreased, eigenvalues start to jump 
off the real axis and form vertical lines as shown in Figure \ref{eabc2} when $\e =10$,
in contrast to the parabolas in the cases of cat's eye and 3D shear. 
When $\e$ is decreased to $\e =0.1$, many eigenvalues move to the right half plane, i.e. there 
are many unstable eigenvalues as shown in Figure \ref{eabc3}. Notice that for the 
full linear NS operator, there should be an infinite tail of negative eigenvalues to the left.
When $\e =0$, the eigenvalues of the Galerkin truncation of linear NS are symmetric with respect 
to the real and imaginary axes as shown in Figure \ref{eabc4}. When $\e =0$, the full linear 
Euler operator at the ABC flow has a continuous spectrum similar to that at the cat's eye 
\cite{Shv06}. That is, the continuous spectrum of the linear Euler at the ABC flow, in any Sobolev space 
$H^s(\mathbb{T}^2)$ where $s$ is a non-negative integer, is a vertical band of width 
$2s \sg$ symmetric with respect to the imaginary axis 
$\{ \la \ : \ |\text{Re}(\la )| \leq s \sg \}$ as shown in Figure \ref{kat7m1}, where 
$\sg >0$ is the largest Liapunov exponent of the vector field given by ABC flow. Thus the width of 
the vertical band is proportional to the scale $s$ of the 
Sobolev space $H^s(\mathbb{T}^2)$ \cite{Shv06}. The union of all such bands for all integers $s \geq 0$ is 
the entire complex plane. The eigenfunctions of the linear NS at the ABC flow  when $\e >0$ 
belong to $H^s(\mathbb{T}^2)$ for all integers $s \geq 0$. As $\e$ is decreased, the eigenvalues 
move into the right half plane. The $\e \ra 0^+$ limiting picture of the eigenvalues of the linear NS 
at the ABC flow is that the 
eigenvalues are dense on the entire plane, in contrast to the left half plane in the case of cat's 
eye as shown in Figure \ref{kat6m1}. That is, all the eigenvalues of the 
linear NS at the ABC flow  condense into the entire plane -- ``condensation''. Thus the possible 
instability hinted by the right half band of the 
continuous spectrum of linear Euler in $H^s(\mathbb{T}^2)$ may be realized by real 
viscous fluids in this case in contrast to the cat's eye case.


\section{Numerical Verification of the Heteroclinics Conjecture for 3D Euler Equations}

\begin{figure}
\includegraphics[width=4.0in,height=4.0in]{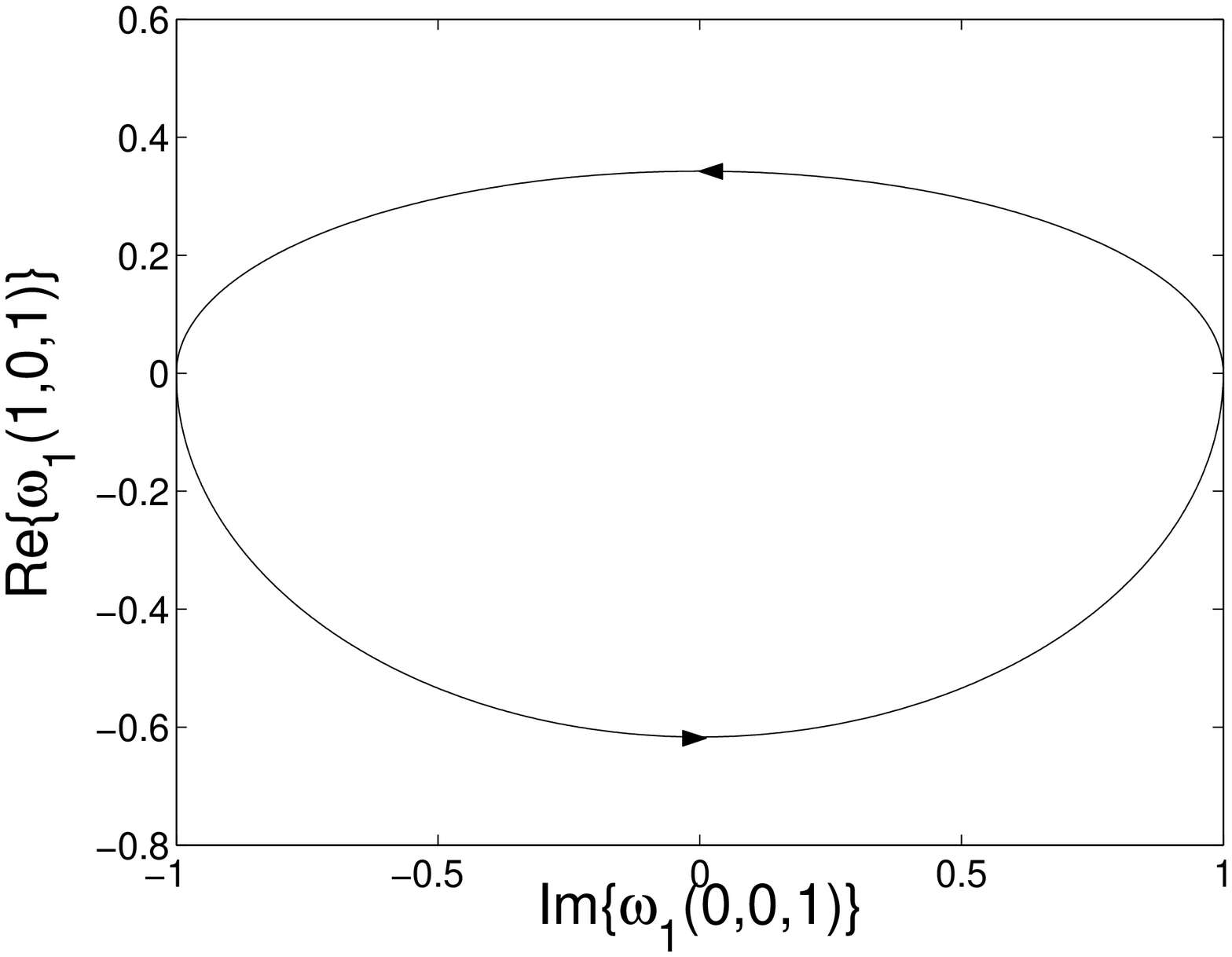}
\caption{An approximate heteroclinic cycle of the Galerkin truncation: $|\k_n| \leq 1$ ($n=1,2,3$) 
of the 3D Euler equations.}
\label{3gh}
\end{figure}

For 3D Euler equations, one can also pose the heteroclinics conjecture.
\begin{itemize}
\item The Heteroclinics Conjecture: There is a heteroclinic orbit of the 3D Euler 
equations that connects $\Om$ (\ref{SABC}) and $-\Om$. 
\end{itemize}
As discussed in the Introduction, the rationality of this conjecture comes from the fact 
that 3D Euler equations have a Lax pair \cite{LY03}.
Below we will verify this conjecture for the Galerkin truncation: $|\k_n| \leq 1$ ($n=1,2,3$)
where $k=(\al \k_1, \be \k_2, k_3)$. Even though this is the smallest Galerkin truncation, 
the dimension of the resulting system is still very large.
For this Galerkin truncation, the fixed point (\ref{SABC})
is still a fixed point. The linearized Galerkin truncation operator at this fixed point 
can be obtained by the corresponding Galerkin truncation the 3D linear Euler operator. 
In this case, the line segment labeled by $\hk = (\al ,0,0)$ ($\e=0$) has a positive eigenvalue 
$\la = 0.5792$, and the corresponding eigenvector $v$ is given by:
\begin{eqnarray*}
& & \om_1 (1,0,-1) = 0.328919 -i \ 0.246347, \
\om_2 (1,0,-1) = -0.246347 -i \ 0.328919 , \\
& & \om_3 (1,0,-1) = 0.230243 -i \ 0.172443  ,
\ \om_1 (1,0,0) = 0  ,  \\
& & \om_2 (1,0,0) = -0.19583 -i \ 0.26147  , \\
& & \om_3 (1,0,0) = 0.183029 -i \ 0.137081  ,  
\ \om_1 (1,0,1) = 0.328919 -i \ 0.246347  , \\
& & \om_2 (1,0,1) = 0.246347 +i \ 0.328919  , 
\ \om_3 (1,0,1) = -0.230243 +i \ 0.172443 , 
\end{eqnarray*}
and all other $\om_\ell (k)$'s are zero. Starting from the initial condition
\begin{equation}
\om = \om^* + 10^{-3} v \ , \label{3IC}
\end{equation}
where $\om^*$ is the Fourier transform of the fixed point (\ref{SABC}), the approximate 
heteroclinic orbit reaches order $\sim 10^{-3}$ neighborhood of $-\om^*$ during the 
time interval [$0, 29.33$]. This approximate heteroclinic orbit is the lower branch of the 
approximate heteroclinic cycle shown in Figure \ref{3gh}.
Notice that the approximate heteroclinic orbit here does not have the extra loop as in 
Figure \ref{fi1}. When more modes are included in the Galerkin truncation, extra loops may be 
generated.

\section{Melnikov Integral and Numerical Simulation of Chaos in 3D Navier-Stokes Equation}

Setting $\e = 0$ in the 3D NS (\ref{3DNS}), one gets the corresponding 
3D Euler equation for which one has the following constants of motion:
\[
E=\int_{\mathbb{T}^3} |u|^2dx\ , \quad H=\int_{\mathbb{T}^3} u \cdot \Om dx 
\]
where $E$ is the kinetic energy and $H$ is the helicity. We will use the constant of motion
\[
G=E-H=\int_{\mathbb{T}^3} |u|^2dx - \int_{\mathbb{T}^3} u \cdot \Om dx 
\]
to build a Melnikov integral for the corresponding 3D Navier-Stokes equation (\ref{3DNS}).
We will try to make use of the Melnikov integral as a measure of chaos and to conduct 
a control of chaos, around the 3D shear flow (\ref{SABC}). The gradient of $G$ in $u$ or $\Om$ is given by
\[
\na_uG=2(u-\Om )\ , \quad \na_\Om G =2\text{ curl}^{-1} (u-\Om )\ ,
\]
where curl $= \na \times \ $. The gradient is zero at the 3D shear flow (\ref{SABC}). We define the 
Melnikov function for the 3D NS (\ref{3DNS}) as
\begin{eqnarray*}
M &=& \frac{\al \be}{16 \pi^3}\int_{-\infty}^{+\infty} \int_{\mathbb{T}^3} \na_\Om G 
[\Dl \Om + f(t,x) +b\tdl (x)] dxdt \\
&=& \frac{\al \be}{16 \pi^3}\int_{-\infty}^{+\infty} \int_{\mathbb{T}^3} 2\text{ curl}^{-1} (u-\Om )
[\Dl \Om + f(t,x) +b\tdl (x)] dxdt \ .
\end{eqnarray*}

Next we conduct numerical simulations on the Galerkin truncation $|\k_n| \leq 1$ ($n=1,2,3$). 
When $\e =0$, the Liapunov exponent $\sg =0$ for all the numerical tests 
that we run. This indicates that there 
is no chaos when $\e =0$. Often the smallest Galerkin truncation is an integrable system 
\cite{Li03e} \cite{Li06e}. In such a circumstance, the Melnikov integral represents the 
leading order term of the distance between the broken heteroclinic orbit and the 
center-stable manifold of the fixed point. But the dimension of the center-stable manifold is 
large. The zero of the Melnikov integral implies that the unstable manifold in which the broken 
heteroclinic orbit lives, intersects with the center-stable manifold. Therefore, there is a 
new heteroclinic orbit which lives in the intersection. Such a heteroclinic orbit does not 
immediately imply the existence of chaos, even though it may lead to some transient chaos 
characterized by finite time positive Liapunov exponent (infinite time positive Liapunov exponent 
is zero). To compute the Melnikov integral, we choose the external force and control as follows
\begin{eqnarray*}
& & f_1 = a \sin t \cos (x_1 +\al x_2)\ , \quad f_2=f_3=0\ , \\
& & \tdl_1 (x)= \sum_{\k} e^{ik\cdot x} \ , \quad \tdl_2=\tdl_3 = 0 \ ,
\end{eqnarray*}
where the sum is over the Galerkin truncation.
Then the Melnikov integral has the expression
\begin{equation}
M = M_1 + a \sqrt{M_2^2+M_3^2} \sin (t_0 +\th )+bM_4 \ ,
\label{3mel}
\end{equation}
where 
\begin{eqnarray*}
& & \sin \th = \frac{M_3}{\sqrt{M_2^2+M_3^2}} \ , \quad \cos \th = \frac{M_2}{\sqrt{M_2^2+M_3^2}} \ , \\
M_1 &=& -\int_{-\infty}^{+\infty} \sum_k \text{Re} \left \{ i \ve_{\ell mn}k_m (i|k|^{-2} 
\ve_{nsr} k_s \om_r -\om_n)\overline{\om_\ell (k)}\right \} dt \ , \\
M_2 &=& \int_{-\infty}^{+\infty} \cos t \ \text{Re} \left \{ i |k|^{-2}\ve_{1 mn}k_m (i|k|^{-2} 
\ve_{nsr} k_s \om_r -\om_n)\right \}_{k=(\al , 0 ,1)} dt \ , \\
M_3 &=& \int_{-\infty}^{+\infty} \sin t \ \text{Re} \left \{ i |k|^{-2}\ve_{1 mn}k_m (i|k|^{-2} 
\ve_{nsr} k_s \om_r -\om_n)\right \}_{k=(\al , 0 ,1)} dt \ , \\
M_4 &=& \int_{-\infty}^{+\infty} \sum_k \text{Re} \left \{ i |k|^{-2}\ve_{1 mn}k_m (i|k|^{-2} 
\ve_{nsr} k_s \om_r -\om_n)\right \} dt \ , 
\end{eqnarray*}
where the sum is over the Galerkin truncation, all the integrals are evaluated along the 
lower heteroclinic orbit in Figure \ref{3gh} for the time interval [$-29.33/2, 29.33/2$],
rather than ($-\infty , \infty$). Direct numerical computation gives that 
\[
M_1 = 645.7 \ , \quad  M_2 = 1.581\ , \quad M_3= 0 \ , \quad 
M_4= 47.86 \ .
\]
When $b =0$ (no control), $M$ has roots when 
\[
|a| >  408.4 \ .
\]
When $\e = 10^{-5}$, $b= 0$, and $T=4\times 10^4 \pi$, we find that 
\[
\begin{array}{lllll}  a=300 & a=400 & a=420 & a=600 & a=800 \cr 
\sg = 0.8 \times 10^{-5} & \sg = 1.3 \times 10^{-4} & \sg = 0.8 \times 10^{-4} & 
\sg = 4.9 \times 10^{-4}  & \sg = 4.9 \times 10^{-4} \ .\cr 
\end{array}
\]
Around $a=400$, $\sg$ has a jump of one order which seems to be in agreement with the 
Melnikov prediction. However, when $a=100$, $\sg = 3.6 \times 10^{-4}$ which shows that 
the Melnikov prediction is not very effective. We do not know the specific reason.

\section{Melnikov Integral and Control of Chaos in 3D Navier-Stokes Equation}

Now we turn on the control ($b \neq 0$). When 
\[
b = - M_1/M_4 \approx  -13.5 \ ,
\]
the Melnikov integral $M$ (\ref{3mel}) has roots for any $a$.

When $\e = 10^{-5}$, $b= -13.5$, and $T=4\times 10^4 \pi$, we find that 
\[
\begin{array}{llll}  a=1 & a=10 & a=100 & a=200 \cr 
\sg = 0.098 & \sg = 0.125 & \sg = 0.095 & 
\sg = 0.083  \ .\cr 
\end{array}
\]
Thus under the control, chaos exists even when $a=1$. Also changing the value of 
$b$, $\sg$ does not change dramatically. Therefore, our control prediction above 
may not be very effective.

\section{Numerical Verification of the Heteroclinics Conjecture for a Line Model}

Returning to the 2D Navier-Stokes equation (\ref{2DNS}), numerical simulations on large 
Galerkin truncations are still challenging to the current computer ability. Here we 
will study a simple line model \cite{Li03e} obtained by a special Galerkin truncation
\cite{Li03e}. Let $p=(1,0)$ and $\hk = (0, \al )$, the line model is given by the 
Galerkin truncation:
\[
\{ \pm p, \ \pm (\hk + n p), \quad \forall n \in \mathbb{Z} \}
\]
We will work in the invariant subspace where $\om_k$'s are real-valued. The governing 
equation of the line model is
\begin{eqnarray}
\dot{\om}_n &=& A_{n-1} \om_* \om_{n-1} - A_{n+1} \om_* \om_{n+1}\non \\
& & +\e [-(n^2+\al^2)\om_n + F_n +b \Dl_n] \ , \label{LM1} \\
\dot{\om}_* &=& -\sum_{n\in \mathbb{Z}}A_{n-1,n} \om_{n-1} \om_n 
+\e [-\om_* + F_* +b \Dl_*]\ , \label{LM2} 
\end{eqnarray}
where $\om_n = \om_{\hk + np}$, $\om_*=\om_p$, similarly for $F$ and $\Dl$ as the Fourier 
transforms of $f$ and $\tdl$, and
\begin{eqnarray*}
& & A_n = 2A(p, \hk + np) = \al \left [ \frac{1}{n^2+\al^2}-1 \right ] \ , \\
& & A_{n-1,n}=2A(\hk +(n-1)p, \hk + np) = \al \left [ \frac{1}{(n-1)^2+\al^2}- 
\frac{1}{n^2+\al^2}\right ] \ .
\end{eqnarray*}
For the line model, verification of the heteroclinics conjecture is relatively easier.
First of all, for the line model ($\e=0$), it can be proved that the fixed point 
$\Om = 2 \cos x_1$ 
has a 1-dimensional local unstable manifold $W^u$. The basic idea of the proof is 
that one can apply the Riesz projections to the spectrum of the linearized line
model operator at the fixed point, and the nonlinear terms have bounded coefficients 
so that they are quadratic in a Banach algebra. For the full 2D Euler equation, 
the difficulty lies at the fact that the nonlinear term is not quadratic in a Banach algebra.

\begin{figure}
\includegraphics[width=4.0in,height=4.0in]{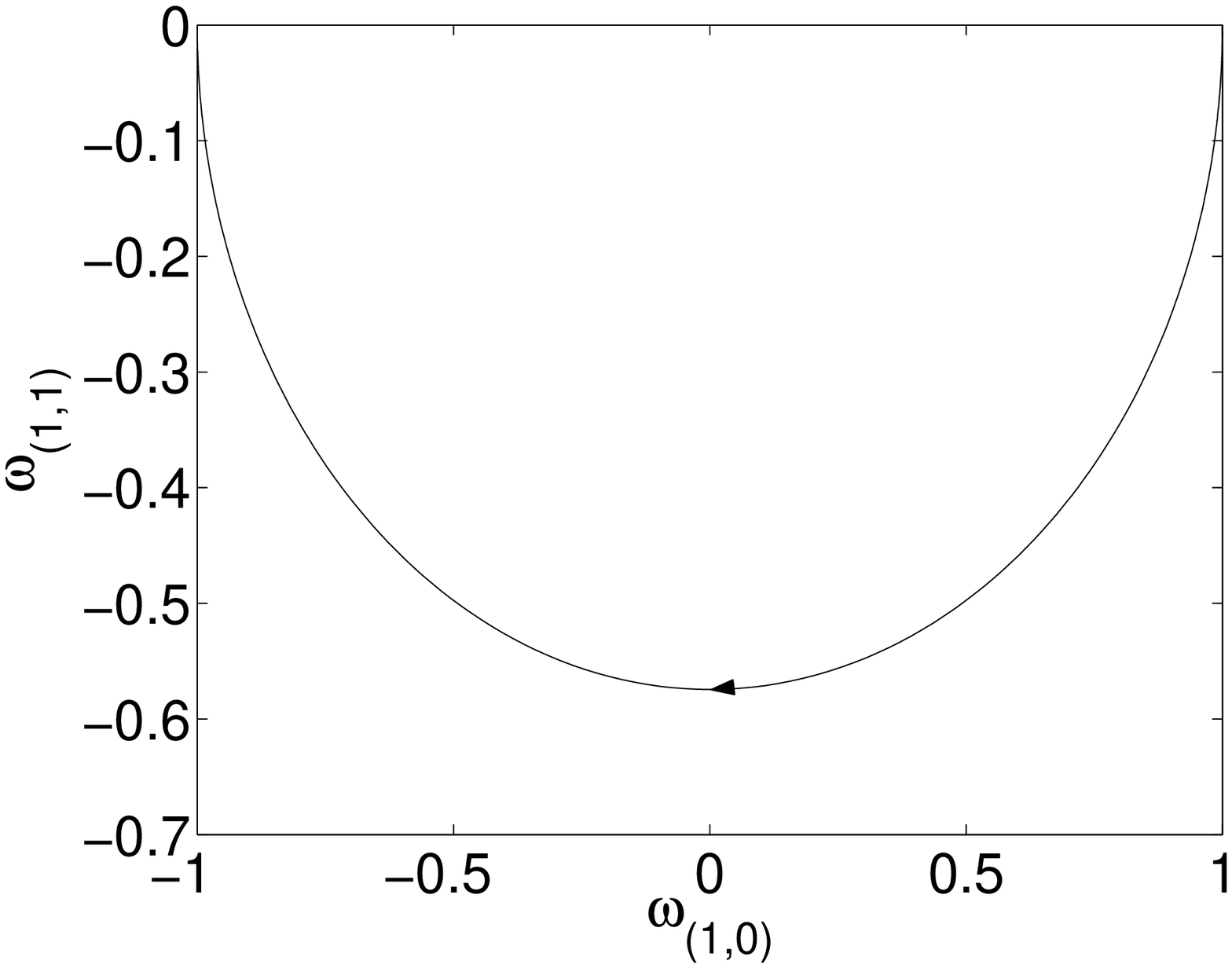}
\caption{Numerically exact heteroclinic orbit of the line model ($\e=0$) for $|n| \leq 32$.}
\label{1L}
\end{figure}

Denote by $\Sg$ the 1 co-dimensional hyperplane
\[
\Sg = \left \{ \om \ | \ \om_{(1,0)} = 0 \right \} \ .
\]
We have the corollary of Theorem \ref{hcthm}.
\begin{corollary}
Assume that $W^u \cap \Sg \neq \emptyset$; then the heteroclinics conjecture is true, i.e.  
there is a heteroclinic orbit that connects $\Om = 2 \cos x_1$ and $-\Om$.
\end{corollary}
For any truncation ($|n| \leq N$) of the line model, we first calculate the 
unstable eigenvector. Then we track the heteroclinic orbit with the 
initial condition provided by the unstable eigenvector. 
Numerically exact heteroclinic orbit is obtained for any $N$ ($|n| \leq N$). That is,
for any $N$ ($|n| \leq N$), it can be verified numerically that 
\[
W^u \cap \Sg \neq \emptyset \ .
\]
For $|n| \leq 32$, the heteroclinic orbit is shown in Figure \ref{1L}.
In comparison with the full 2D Euler equation, the hyperplane $\Sg$ here is only 
$1$ co-dimensional. This is the simplest nontrivial case to study the intersection 
$W^u \cap \Sg$.
We also conduct calculations on the Liapunov exponents. When $\e =0$, $|n| \leq 32$, 
$\sg =0$ for all the computed time intervals, which means that there is no chaos.
This is true for any $N$ ($n \leq N$) and any computational time interval. This 
indicates that the line model may be integrable when $\e =0$. From these facts, it is clear 
that the line model is a good starting point for a rigorous analysis. For instance, it is 
hopeful to make the Melnikov integral theory rigorous.

\section{Melnikov Integral and Numerical Simulation of Chaos in the Line Model}

For the line model, the kinetic energy and enstrophy are still invariants when $\e=0$. 
Choosing the same external force (\ref{EF}) and control (\ref{SC}), we have the Melnikov 
integral which is the same with that of 2D NS except that the Fourier modes summation is over 
the line model,
\begin{equation}
M_0 = M_1 + a \sqrt{M_2^2+M_3^2} \sin (t_0 +\th )+ bM_c\ .
\label{Lmel}
\end{equation}
For the truncation $|n| \leq 32$, we evaluate these integrals along the heteroclinic orbit
in Figure \ref{1L}, and obtain that
\[
M_1 = -6.0705, \ M_2 = -0.10665, \ M_3 = 0, \ M_c = 11.9728
\]
For the case of no control ($b=0$), when 
\[
|a| > 56.92 
\]
the Melnikov integral $M$ has roots. 

We conduct some numerical simulations on 
the (transient) chaos. When $\e = 10^{-3}$, and $T=2\times 10^3 \pi$, we find that 
\[
\begin{array}{ll}  a \leq 200 & a=400 \cr 
\sg < 0 & \sg = 7.2 \times 10^{-4}  \ .\cr 
\end{array}
\]
According to the roots of the Melnikov integral, when $|a| > 56.92$, the broken
heteroclinic orbit may re-intersect with certain center-stable manifold. According 
the above Liapunov exponent result, transient chaos is generated when $a=400$ 
which may be due to the generation of new heteroclinic cycles.
As can be expected, the Melnikov prediction performs better here for the line model.
When $\e =0$, there is no chaos. When $\e > 0$, we see that transient chaos 
appears when $|a|$ is greater than certain threshold which is in the range 
$|a| > 56.92$ predicted by the Melnikov integral.

\section{Melnikov Integral and Control of Chaos in the Line Model}

Now we turn on the control ($b \neq 0$). When 
\[
b=-M_1 /M_c = 0.50702425
\]
the Melnikov integral $M$ has roots for any $a$. To test the effectiveness of the 
control, we set $b$ to the above value and conduct some numerical simulations on 
the (transient) chaos.
When $\e = 10^{-3}$, $b= 0.50702425$, and $T=2\times 10^3 \pi$, we find that 
\[
\begin{array}{lllll}  a =1 & a=10 & a=50 & a=200 & a=400  \cr 
\sg = -12.6 \times 10^{-4}  & \sg = 2 \times 10^{-4} & \sg = 0.2 \times 10^{-4} & 
\sg = 2 \times 10^{-4} & \sg = 0.87 \times 10^{-4} \ .\cr 
\end{array}
\]
The control clearly enhanced the transient chaos. The control effectively pushed 
the threshold of $a$ backward from $400$ to $10$ for the generation of transient chaos.
This shows that the Melnikov integral control performs better here for the line model.

\section{Numerical Verification of the Heteroclinics Conjecture for a Two Lines Model}

To gain an understanding of the effect of the other modes $np$ ($|n|\geq 2$) on the 
line model, we introduce the two lines model which is the Galerkin truncation: 
\[
\{ (k_1,k_2), \quad |k_2|\leq 1 \}.
\]
We also work in the invariant subspace where $\om_k$'s are real-valued. 

For the two lines model, one can derive the governing equations in the 
physical variables. Let 
\[
\Om = \om (t, x) + e^{i\al y} q(t,x) + e^{-i\al y} \bq (t,x) \ ,
\]
where $\om$ is real-valued, $q$ is complex-valued (the Fourier transform of $q$ is 
real-valued), and 
\[
\int_0^{2\pi } \om (t, x) dx = 0 \ .
\]
Let
\[
f +b\tdl = \eta (t, x) + e^{i\al y}F(t,x) + e^{-i\al y} \bar{F}(t,x) \ ,
\]
where $\eta$ is real-valued, $F$ is complex-valued (the Fourier transform of $F$ is 
real-valued), and 
\[
\int_0^{2\pi } \eta (t, x) dx = 0 \ .
\]
Substituting the above expressions into the 2D NS (\ref{2DNS}), and 
ignoring the terms involving $e^{i2\al y}$ and $e^{-i2\al y}$, one gets
the two lines model in the physical variables,
\begin{eqnarray}
& & i\pa_t q + \al \left [ (\pa_x \om ) (\pa_x^2 -\al^2)^{-1} 
-(\pa_x^{-1}\om ) \right ] q = i \e \left [ (\pa_x^2 -\al^2) q + 
F \right ] \ ,  \label{me1} \\
& & \pa_t \om + i \al \pa_x \left [ q (\pa_x^2 -\al^2)^{-1} \bq - 
\bq (\pa_x^2 -\al^2)^{-1} q \right ] = \e \left [ \pa_x^2 \om + 
\eta \right ] \ . \label{me2} 
\end{eqnarray}
Introducing $\th = \pa_x^{-1}\om$, $\vphi = (\pa_x^2 -\al^2)^{-1} q$, 
and $h = \pa_x^{-1} \eta$, one gets
\begin{eqnarray}
& & i\pa_t q + \al (\vphi \pa^2_x \th - \th q ) = i \e \left [ 
(\pa_x^2 -\al^2) q + F \right ] \ ,  \label{eme1} \\
& & \pa_t \th + i \al ( q \bar{\vphi} - \bq \vphi ) = \e \left [ 
\pa_x^2 \th + h \right ] \ , \label{eme2} \\
& & \quad (\pa_x^2 -\al^2) \vphi = q \ . \label{eme3}
\end{eqnarray}
When $\e =0$, then Kinetic energy and enstrophy
\[
E_0 = \int_0^{2\pi} \left [\th^2 +2 \al^2 |\vphi|^2 +2|\pa_x \vphi |^2 
\right ] dx \ , \quad E_1 = \int_0^{2\pi} \left [ \om^2 + 2 |q|^2 
\right ] dx \ ,
\]
are still invariants. 

\begin{figure}
\includegraphics[width=4.0in,height=4.0in]{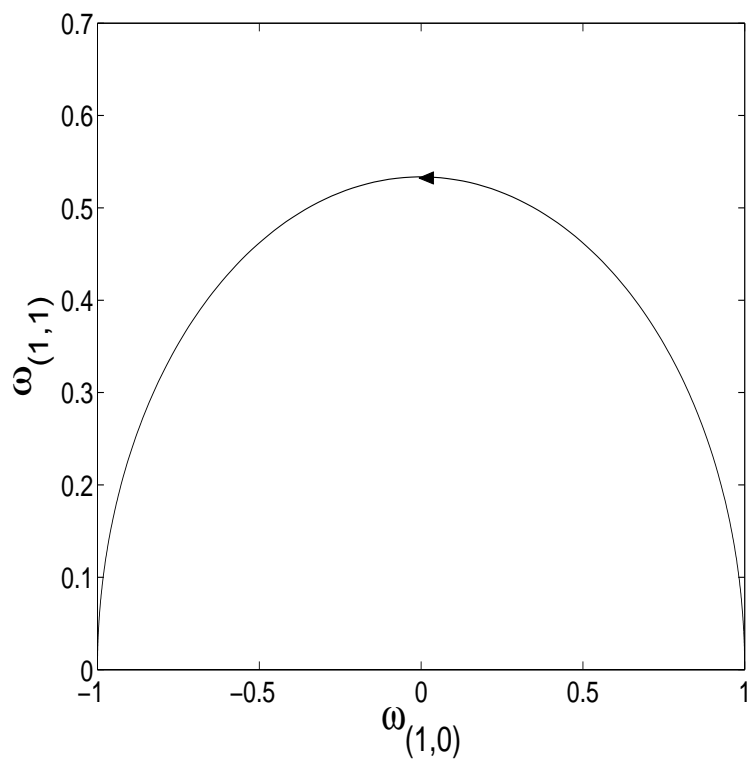}
\caption{Numerically exact heteroclinic orbit of the two lines model for $|k_1| \leq 2$.}
\label{2L2b}
\end{figure}

\begin{figure}
\includegraphics[width=4.0in,height=4.0in]{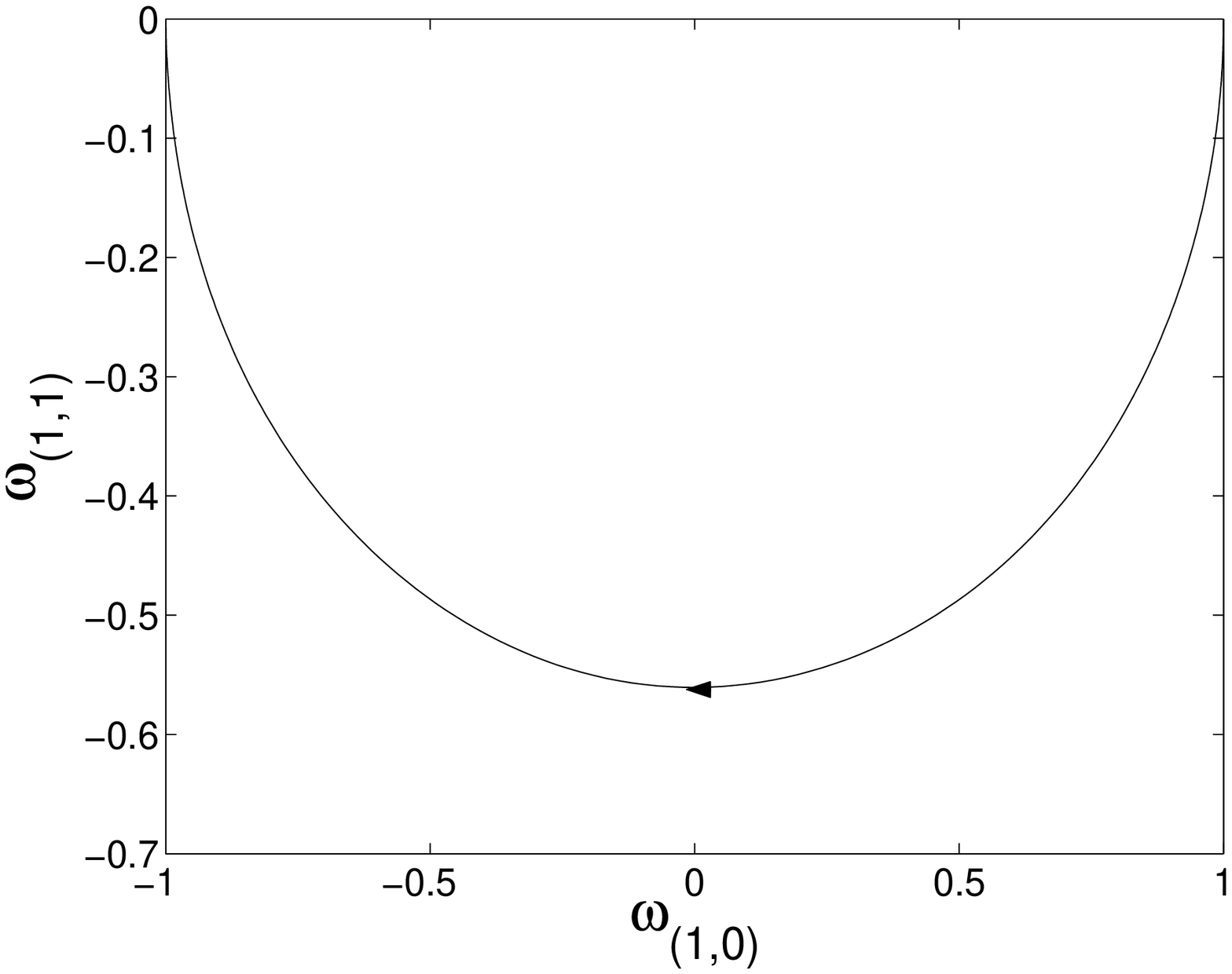}
\caption{Approximate heteroclinic orbit of the two lines model for $|k_1| \leq 4$.}
\label{2L4b}
\end{figure}

\begin{figure}
\includegraphics[width=4.0in,height=4.0in]{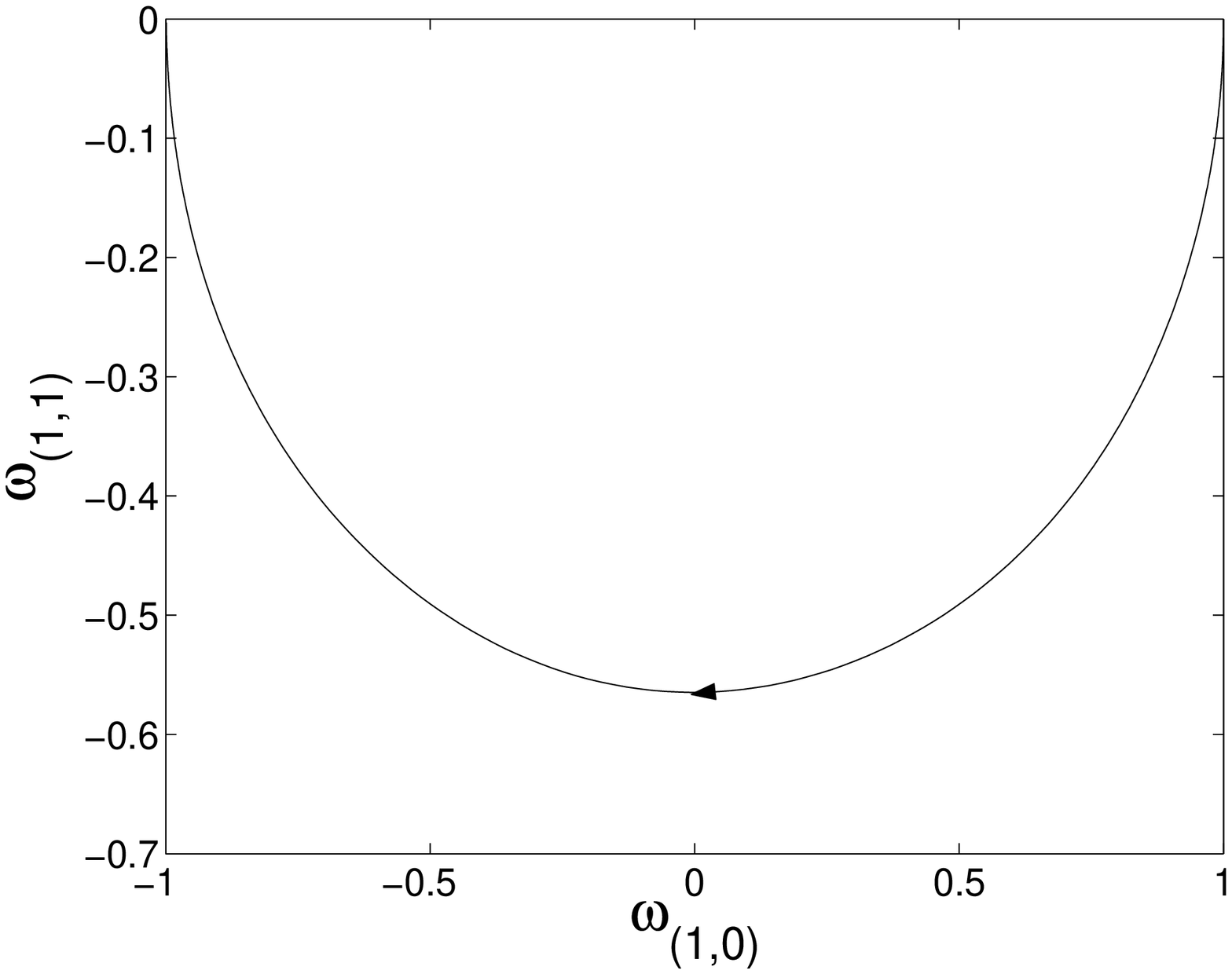}
\caption{Approximate heteroclinic orbit of the two lines model for $|k_1| \leq 16$.}
\label{2L16b}
\end{figure}

Denote by $\Sg$ the hyperplane
\[
\Sg = \left \{ \om \ | \ \om_k = 0 \ , \quad \text{ whenever } k_2=0 \right \} \ .
\]
For the two lines model, when $\e =0$, existence of a local unstable manifold 
for the fixed point $\Om = 2 \cos x_1$ is an open problem due to the fact that the 
coefficients of the nonlinear terms are not bounded, i.e. the nonlinear terms are 
not quadratic in a Banach algebra. Then we have the corollary of Theorem \ref{hcthm}.
\begin{corollary}
Assume that the fixed point $\Om = 2 \cos x_1$ has a 1-dimensional 
local unstable manifold $W^u$, 
and $W^u \cap \Sg \neq \emptyset$; then the heteroclinics conjecture is true, i.e.  
there is a heteroclinic orbit that connects $\Om = 2 \cos x_1$ and $-\Om$.
\end{corollary}
For any truncation ($|k_1| \leq N$) of the two lines model, we first calculate the 
unstable eigenvector. Then we track the heteroclinic orbit with the 
initial condition provided by the unstable eigenvector.
For $|k_1| \leq 2$, numerically exact heteroclinic orbit is obtained. That is,
it can verified numerically that 
\[
W^u \cap \Sg \neq \emptyset \ .
\]
Figure \ref{2L2b} shows the numerically exact heteroclinic orbit. For $|k_1| \leq 4$,
\[
\text{Distance } (W^u, \Sg ) \approx 0.0086 \ .
\]
Figure \ref{2L4b} shows the approximate heteroclinic orbit. For $|k_1| \leq 16$,
\[
\text{Distance } (W^u, \Sg ) \approx 0.012 \ .
\]
Figure \ref{2L16b} shows the corresponding approximate heteroclinic orbit.
Unlike the line model, here we do not always get numerically exact heteroclinic orbits. 
This is due to the influence of the modes ($k_1,0$).

\section{Melnikov Integral and Numerical Simulation of Chaos in the Two Lines Model}

When $\e =0$, there is very weak chaos for the computational interval 
$t \in [0, 4\times 10^4 \pi ]$:
\[
\begin{array}{ccc} |n| \leq 4 & |n| \leq 8 & |n| \leq 16 \cr
\sg = 9.5 \times 10^{-4} & \sg = 8.2 \times 10^{-4} & \sg = 8.5 \times 10^{-4} \ .\cr 
\end{array}
\]

For the two lines model, the kinetic energy and enstrophy are still invariants when $\e=0$. 
Choosing the same external force (\ref{EF}) and control (\ref{SC}), we have the Melnikov 
integral which is the same with that of 2D NS except that the Fourier modes summation is over 
the two lines model,
\begin{equation}
M_0 = M_1 + a \sqrt{M_2^2+M_3^2} \sin (t_0 +\th )+ bM_c\ .
\label{2Lmel}
\end{equation}
For the truncation $|k_1| \leq 16$, we evaluate these integrals along the heteroclinic orbit
in Figure \ref{2L16b}, and obtain that
\[
M_1 = -4.9\ , \quad M_2 =-0.0948\ , \quad  M_3 = 0 \ , \quad  M_c = 12.2498\ .
\]
For the case of no control ($b=0$), when 
\[
|a| > 51.688 \ , 
\]
the Melnikov integral $M$ has roots. We conduct some numerical simulations on 
the (transient) chaos. When $\e = 10^{-3}$, and $T=2\times 10^3 \pi$, we find that 
\[
\begin{array}{ll}  a \leq 200 & a=400 \cr 
\sg < 0 & \sg = 1.27 \times 10^{-2}  \ .\cr 
\end{array}
\]
According to the roots of the Melnikov integral, when $|a| > 51.688$, the broken
heteroclinic orbit may re-intersect with certain center-stable manifold. According 
the above Liapunov exponent result, strong transient chaos is generated when $a=400$ 
which may be due to the generation of new heteroclinic cycles.
Thus the result is almost the same as that of the line model.
As can be expected, the Melnikov prediction also performs well here for the two lines model.
When $\e =0$, there is weak transient chaos. 
When $\e > 0$, we see that strong transient chaos 
appears when $|a|$ is greater than certain threshold which is in the range 
$|a| > 51.688$ predicted by the Melnikov integral. In comparison with the line model,
here the transient chaos seems very strong. On the other hand, when $\e =0$, the two 
lines model here still has weak chaos, in contrast to the fact of no chaos at all for 
the line model. 

\section{Melnikov Integral and Control of Chaos in the Two Lines Model}

Now we turn on the control ($b \neq 0$). When 
\[
b=-M_1 /M_c = 0.4 \ ,  
\]
the Melnikov integral $M$ has roots for any $a$. To test the effectiveness of the 
control, we set $b$ to the above value and conduct some numerical simulations on 
the (transient) chaos. When $\e = 10^{-3}$, $b=0.4$, and $T=2\times 10^3 \pi$, 
we find that 
\[
\begin{array}{lllll}  a \leq 400 & a=500 & a=600 & a=700 & a=800 \cr 
\sg < 0 & \sg = 0.97 \times 10^{-3} & \sg = 3.0 \times 10^{-4} & \sg = 4.6 \times 10^{-2} 
& \sg = 5.0 \times 10^{-2}  \ .\cr 
\end{array}
\]
The control seems to tame the transient chaos, in contrast to the line model. The control 
pushed the threshold of $a$ forward from $400$ to $700$ for the generation of strong 
transient chaos.

\section{Conclusion and Discussion}

Through a combination of analytical and numerical studies, we now have a better 
understanding on the zero viscosity limit of the spectra of linear NS operators. 
We proposed and numerically studied the so-called heteroclinics conjecture for both 
2D and 3D Euler equations. We also proposed the Melnikov integral as a tool for 
predicting and controling chaos.

Our numerical verification on the heteroclinics conjecture was limited by 
resonable computing time. We realized that increasing the size of the Galerkin 
truncations can quickly reach the limit of our computer ability. We did not try to utilize 
today's supercomputer due to the fact that Galerkin truncations are essentially 
singular perturbations of Euler equations. We believe that analysis is the key to 
a better understanding of the heteroclinics conjecture.

We realized through our numerical simulations that Liapunov exponent performs very 
well as a measure of (even transient) chaos. Our Melnikov prediction for NS equations 
is of course not as rigorous and effective as that for sine-Gordon system. In fact,
it is a rough indicator for predicting and controling chaos. Nevertheless, we 
believe that the Melnikov integral theory for NS equations has a lot of potential 
especially in the current circumstance that there is no effective tools dealing 
with chaos in NS equations.

We also believe that both the line and the two lines models have great potential
in future analytical studies on modeling the dynamics of 2D NS equations.

\section{Appendix A: Melnikov Integral and Control of Chaos in a Sine-Gordon Equation}

Consider the sine-Gordon equation \cite{Li04c}
\begin{equation}
u_{tt}=\frac{9}{16} u_{xx} + \sin u +\e \left [ -a u_t + \left ( 1+b\tdl (x) \right )
\cos t \sin^3 u \right ]\ ,
\label{PSG}
\end{equation}
which is subject to periodic boundary condition and odd constraint
\begin{equation}
u(t, x+2\pi ) = u(t, x)\ , \quad u(t, -x) = - u(t, x)\ ,
\label{obc}
\end{equation}
where $u$ is a real-valued function of two real variables ($t,x$), $\e$ is a small 
perturbation parameter, $a> 0$ is the damping coefficient, $b \tdl (x)$ is the spatially 
localized control, $\tdl (x)$ is an even and $2\pi$-periodic function of $x$, and $b$ is 
the control parameter. The system (\ref{PSG}) is invariant under the transform $u \ra -u$.

The natural phase space for (\ref{PSG}) is $(u,u_t) \in H^{n+1}\times H^n$ ($n \geq 0$)
where $H^n$ is the Sobolev space on [$0,2\pi$]. Let $P$ be the Poincar\'e period-$2\pi$
map of (\ref{PSG}) in $H^{n+1}\times H^n$. Without the control ($b=0$), we have the 
following chaos theorem \cite{Li04c} \cite{Li05c}.
\begin{theorem}[\cite{Li04c}]
There exists a constant $a_0 >0$, when $\e$ is sufficiently small, for any $a \in \left [ 
\frac{1}{100} a_0, a_0 \right ]$ there exists a symmetric pair of homoclinic orbits 
$h_\pm$ ($h_-=-h_+$) asymptotic to $(u,u_t)=(0,0)$. In the neighborhood of $h_\pm$, there 
exists chaos to the sine-Gordon equation (\ref{PSG}) in the following sense: There is a 
Cantor set $\Xi$ of points in $H^{n+1}\times H^n$ ($n \geq 0$), which is invariant under 
an iterated Poincar\'e period-$2\pi$ $P^K$ for some $K$. The action of $P^K$ on $\Xi$ 
is topologically conjugate to the Bernoulli shift on two symbols $0$ and $1$. 
\label{CT}
\end{theorem}
In the product topology, the Bernoulli shift has the property of sensitive dependence 
upon initial data - the signature of chaos. 

When we turn on the control ($b \neq 0$), we hope to find values of $b$ such that the 
chaos in Theorem \ref{CT} is controlled (tamed - annihilated or less chaotic, 
enhanced - more chaotic). Our main tool is the Melnikov function. To build such a function,
we need results from integrable theory. When $\e =0$, the fixed point ($u=0$) of the 
sine-Gordon equation (\ref{PSG}) has a figure eight connecting to it \cite{Li04c}:
\begin{equation}
u = \pm 4 \arctan \left [ \frac{\sqrt{7}}{3} \text{ sech }\tau \sin x \right ] \ , 
\label{f8}
\end{equation}
where $\tau = \frac{\sqrt{7}}{4} (t-t_0)$ and $t_0$ is a real parameter. Along this 
figure eight, a Melnikov vector has the expression \cite{Li04c}:
\begin{equation}
\frac{\pa F_1}{\pa u_t} = \pm \frac{7\pi}{12\sqrt{2}} \text{ sech }\tau \tanh \tau 
\sin x \left [ \frac{9}{16}+\frac{7}{16}\text{ sech}^2\ \tau \sin^2 x \right ]^{-1}\ ,
\label{MV}
\end{equation}
where $F_1$ is a constant of motion. When $\e \neq 0$, the Melnikov function for 
(\ref{PSG}) is given by \cite{Li04c}:
\[
M(t_0,a,b) = \int_{-\infty}^{+\infty} \int_0^{2\pi} \frac{\pa F_1}{\pa u_t} 
\left [ -au_t + \left ( 1+b\tdl (x) \right )\cos t \sin^3 u \right ] dx dt \ ,
\]
where $u$ and $\frac{\pa F_1}{\pa u_t}$ are given in (\ref{f8}) and (\ref{MV}). 
Using the odd and even property of (\ref{f8}) and (\ref{MV}) in $t$ and $x$, we 
obtain
\begin{equation}
M(t_0,a,b) = -a M_a +\sin t_0 \ (M_0 +bM_b)\ ,
\label{MF}
\end{equation}
where 
\begin{eqnarray*}
M_a &=& \int_{-\infty}^{+\infty} \int_0^{2\pi} \frac{\pa F_1}{\pa u_t} u_t \ dx dt \ , \\
M_0 &=& -\int_{-\infty}^{+\infty} \int_0^{2\pi} \frac{\pa F_1}{\pa u_t}\sin 
\frac{4}{\sqrt{7}}\tau \ \sin^3 u \ dx dt \ , \\
M_b &=& -\int_{-\infty}^{+\infty} \int_0^{2\pi} \frac{\pa F_1}{\pa u_t}\tdl (x) \sin 
\frac{4}{\sqrt{7}}\tau \ \sin^3 u \ dx dt \ .
\end{eqnarray*}
In the phase space $H^{n+1}\times H^n$ ($n \geq 0$), $(u,u_t)=(0,0)$ is a saddle 
point under the Poincar\'e period-$2\pi$ map of (\ref{PSG}) with one-dimensional 
unstable manifold $W^u$ and one-codimensional stable manifold $W^s$. The Melnikov function
$\e M(t_0,a,b)$ is the leading order term of the distance between $W^u$ and $W^s$. For the 
entire rigorous theory, see \cite{Li04}. When $|aM_a| < |M_0+bM_b|$, the roots of $M$ 
are given by
\begin{equation}
\sin t_0 = \frac{a M_a}{M_0 +bM_b} \ .
\label{rts}
\end{equation}
Near these roots, $W^u$ and $W^s$ intersect. This leads to the existence of a symmetric 
pair of homoclinic orbits and chaos in Theorem \ref{CT}. When $|aM_a| > |M_0+bM_b|$, 
i.e.
\begin{equation}
-a|M_a| - M_0 <bM_b< a|M_a|- M_0 \ ,
\label{T1}
\end{equation}
the Melnikov function is not zero, and we have the following theorem.
\begin{theorem}
When the control parameter $b$ satisfies (\ref{T1}), the chaos in Theorem \ref{CT} 
disappears.
\label{TT}
\end{theorem}
\begin{proof}
When the  control parameter $b$ satisfies (\ref{T1}), the Melnikov function is not zero 
for any $t_0$, and $W^u$ and $W^s$ do not intersect. Thus the pair of homoclinic orbits
and the corresponding chaos in Theorem \ref{CT} disappear.
\end{proof}
Theorem \ref{TT} only claims that the chaos in Theorem \ref{CT} disappears. This 
does not mean that there is no chaos in the entire phase space $H^{n+1}\times H^n$ 
($n \geq 0$). An important point here is that by manipulating the localized control 
$b \tdl (x)$, one can change the Melnikov function which leads to the disappearance of 
the non-localized (in $x$) chaos. The control condition (\ref{T1}) is also interesting: 
It is not true that the larger the control parameter $b$ is, the better the taming is. 
In fact, when $b$ is large enough, the chaos will reappear.

When $|aM_a| < |M_0+bM_b|$, the Melnikov function (\ref{MF}) has roots (\ref{rts}), and 
Theorem \ref{CT} holds. As a function of $t_0$, the Melnikov function $M$ has the maximal 
absolute value (the $L^\infty$ norm),
\[
M_*(a,b) = a|M_a|+|M_0+bM_b|\ ,
\]
for $t_0 \in [0,2\pi ]$. $M_*$ is the leading order term of the maximal distance between 
$W^u$ and $W^s$. Notice that $W^u$ and $W^s$ intersects near the $t_0$ given by (\ref{rts}).
So the larger the $M_*$ is, the more violent the chaos is. Thus $M_*(a,b)$ serves as a 
measure of the strength of the chaos. By changing the control parameter $b$, we can adjust 
the strength $M_*$ of the chaos - enhancing or decreasing.

\section{Appendix B: The Lagrange Flow Induced by a Solution to the 2D Euler Equation 
Is Always Integrable}

It is well-known that 2D Euler equation is globally well-posed \cite{Kat75} \cite{Kat86}.
For any solution to the 2D Euler equation, let $\Psi = \Psi (t,x_1,x_2)$ be the 
corresponding stream function. Then the Lagrange flow induced by the solution is given by
\begin{equation}
\frac{dx_1}{dt} = - \frac{\pa \Psi}{\pa x_2} \ , \quad
\frac{dx_2}{dt} =  \frac{\pa \Psi}{\pa x_1} \ .
\label{LF}
\end{equation}
\begin{theorem}
The Lagrange flow (\ref{LF}) induced by a solution to the 2D Euler equation is always 
integrable.
\end{theorem}
\begin{proof}
Assume that $\Psi (t,x_1,x_2)$ is not a steady state, i.e. 
it depends upon $t$ (in this case $\Dl \Psi$ is functionally independent of 
$\Psi$, otherwise, $\Psi (t,x_1,x_2)$ would be a steady state).
Introducing the new Hamiltonian $H =\Psi (\th , x_1,x_2)-\psi$, 
and converting (\ref{LF}) into an autonomous system
\begin{equation}
\frac{dx_1}{dt} = - \frac{\pa H}{\pa x_2} \ , \quad
\frac{dx_2}{dt} =  \frac{\pa H}{\pa x_1} \ , \quad 
\frac{d\th}{dt} = - \frac{\pa H}{\pa \psi} \ , \quad
\frac{d\psi}{dt} =  \frac{\pa H}{\pa \th} \ . \quad 
\label{LF1}
\end{equation}
Notice that the vorticity $\Om = \Dl \Psi$ is another constant of motion of (\ref{LF1}) 
besides $H$:
\[
\frac{d}{dt} \Dl \Psi = \pa_\th \Dl \Psi -\pa_{x_1}\Dl \Psi  \pa_{x_2}\Psi 
+ \pa_{x_2}\Dl \Psi  \pa_{x_1}\Psi = 0\ .
\] 
Since $\Dl \Psi$ is independent of $\psi$, $\Dl \Psi$ and $H$ are 
functionally independent. Thus (\ref{LF1}) is integrable in the Liouville sense. 
In the case that $\Psi$ is independent of $t$ (i.e. a steady state), then 
(\ref{LF}) is an autonomous system, thus also integrable in the Liouville sense.
\end{proof}
A common way to obtain steady states of 2D Euler equation is by solving 
\[
\Dl \Psi = f(\Psi )
\]
where $f(\Psi )$ is an arbitrary function of $\Psi$, i.e. $\Dl \Psi$ and $\Psi$ are 
functionally dependent.

\end{document}